\documentclass{article}
\usepackage{arxiv}
\usepackage{authblk}
{\end{list}}

\def\E{{\rm E}}

\def\bi{{\bf i}}

\def\bm{{\bf m}}

\def\bX{{\bf X}}
\def\bx{{\bf x}}
\def\bY{{\bf Y}}

\def\bz{{\bf z}}

\def\bbeta{\mbox{\boldmath $\beta$}}

\def\bmu{\mbox{\boldmath $\mu$}}

\def\b1{{\bf 1}}

\def\blot{\quad {$\vcenter{\vbox{\hrule height.4pt
             \hbox{\vrule width.4pt height.9ex \kern.9ex \vrule 
width.4pt}
             \hrule height.4pt}}$}}
\usepackage[ruled, vlined]{algorithm2e}
\usepackage{amsmath}
\usepackage{amsfonts}
\usepackage{amssymb}
\usepackage{amsbsy}
\usepackage{amsthm}
\usepackage{comment}
\usepackage{multirow}
\theoremstyle{plain}
\newtheorem{theorem}{Theorem}[section]

\theoremstyle{definition}

\theoremstyle{definition}

\theoremstyle{remark}

\usepackage{tikz}
\newcommand*\circled[1]{\tikz[baseline=(char.base)]{
            \node[shape=circle,draw,inner sep=2pt] (char) {#1};}}
\allowdisplaybreaks

\begin{document}
\title{\Large Stochastic Simulation Uncertainty Analysis to Accelerate Flexible Biomanufacturing Process Development}

\author[  ~,1]{Wei Xie\thanks{Corresponding author: w.xie@northeastern.edu}}
\author[2]{Russell R. Barton}
\author[3]{Barry L. Nelson}
\author[1]{Keqi Wang}

\begin{center}
\affil[1]{Department of Mechanical and Industrial Engineering, Northeastern University, 
MA, 
USA}
\affil[2]{Supply Chain and Information Systems, Smeal College of Business, 
The Pennsylvania State  University, 
PA,  
USA} 
\affil[3]{Department of Industrial Engineering and Management Sciences, 
IL,  
USA} 
\end{center}




\maketitle
\begin{abstract}
Motivated by critical challenges and needs from biopharmaceuticals manufacturing, we propose a general metamodel-assisted stochastic simulation uncertainty analysis framework to accelerate the development of a simulation model with modular design for flexible production processes. There are often very limited process observations. 
Thus, there exist both simulation and model uncertainties in the system performance estimates. In biopharmaceutical manufacturing, model uncertainty often dominates.
The proposed framework can produce a confidence interval that accounts for simulation and model uncertainties by using a metamodel-assisted bootstrapping approach. Furthermore, a variance decomposition is utilized to estimate the relative contributions from each source of model uncertainty, as well as simulation uncertainty. This information can be used to improve the system mean performance estimation.
Asymptotic analysis provides theoretical support for our approach, while the empirical study demonstrates that it has good finite-sample
performance.

\end{abstract}
 

\section{Introduction}
\label{sec:introduction}

While the biopharmaceutical industry has developed various innovative 
bio-drugs for severe diseases, such as 
cancers, autoimmune disorders, and infectious diseases, the current manufacturing systems are unable to rapidly produce new and existing drugs when needed, largely due to 
critical challenges, including 
{high complexity, high variability, and very limited process data}. 
Biotherapeutics are manufactured in living organisms (e.g., cells) whose biological processes are very complex. Manufacturing process typically consists of multiple integrated unit operations.  
There is often very limited data, i.e., having 3--20 process observations is typical in biomanufacturing \cite{OBrien_2021}, reflecting the high cost and long time needed to run lab experiments. Also, the more personalized nature of emerging bio-drugs (e.g., cell and gene therapies) 
makes it difficult to collect extensive data on every possible variety of drugs
and every protein therapy can be unique, which often forces R\&D efforts to work with just 3--5 batches.



\begin{sloppypar}
Simulation 
can facilitate the development of flexible production systems with modular design. 
\textit{Hybrid (``mechanistic+statistical") simulation models} can support interpretable and robust decision making, while requiring much less data than purely data-based models.
The mechanistic model parameters (such as cell growth rate, oxygen and nutrient uptake rates) 
can facilitate the learning of underlying biological/physical/chemical (a.k.a. \textit{biophysicochemical}) mechanisms.
Thus, in this paper, we suppose that the model family or structure, 
built on mechanism prior knowledge, is given.
The model parameters are estimated from very limited real-world data, which introduces \textit{model uncertainty.} 
When we create a simulation model to predict the performance of a real system, there exist the errors induced by both simulation estimation uncertainty and process model uncertainty. 
\end{sloppypar}


\begin{sloppypar}
In the biomanufacturing literature, modeling of bioprocess dynamics while considering different sources of uncertainty (e.g., batch-to-batch variations, measurement errors, and model uncertainty)
is critical \cite{rodriguez2020digital}. 
Model uncertainty quantification can be divided into frequentist and Bayesian approaches. 
In frequentist inference, model parameter estimation uncertainty is typically quantified via a confidence interval or standard deviation \cite{moller2020model,wang2019stochastic}. In Bayesian inference, 
posterior distributions
are used to quantify and update model uncertainty \cite{hernandez2019predicting, xie2022interpretable}. 

This study is directly related to the existing frequentist and Bayesian approaches on uncertainty quantification and sensitivity analysis; see  recent reviews in \cite{corlu2020stochastic,borgonovo2016sensitivity}. 
The Bayesian approaches typically use the posterior distributions of inputs given the real-world data to quantify the input distribution uncertainty; see for example 
\cite{Zouaoui_Wilson_2003, Zouaoui_Wilson_2004, biller2011accounting}. 
Direct bootstrapping, as frequentist approach, quantifies the impact of input uncertainty using bootstrap resampling of the input data and runs simulations at each bootstrap resample point to estimate the impact on the system mean \cite{barton2001resampling, barton2007presenting}. Compared with the Bayesian approaches, the direct bootstrap can be adapted to any input process without additional analysis (e.g., posterior distribution derivation).
The metamodel-assisted bootstrapping approach is further introduced by \cite{barton_nelson_xie_2011}. In this framework, the uncertainty is propagated to the output mean by a metamodel, which can be constructed using simulation results from a small number of runs. Thus, this method does not need substantial computational effort.

\end{sloppypar}


Built on \cite{barton_nelson_xie_2011}, \textit{we propose a metamodel-assisted uncertainty quantification and sensitivity analysis
(UQ\&SA)  framework 
to accelerate the development of 
flexible manufacturing process with modular design.}
As a result we can form 
a confidence interval (CI) quantifying the overall estimation uncertainty of the system's mean performance. Specifically, bootstrap resampling of the real-world data is used to approximate the model uncertainty. Then, a Gaussian process (GP) metamodel is used to propagate the heterogeneous process model uncertainty to the output mean response.
Since model uncertainty typically dominates in the biopharmaceutical manufacturing processes, we further develop sensitivity analysis to quantify the contribution from each source of model uncertainty.
The key contributions of this study are threefold.
\begin{itemize}
    \item First, we introduce a metamodel-assisted uncertainty quantification (UQ) and sensitivity analysis (SA) framework for hybrid model based simulations. The proposed algorithm can delivery a percentile CI 
    of system mean response, accounting for both model and simulation uncertainties. A further sensitivity analysis can provide the relative contribution from each source of uncertainty.
Differing with existing simulation studies in the literature that typically consider the simulation model as a black-box (see for example the review paper \cite{corlu2020stochastic}), hybrid model based simulation can leverage existing mechanistic models, facilitate mechanism learning, and support interpretable decision making. 
\item Second, 
under the assumption that the unknown mean response surface is a realization of GP
, which is a useful representation in many problems,
we provide a systematic asymptotic analysis on the proposed GP metamodel assisted UQ and SA framework, including
(1) the asymptotic consistency of the proposed CI; and 
(2) the asymptotic consistency of variance estimators quantifying each source of model uncertainty and simulation uncertainty.
\item Third, we provide a comprehensive empirical study to show that the proposed framework has promising finite sample performance, especially under  situations with very limited real-world data.
\end{itemize}

Some existing simulation methodologies can be integrated into the proposed framework to support 
extensions for computational saving and system risk performance assessment, such as measured by quantiles. Considering the total simulation cost required to achieve consistent estimation of model uncertainty when using the conventional bootstrap resampling techniques, \cite{lam2018subsampling,lam2022subsampling} proposed the subsampling techniques
as a computational saver to promote the computational efficiency. 
In addition, the proposed UQ and SA framework can be extended to system quantile performance measure through GP based percentile regression; see for example \cite{zhang2022pooled, xie2018metamodel,xie2017factor}.




The remainder of the paper is organized as follows.
We present the problem description in Section~\ref{sec:problem_Description} and give a brief review of the metamodel-assisted bootstrapping approach in Section~\ref{sec:MMABS}. In Section~\ref{Sec:varDecomp}, we provide an algorithm to build an interval quantifying the overall estimation uncertainty of system mean performance, accounting for both model and simulation uncertainties. Then, we provide a variance decomposition approach to estimate the relative contribution from each source of model uncertainty, as well as simulation uncertainty. We provide an empirical study in Section~\ref{sec:empirical}   
and conclude the paper in Section~\ref{sec:conclusion}. 
All proofs are provided in the 
Appendix.

\section{Problem Description and Proposed Framework}
\label{sec:problem_Description}

\begin{figure*}[h!]
	\centering
	\includegraphics[width=0.9\textwidth]{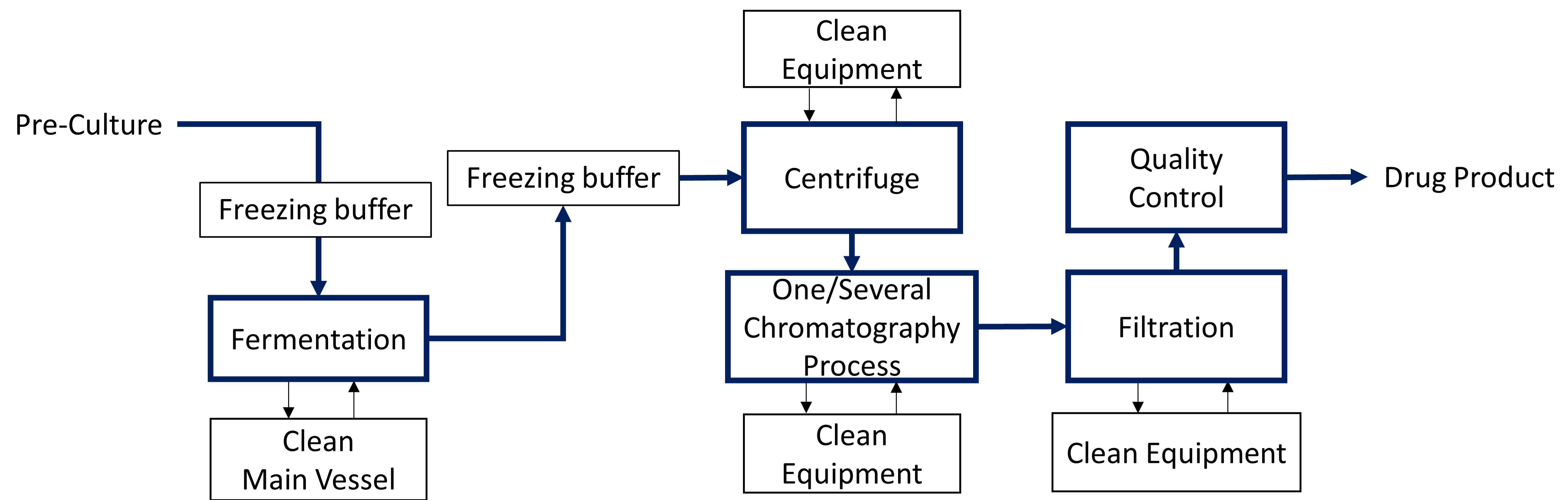}
	\caption{An illustrative example of integrated biomanufacturing process.} \label{fig_api}
\end{figure*}

A typical biomanufacturing system consists of multiple unit operations, including upstream fermentation for drug substance production and downstream purification to meet quality requirements \cite{doran2013bioprocess}. 
It can consist of numerous unit operations; see an example illustrated in Figure~\ref{fig_api}. Operations typically include (1) fermentation, (2) centrifugation, (3) chromatography, (4) filtration, and (5) quality control.
Operation unit (1) belongs to upstream cell culture and target drug substance production process, and (2)--(5) belong to downstream purification process.

To guide reliable and interpretable decision making, a simulation model can be developed based on hybrid models 
of modules defined according to bioprocess biophysicochemical mechanisms, dynamics, and interdependence of mechanistic parameters.
Given very limited real-world data, we take existing mechanistic models 
as prior knowledge on the structure of mechanism relationships
and create \textit{parametric hybrid models}. It can leverage the advantages of mechanistic and statistical models to facilitate mechanism learning and improve sample efficiency and decision interpretability.

The fermentation is the most critical operation unit in the production process and it determines the generation of target drug substance (such as protein monoclonal antibodies or mAbs) and impurities. 
Here we use a simple example of fermentation on protein production to illustrate bioprocess hybrid modeling. 
Specifically, the target protein and biomass generation in the exponential-growth phase of fermentation process can be modeled with the {cell-growth kinetics} mechanism \cite{doran2013bioprocess}.
Built on it, we construct a hybrid model capturing bioprocess dynamics and variation, i.e.,
\begin{equation}
   X_t = X_0 \cdot e^{\gamma t} + \epsilon_P,
    \label{eq.hybridFermentation}
\end{equation}
where 
$X_{t}$ represents the biomass concentration at time $t$ and the growth rate, denoted by $\gamma$, depends on biological properties of working cells and culture environments. We model batch-to-batch variation on: 
(1) the specific growth rate as $\gamma \sim N(\mu_\gamma,\sigma_\gamma^2)$;
and (2) raw materials or initial concentration of seed cells as $X_0\sim N(\mu_0,\sigma^2_0)$.
In addition, we model the measurement error or residual as 
$\epsilon_P \sim N(0,\sigma^2_P)$ to capture the integrated impact from ignored factors.
Larger variance from the residual indicates less understanding on underlying bioprocessing mechanisms obtained from the existing exponential growth mechanistic model. 
Thus, the distributions of residual 
$\epsilon_P$, protein growth rate $\gamma_P$, and raw materials $X_0$ uniquely characterize the hybrid model of target protein accumulation during the fermentation process.

The normality assumption is often used in the biopharmaceutical literature to model batch-to-batch variations and measurement errors since they are often induced by many underlying factors; see for example \cite{mockus2015batch}. 
In addition, there is often very limited data. 
In our previous study, we used real-world fermentation process data with 
the size of 8 batches to conduct the hypothesis test which validates the normality assumption \cite{xie2022interpretable}.

An integrated biomanufacturing system is often composed of multiple interconnected modules. 
Suppose that the simulation model is a function of $L$ parametric multivariate and univariate models 
$F\equiv\{F_1,F_2,\ldots,F_L\}$ 
characterizing the underlying bioprocess dynamics and variations.
Each $\ell$-th model $F_\ell$ can be uniquely characterized by $h_\ell$ unknown parameters.
In the simple fermentation example mentioned above in (\ref{eq.hybridFermentation}), the variation of residual $\epsilon_P$ is characterized by model $F_1$ specified by  parameter $\sigma^2_P$; the batch-to-batch variation on the growth rate $\gamma$ is characterized by model $F_2$ specified by parameters $\{\mu_\gamma, \sigma^2_\gamma\}$ and the raw material uncertainty is characterized by model $F_3$ specified by parameters $\{\mu_0, \sigma^2_0\}$.

Each $h_\ell$-parameter distribution is uniquely specified by its first (finite) $h_\ell$ moments, which is true for the distributions that are most often used in stochastic simulation. 
The moments are chosen as the input
variables for the metamodel of the system response surface because when they are close, the corresponding distributions will be similar and therefore generate similar outputs. 
Let $\mathbf{x}_{[\ell]}$ denote an $h_\ell\times 1$ vector of the first $h_\ell$ moments for the $\ell$-th model and $d=\sum_{\ell=1}^L h_\ell$. 
Then, by stacking $\bx_{[\ell]}$ with $\ell=1,2,\ldots,L$ together, we have a $d\times 1$ dimensional input vector, denoted by
$\bx$. Notice that $F=\{F_1,F_2,\ldots,F_L\}$ is completely characterized by the collection of model moments $\bx=(\mathbf{x}_{[1]}, \mathbf{x}_{[2]}, \ldots, \mathbf{x}_{[L]})^\top$. 

\begin{sloppypar}
The output from the $j$-th replication of a simulation with model moments $\mathbf{x}$ can be written as 
\begin{equation}
Y_j(\mathbf{x})=\mu(\mathbf{x})+\epsilon_j(\mathbf{x})
\label{eq.Y_x}
\end{equation}
where $\mu(\mathbf{x}) = \E[Y_j(\mathbf{x})]$ denotes the unknown expected performance (e.g., productivity of protein drug substance) 
and $\epsilon_j(\mathbf{x})$ represents the simulation error with mean zero. The simulation output depends on the choice of process models.
Let $\Psi \equiv\{\bx\in\Re^d{:} \mbox{ the 
random variable } Y(\bx) \mbox{ is defined and $\mu(\bx)$ is
finite}\}$ denote the region of interest. We assume $\mu(\bx)$ is continuous for $\bx\in\Psi$.
\end{sloppypar}

\begin{sloppypar}
The underlying ``correct'' process models, denoted by $F^c\equiv\{F^c_1,F^c_2,\ldots,F^c_L\}$, specified by the moments, $\bx_c=(\mathbf{x}_{[1],c}, \mathbf{x}_{[2],c}, \ldots, \mathbf{x}_{[L],c})$, are unknown and are estimated from a finite sample of real-world data.
Suppose that the set of true parameters $\bx_c$ is in the interior of $\Psi$. Our goal is to find
a $(1-\alpha)100\%$ 
CI, denoted by $[Q_L,Q_U]$, such that 
\begin{equation}
\mbox{Pr}\{\mu(\mathbf{x}_c)\in [Q_L,Q_U]\} = 1-\alpha,
\label{eq.CI_1}
\end{equation}
which quantifies the overall estimation uncertainty of system mean performance, accounting for simulation and model uncertainties.
Then, 
if this interval is too wide,
we further develop a variance decomposition to quantify the contribution from each source of model uncertainty, which can guide more data collection and improve the system mean response estimation. 
\end{sloppypar}

The true moments $\bx_c$ are unknown and estimated based on a finite sample $\mathbf{Z}_{\mathbf{m}}$ from $F^c$. Let $m_\ell$ denote the number of i.i.d.\ real-world observations available from the $\ell$-th model, i.e., $\mathbf{Z}_{\ell,m_\ell}\equiv\left\{Z_{\ell,1},Z_{\ell,2},\ldots,Z_{\ell,m_\ell} \right\}$ with $Z_{\ell,i}\stackrel{i.i.d}\sim F^c_\ell$, $i=1,2,\ldots,m_\ell$. 
Let $\mathbf{Z}_{\mathbf{m}}=\{\mathbf{Z}_{\ell,m_\ell}, \ell=1,2,\ldots,L\}$ be the collection of samples from all $L$ model distributions in $F^c$, where $\mathbf{m}=(m_1,m_2,\ldots,m_L)$. 
Let $\mathbf{X}_{\mathbf{m}}$ be a $d\times 1$ dimensional moment estimator that is a function of $\mathbf{Z}_{\mathbf{m}}$ written as
$\mathbf{X}_{\mathbf{m}}=\mathbf{X}(\mathbf{Z}_{\mathbf{m}})$.
Specifically, $\mathbf{X}_{\ell,m_\ell}=\mathbf{X}_\ell(\mathbf{Z}_{\ell,m_\ell})$
and
$\mathbf{X}_{\mathbf{m}}^T=(\mathbf{X}_{1,m_1}^T,\mathbf{X}_{2,m_2}^T,\ldots,
\mathbf{X}_{L,m_L}^T)$. Let $F_{\mathbf{X}_{\mathbf{m}}}^c$ represent the true, unknown distribution of $\mathbf{X}_{\mathbf{m}}$. 
Therefore, the impact of model uncertainty is captured by the sampling
distribution of $\mu(\mathbf{X}_{\mathbf{m}})$ with $\mathbf{X}_{\mathbf{m}}\sim
F_{\mathbf{X}_{\mathbf{m}}}^c$. The real-world data are a particular realization of $\mathbf{Z}_{\mathbf{m}}$, say $\mathbf{z}_{\mathbf{m}}^{(0)}$.
Given a finite sample of real-world data $\mathbf{z}_{\mathbf{m}}^{(0)}$, we use bootstrap resampling to approximate $F_{\mathbf{X}_{\mathbf{m}}}^c$ and a metamodel to represent $\mu(\bx)$.
Notice that the components of the
moment estimator $\mathbf{X}_{\mathbf{m}}$ can be
statistically dependent.


Suppose each experiment is 
expensive. The proposed metamodel-assisted bootstrapping uncertainty analysis framework can accelerate the development of a simulation model 
for a flexible and integrated real manufacturing system with modular design.
Since the underlying response surface $\mu(\cdot)$ is unknown, we model our prior belief about $\mu(\cdot)$ by a Gaussian Process (GP). Given a set of stochastic simulation outputs, the GP-based belief is updated by a posterior distribution, 
denoted by $M_p(\cdot)$. When we use this metamodel to propagate the sampling distribution of $\mathbf{X}_{\mathbf{m}}$ to the output mean, it introduces the {simulation uncertainty} induced by finite simulation runs (i.e., finite design points and finite run length in each simulation run). \textit{Thus, the estimation uncertainty of underlying system mean performance $\mu(\mathbf{x}_c)$ is characterized by the compound random variable, $M_p(\mathbf{X}_{\mathbf{m}})$, accounting for both model and simulation uncertainties.} Based on the variability of $M_p(\mathbf{X}_{\mathbf{m}})$, we can construct an interval estimator $[Q_L,Q_U]$ in 
(\ref{eq.CI_1}) to quantify the overall estimation uncertainty of real system mean response $\mu(\mathbf{x}_c)$.

\begin{sloppypar}
 {We further develop a variance decomposition measuring the contributions to $\mbox{Var}[M_p(\mathbf{X}_{\mathbf{m}})]$ from simulation uncertainty quantified by GP
$M_p(\cdot)$
and model uncertainty quantified by the sampling distribution of $\mathbf{X}_{\mathbf{m}}^T=(\mathbf{X}_{1,m_1}^T,\mathbf{X}_{2,m_2}^T,\ldots, \mathbf{X}_{L,m_L}^T)$.} 
Therefore, if this interval is too wide, our study can guide further data collection 
to efficiently update the simulation model to faithfully represent the real system and improve the estimation accuracy of $\mu(\mathbf{x}_c)$.
\end{sloppypar}

If the simulation uncertainty dominates, we will allocate more computational resource to improve our knowledge on the mean response surface $\mu(\cdot)$.
However, in biopharmaceutical manufacturing with high stochasticity and very limited process observations, model uncertainty often dominates.
The distribution of model uncertainty depends on heterogeneous process observations, 
as well as the complexity of the underlying mechanisms and inherent stochasticity at each part of the integrated biomanufacturing system. Thus, if certain model uncertainty, say $\mathbf{X}_{\ell,m_\ell}$ with $\ell=1,2,\ldots,L$, dominates the system performance estimation uncertainty, it will guide us collecting the additional real-world data 
there to improve the simulation model. 

\section{Metamodel-Assisted Bootstrapping for Uncertainty Quantification}
\label{sec:MMABS}

We introduce the metamodel-assisted bootstrapping and provide the algorithm for uncertainty analysis. 
Basically, we first find the space-filling design points covering the most likely bootstrap samples of model moments, denoted by 
$\widehat{\mathbf{X}}_{\mathbf{m}}^{(b)}
$ with $b=1,2,\ldots,B$, quantifying the model uncertainty. Then, we run simulations and construct the GP or stochastic kriging (SK)  metamodel for the mean response surface $\mu(\cdot)$ quantifying  simulation uncertainty in Section~\ref{subsec:GP}. This metamodel is used to propagate the model uncertainty to output mean. We introduce the metamodel-assisted bootstrapping in Section~\ref{subsec:MAB-UQ} to construct an interval of $\mu(\mathbf{x}_c)$ accounting for both simulation and model uncertainties, and show its asymptotic consistency in Section~\ref{subsec:CI_consistency}.


\subsection{Stochastic Kriging Metamodel}
\label{subsec:GP}

Since the outputs from simulations include simulation variability that often changes significantly across the design space of process models 
specified by moments $\mathbf{x}$, SK is introduced to distinguish the uncertainty about the response surface from the simulation uncertainty \cite{ankenman_nelson_staum_2010, kleijnen2017regression}.
Suppose that the underlying unknown 
response surface can be thought of as a realization of a stationary GP. 
The simulation output $Y$ is modeled as,
\begin{equation}
\label{eq:sk}
Y_j(\mathbf{x})=\mathbf{\beta}_0+W(\mathbf{x})+\epsilon_j(\mathbf{x})
\end{equation}
where $\bx$ denotes a $d\times 1$ vector of model moments.  
SK uses a mean-zero, second-order stationary GP $W(\bx)$ to account for the spatial dependence of the response surface. Thus, the uncertainty about the true response surface $\mu(\bx)$ is represented by a GP $M(\bx)\equiv\mathbf{\beta}_0+W(\mathbf{x})$ (note that $\beta_0$ can be replaced by a more general trend term $\mathbf{f}(\bx)^\top \bbeta$). 
For many, but not all, simulation settings the output is an average of a large number of more basic outputs, so a normal approximation can be applied: $\epsilon(\bx)\sim \mbox{N}(0,\sigma^2_{\epsilon}(\bx))$.  

In SK, the covariance between $W(\mathbf{x})$ and $W(\mathbf{x}^\prime)$ quantifies how knowledge of the surface at some design points affects the prediction of the surface. A parametric form of the spatial covariance, denoted by $\Sigma(\mathbf{x},\mathbf{x}^\prime)=\mbox{Cov}[W(\mathbf{x}),W(\mathbf{x}^\prime)] 
=\tau^2r(\mathbf{x}-\mathbf{x}^\prime)$, is typically assumed where $\tau^2$ denotes the variance and $r(\cdot)$ is a correlation function that depends only on the distance $\mathbf{x}-\mathbf{x}^\prime$. 
Based on our previous study \cite{xie_nelson_staum_2010}, we use the
product-form Gaussian correlation function 
$  r(\mathbf{x}-\mathbf{x}^\prime)=
\exp (-\sum_{j=1}^d \theta_j(x_j-x^\prime_j)^2 )
$
for the empirical evaluation in Section~\ref{sec:empirical}.
Let $\pmb{\theta}=(\theta_1,\theta_2,\ldots,\theta_d)$ represent the correlation parameters. 
Thus, the prior knowledge of the response surface $\mu(\mathbf{x})$ is represented by a Gaussian process, i.e., $ M(\bx)\sim \mathrm{GP}(\mathbf{\beta}_0,\tau^2 r(\mathbf{x}-\mathbf{x}^\prime)).$

\begin{sloppypar}
To reduce the uncertainty about $\mu(\bx)$, we choose an experiment design consisting of pairs $\mathcal{D} \equiv\{(\bx_i,n_i),i=1,2,\ldots,k\}$ at which to run simulations and collect observations, where $(\bx_i,n_i)$ denotes the location and the number of replications, respectively, at the $i$-th design point. The design that we recommend is described in more detail in \ref{appendix:ExperimentDesign}. 
The
simulation outputs at $\mathcal{D}$ are $\mathbf{Y}_\mathcal{D}\equiv \left\{ 
(Y_1(\mathbf{x}_i),Y_2(\mathbf{x}_i),\ldots,Y_{n_i}(\mathbf{x}_i));
i=1,2,\ldots,k \right\}$ and the sample mean at design point $\bx_i$
is $\bar{Y}(\mathbf{x}_i)=\sum_{j=1}^{n_i}Y_j(\mathbf{x}_i)/n_i$.  
Let the sample means at all $k$ design points be 
$\bar{\mathbf{Y}}_\mathcal{D}=(\bar{Y}(\mathbf{x}_1),\bar{Y}(\mathbf{x}_2),
\ldots,\bar{Y}(\mathbf{x}_k))^T$.
Set the simulations at different design points independent. Then, the variance of 
$\bar{\mathbf{Y}}_\mathcal{D}$ is represented by a $k\times k$ diagonal matrix
$C=\mbox{diag}\left\{\sigma^2_{\epsilon}(\bx_1)/n_1,\sigma^2_{\epsilon}(\bx_2)/n_2,
\ldots,\sigma^2_{\epsilon}(\bx_k)/n_k \right\}$. 
\end{sloppypar}

Let $\Sigma$ be the $k\times k$ spatial covariance matrix of the design points and let $\Sigma(\bx,\cdot)$ be the $k\times 1$ spatial covariance vector between the design points and a fixed prediction point $\bx$.  If the parameters $(\tau^2,\pmb{\theta}, C)$ are known, then the metamodel or simulation uncertainty can be characterized by a refined GP $M_p(\bx)$ that denotes the conditional distribution of $M(\bx)$ given simulation outputs $\bar{\mathbf{Y}}_\mathcal{D}$,
\begin{equation}
M_p(\bx)\sim \mathrm{GP}(m_{p}(\bx),\sigma^2_{p}(\bx))
\label{eq.posterior}
\end{equation}
where 
the minimum mean squared error (MSE) linear unbiased predictor is
\begin{equation}
m_{p}(\bx)=\widehat{\beta}_0
+\Sigma(\bx,\cdot)^\top(\Sigma+C)^{-1}
(\bar{\mathbf{Y}}_\mathcal{D}-\widehat{\beta}_0\cdot 1_{k\times 1}),
\label{eq.predictor1}
\end{equation}
and the corresponding variance is 
\begin{equation}
\begin{split}
\label{eq.MSE1}
\sigma^2_{p}(\bx) =
\tau^2-\Sigma(\bx,\cdot)^\top(\Sigma+C)^{-1}\Sigma(\bx,\cdot)   \\
+\mathbf{\eta}^\top[1_{k\times 1}^\top(\Sigma+C)^{-1}1_{k\times 1} ]^{-1}\mathbf{\eta}  
\end{split}
\end{equation}
where $\widehat{\beta}_0=[1_{k\times 1}^\top(\Sigma+C)^{-1}1_{k\times 1}]^{-1}1_{k\times 1}^\top(\Sigma+C)^{-1}\bar{\mathbf{Y}}_\mathcal{D}$
and $\mathbf{\eta}=1-1_{k\times 1}^\top(\Sigma+C)^{-1}\Sigma(\bx,\cdot)$ \cite{ankenman_nelson_staum_2010}. 
The spatial correlation parameters $\tau^2$ and $\pmb{\theta}$ are estimated by using MLEs. 
The sample variance is used as an estimate for the simulation variance at design points $C$. By plugging $(\widehat{\beta}_0, \widehat{\tau}^2, \widehat{\pmb{\theta}}, \widehat{C})$ into Equations~(\ref{eq.predictor1}) and (\ref{eq.MSE1}), we can obtain the estimated mean $\widehat{m}_p(\mathbf{x})$ and variance $\widehat{\sigma}_p^2(\mathbf{x})$. Thus, the metamodel we use is $\widehat{\mu}(\bx)=\widehat{m}_p(\bx)$ with marginal variance estimated by $\widehat{\sigma}^2_p(\bx)$.

\textcolor{black}{\cite{ankenman_nelson_staum_2010} demonstrate that $\widehat{m}_p(\bx)$ is still an unbiased predictor even with the plug-in estimator $\widehat{C}$, and 
the variance inflation of $\sigma^2_p(\bx)$ caused by using $\widehat{C}$ is typically small. 
}
In the asymptotic analysis, we assume that the parameters $(\tau^2,\pmb{\theta},C)$ are known. This is necessary (and common in the kriging literature) because including the effect of parameter estimation is mathematically intractable. Further, there is both theoretical and empirical evidence that in many cases prediction accuracy is minimally affected by using estimated hyperparameters; see \cite{wang2021hyperparm}.


\subsection{Metamodel-Assisted Bootstrapping for Uncertainty Quantification}
\label{subsec:MAB-UQ}

The proposed metamodel-assisted bootstrapping can provide a CI for the true mean performance, which accounts for both model and simulation uncertainties. Since $m_p(\bx)$ is an unbiased predictor under the GP assumption, $\sigma^2_p(\bx)=0$ for all $\bx$ would imply that there is no simulation uncertainty due either to a finite number of design points $\bx_i$ or finite number of replications $n_i$; that is, $m_p(\bx) =
\mu(\bx)$.  Unfortunately, 
if the budget is tight relative to the complexity of the true response surface, then the effect of simulation uncertainty could be substantial, resulting in significant undercoverage of the confidence interval of \cite{barton_nelson_xie_2011} as we show in Section~\ref{sec:empirical}. The new interval introduced here does not suffer this degradation, and therefore is robust to the amount of simulation effort that can be expended. 

The kriging literature is the foundation for our work; see for instance \cite{Santer_2003}. 
Kriging models uncertainty about the function as a GP $M(\cdot)$ by assuming $\mu(\cdot)$ is a realization of $M(\cdot)$.  An interval constructed to cover the conditional distribution of $M(\bx_0)$ given the values at the design points is often interpreted as a CI for $\mu(\bx_0)$; see for example \cite{Picheny_2010}. The success of this paradigm is not because the function of interest is actually random---it is not---but because in many problems the conditional GP appears to be a robust characterization of the remaining response-surface uncertainty.

We adopt the kriging paradigm but with two key differences: our prediction point $\bx_c$ is 
unknown and must be
estimated from real-world data, and our function $\mu(\cdot)$ can only be evaluated in the presence of stochastic simulation noise. Given the
simulation outputs $\bar{\mathbf{Y}}_{\mathcal{D}}$, the remaining uncertainty about $\mu(\cdot)$ is characterized by the conditional GP
$M_p(\cdot)$. To account for the impact from both model and simulation uncertainties, we construct an interval $[C_L,C_U]$ covering
$M_p(\bx_c)$ with probability $(1-\alpha)100\%$, i.e.,
\begin{equation}
 \mbox{Pr}\{M_p(\bx_c)\in [C_L,C_U]\} = 1-\alpha. 
\label{eq.CI_3}
\end{equation} 
Since the conditional coverage is $1-\alpha$, the unconditional coverage of $M(\bx_c)$ is $1-\alpha$ as well. \textit{The
revised objective~(\ref{eq.CI_3}) is connected to our objective~(\ref{eq.CI_1}) through the assumption that the function $\mu(\cdot)$ is a realization of the GP $M(\cdot)$.} A procedure that delivers an interval satisfying~(\ref{eq.CI_3}) will be a good approximation for a CI procedure satisfying~(\ref{eq.CI_1}) if $M_p(\cdot)$ faithfully represents the remaining uncertainty about $\mu(\cdot)$.  This is clearly an approximation because in any real problem $\mu(\cdot)$ is a fixed function, therefore we refer to $[C_L,C_U]$ as an approximation for the CI (ACI).

\begin{sloppypar}

Based on a hierarchical approach, we propose Algorithm~\ref{algo:ACI} to build $(1-\alpha)100\%$ bootstrap percentile ACIs to achieve~(\ref{eq.CI_3}). In this procedure, Step~1 provides an experiment design to build a SK
metamodel, which is central to the metamodel-assisted bootstrapping approach.  Since the system model uncertainty is quantified with bootstrapped samples,
we want the metamodel to correctly predict the responses at these sample points $\widehat{\mathbf{X}}_{\mathbf{m}}\sim
\widehat{F}_{\mathbf{X}_{\mathbf{m}}}(\cdot|\mathbf{z}_{\mathbf{m}}^{(0)})$. Thus, the metamodel needs to be accurate and precise in a design space
that covers the ``most likely" bootstrap moment
 estimates, which can be achieved by the experiment design proposed by
\cite{barton_nelson_xie_2011}. 
Specifically, they find the smallest ellipsoid denoted by $E$ that covers the most likely bootstrap resampled moments and then generate a space-filling design that covers $E$; see the details in Appendix \ref{appendix:ExperimentDesign}.
\end{sloppypar}

\begin{algorithm*} 
\DontPrintSemicolon
\KwIn{Given real-world data $\mathbf{z}_\mathbf{m}^{(0)} = \{\mathbf{z}_{\ell,m_\ell}^{(0)}, \ell=1,2,\ldots,L\}$}
\KwOut{Estimated CI and ACI quantifying the overall estimation uncertainty of $\mu(\mathbf{x}_c)$; Estimated model variance $\widehat{\sigma}^2_I$,  simulation variance $\widehat{\sigma}^2_M$ and uncertainty contribution $\widehat{s}_\ell$ from $\ell$-th model. }
\SetKwBlock{Begin}{Function of Uncertainty Quantification (UQ):}{end Function:}
\Begin() 
{   
    \textbf{Step 1:} Build the design space covering the most likely bootstrap moment estimates of process models, and choose a space-filling experiment design $\mathcal{D}=\{(\bx_i,n_i), i=1,2,\ldots, k\}$ as described in  Appendix \ref{appendix:ExperimentDesign}.\\
    \textbf{Step 2:} Run simulations at design points to obtain outputs $\mathbf{Y}_\mathcal{D}$. Compute the sample average $\bar{Y}(\bx_i)$ and sample variance $S^2(\bx_i)$ of the simulation outputs, $i=1,2,\ldots, k$.  \textcolor{black}{Fit the SK metamodel parameters $(\beta_0, \tau^2, \pmb{\theta}, C)$} to obtain $\widehat{m}_p(\bx)$ and $\widehat{{\sigma}}_{p}^2(\mathbf{x})$ using $\left(\bar{Y}(\bx_i), S^2(\bx_i), \bx_i\right)$, $i=1,2,\ldots, k$. \\
    \textbf{Step 3:}  \For {$b = 1 \mbox{ to } B$}
    {
    \textbf{Step 3(a)}: 
    Draw $m_\ell$ samples with replacement from $\mathbf{z}_{\ell,m_\ell}^{(0)}$, denoted by $\mathbf{Z}_{\ell,m_\ell}^{(b)}$, and 
    calculate the corresponding $h_\ell\times 1$ vector of bootstrap moment estimates denoted by $\widehat{{\mathbf{X}}}_{\ell,m_\ell}^{(b)}={\bX}_\ell (\mathbf{Z}_{\ell,m_\ell}^{(b)})$ for $\ell=1,2,\ldots,L$. Then stack the results for all $L$ processes to obtain a $d\times 1$ vector $\widehat{{\mathbf{X}}}_{\mathbf{m}}^{(b)}$. \\
    \textbf{Step 3(b)}: Let $\widehat{\mu}_b\equiv\widehat{m}_p(\widehat{\mathbf{X}}_{\mathbf{m}}^{(b)})$. \\
    \textbf{Step 3(c)}: Draw $\widehat{M}_b \sim \mbox{N}\left( \widehat{m}_p (\widehat{\mathbf{X}}_{\mathbf{m}}^{(b)}), \widehat{\sigma}_{p}^2(\widehat{\mathbf{X}}_{\mathbf{m}}^{(b)}) \right)$. 
    }
    \textbf{Return} \textbf{(1)} Estimated $(1-\alpha)100\%$ bootstrap percentile CI and ACI; 
    \textbf{(2)} Estimated model variance and simulation variance,
    \begin{gather*}
    \begin{aligned}
    \mbox{CI}_0 &= \left[
            \widehat{\mu}_{(\lceil B\frac{\alpha}{2}
            \rceil)},\widehat{\mu}_{(\lceil B(1-\frac{\alpha}{2})\rceil)} \right], 
    &\mbox{CI}_+ &= \left[
            \widehat{M}_{(\lceil B\frac{\alpha}{2} \rceil)},\widehat{M}_{(\lceil 
            B(1-\frac{\alpha}{2})\rceil)}\right], 
    \\
   \widehat{\sigma}^2_I &=\sum_{b=1}^B(\textcolor{black}{\widehat{\mu}}_b-\widehat{{\mu}})^2/(B-1), 
    &\widehat{\sigma}^2_M &= \sum_{b=1}^B\textcolor{black}{\widehat{\sigma}}_p^2 (\widehat{\mathbf{X}}^{(b)}_{\mathbf{m}})/B, 
    \end{aligned}
    \end{gather*}
    where $\widehat{\mu}_{(1)} \le \widehat{\mu}_{(2)} \le \cdots \le
    \widehat{\mu}_{(B)}$ and 
    $\widehat{M}_{(1)} \le \widehat{M}_{(2)} \le \cdots \le
    \widehat{M}_{(B)}$ are the sorted values, and $\widehat{{\mu}}=\sum_{b=1}^B\textcolor{black}{\widehat{\mu}}_b/B$.
}
\vspace{0.1in}

\SetKwBlock{Begin}{Function of Sensitivity Analysis (SA):} 
{end Function:}
\Begin() 
{   
    \For {each $ \mathcal{J} \subseteq \mathcal{L}$ with $\mathcal{L} = \{1, 2, \ldots, L\}$}
    {
        \textbf{Step 4:} Generate bootstrap samples 
        $\widehat{\mathbf{X}}_{\mathcal{J}}^{(b)}$ for $b = 1,2,\ldots, B'$
        and obtain simulation output prediction $\widehat{m}_p({\mathbf{x}}_{-\mathcal{J}}^{(0)},\widehat{\mathbf{X}}_{\mathcal{J}}^{(b)})$. \\
        \textbf{Step 5:} Estimate the cost function $c(\mathcal{J})$ by (\ref{eq.c_hat}).\\
    }
        \textbf{Return} Estimated $\ell$th model uncertainty contribution $\widehat{s}_\ell$ through (\ref{eq.SV}) with $\ell = 1,2,\ldots,L$.\\
}
\caption{Metamodel-Assisted Bootstrap for Uncertainty Quantification and Sensitivity Analysis}
\label{algo:ACI}
\end{algorithm*}

Based on the experiment design provided in Step~1, we run simulations and construct a metamodel in Step~2 \textcolor{black}{by fitting $(\beta_0, \tau^2, \pmb{\theta}, C)$}. Given the metamodel, we predict the simulation's mean responses at different model settings corresponding to bootstrap resampled moments. The bootstrap resampled moments are 
drawn from the bootstrap distribution denoted by $\widehat{F}_{\mathbf{X}_\mathbf{m}}(\cdot|\mathbf{z}_{\mathbf{m}}^{(0)})$. In Step~3(a), we generate bootstrapped model moments.
Then, we return a $(1-\alpha)100\%$ interval estimators as shown in Algorithm~\ref{algo:ACI}. Notice that Step~3(b) accounts for the model uncertainty and Step~3(c) accounts for the simulation uncertainty. Thus, this procedure provides two types of intervals: 
\textbf{(a)} $\mbox{CI}_0$, proposed in \cite{barton_nelson_xie_2011},
returns an estimate of $[Q_L,Q_U]$ in eq.~(\ref{eq.CI_1}) by
assuming $\widehat{m}_p(\bx) = \mu(\bx)$; that is, it only accounts
for model uncertainty and will be in error if there is substantial
simulation uncertainty.
\textbf{(b)} $\mbox{CI}_+$ returns an estimate of $[C_L,C_U]$ in
eq.~(\ref{eq.CI_3}). This ACI accounts for both model and
simulation uncertainty.
As the simulation uncertainty decreases, $\mbox{CI}_0$ and $\mbox{CI}_+$ become closer and closer to each other. Before evaluating the finite-sample performance of $\mbox{CI}_+$ in Section~\ref{sec:empirical}, we establish its asymptotic consistency \textcolor{black}{for objective~(\ref{eq.CI_3})} in Section~\ref{subsec:CI_consistency}.
Then, in Steps~4 and 5, variance decomposition is developed to quantify the contribution from each source of uncertainty, which will be studied in Section~\ref{Sec:varDecomp}.

\subsection{Asymptotic Consistency Study on Interval $\mbox{CI}_+$}
\label{subsec:CI_consistency}

In this section, we show that the ACI $\mbox{CI}_+$ provided in Algorithm~\ref{algo:ACI} satisfies Equation~(\ref{eq.CI_3})
asymptotically. \textit{The asymptotic consistency of this
\textcolor{black}{interval} is proved under the assumption that the true response surface $\mu(\bx)$ is a realization of a GP \textcolor{black}{with all
parameters known except $\beta_0$}.} 
Under this assumption, $M_p(\bx)$ characterizes the remaining simulation uncertainty after observing $\bar{{\mathbf{Y}}}_\mathcal{D}$. Since the model uncertainty is asymptotically correctly quantified by the bootstrap moment estimator $\widehat{\mathbf{X}}_{\mathbf{m}}$, the distribution of $M_p(\widehat{\mathbf{X}}_{\mathbf{m}})$ accounts for both model and simulation uncertainties.
Theorem~\ref{thm:consistency} shows that this interval satisfies objective~(\ref{eq.CI_3}) asymptotically. 
The detailed proof is provided in Appendix \ref{app:consistencyCI+}. 

\begin{sloppypar}
\begin{theorem} \label{thm:consistency}
Suppose that Assumptions~($\star$) in Appendix \ref{sec:assumptions}
hold. Then the interval $\mbox{CI}_+ = [M_{(\lceil B\frac{\alpha}{2} \rceil)}, M_{(\lceil B(1-\frac{\alpha}{2})\rceil)}]$ is asymptotically consistent, 
\begin{equation}
\lim_{m\rightarrow\infty} \lim_{B\rightarrow \infty} 
\Pr\{M_{(\lceil B\alpha/2 \rceil)} \leq M_p(\mathbf{x}_c) 
\leq M_{(\lceil B(1-\alpha/2)\rceil)} \} = 1-\alpha.
\label{eq.11_1}
\end{equation} 
\end{theorem}
\end{sloppypar}

\section{Variance Decomposition for Uncertainty Analysis}
\label{Sec:varDecomp}


In a practical setting, what is the next step if the interval $\mbox{CI}_+$ is so wide that we
are uncomfortable making decisions based on estimates with that level of error? We suggest gaining some sense of the relative contribution from each source of uncertainty as a guide toward either collecting more real-world process data to reduce the model uncertainty or running more simulations to improve the system mean response estimation at any given models $F$.  
The overall estimation variance of system true performance $\mu(\mathbf{x}_c)$ is quantified by $\mbox{Var}[M_p(\mathbf{X}_m)]$.
In Section~\ref{subSec:varDec}, we propose a variance decomposition approach to quantify the contribution from simulation and model uncertainties.
Compared with the existing studies on estimating the relative contributions, such as \cite{
SongNelson_2013}, 
our variance decomposition does not require the homogeneity assumption, i.e., the simulation noise has a constant variance.
Since the effect of model uncertainty is induced by the complex interactions of estimation uncertainties from $L$ models
$(F_1,F_2,\ldots,F_L)$, we further decompose it by using 
Shapley value (SV) based global sensitivity analysis to correctly quantify the contribution from each source of model uncertainty in Section~\ref{subSec:varDecModelUncertainty}. 
This information can provide a guide on which model $F_\ell$ to collect more real-world data and improve the system mean performance estimation.
Then, we provide the asymptotic consistency study over the variance component estimation for each source of uncertainty in Section~\ref{subsec:VarianceComponentAsymptoticConsistency}.

\subsection{Simulation and Model Uncertainty Contribution Quantification}
\label{subSec:varDec}

Suppose that the parameters $(\tau^2,\pmb{\theta},C)$ are known, the simulation uncertainty can be characterized by a GP, and the simulation error follows a normal distribution. 
Then given the simulation outputs $\bar{\mathbf{Y}}_\mathcal{D}$, the simulation uncertainty is characterized by a GP, i.e., $M_p(\bx)\sim
\mbox{N}(m_p(\mathbf{x}),\sigma^2_{p}(\mathbf{x}))$.  Conditional on
$\bar{\mathbf{Y}}_\mathcal{D}$, both $m_p(\mathbf{x})$ and
$\sigma^2_{p}(\mathbf{x})$ are fixed functions.  For notation simplification, all of following derivations are conditional on the simulation outputs $\bar{\mathbf{Y}}_\mathcal{D}$, but we will suppress the ``$|\bar{\mathbf{Y}}_\mathcal{D}$''. 

To quantify the relative contribution of model and simulation uncertainties, we decompose the total variance of
$M_p({\mathbf{X}}_{\mathbf{m}})$ into two parts: 
\textcolor{black}{
\begin{eqnarray}
\lefteqn{ \sigma^2_T \equiv \mbox{Var}[M_p(\mathbf{X}_{\mathbf{m}})] }
\nonumber \\
&=&\mbox{E}\{\mbox{Var}[M_p(\mathbf{X}_{\mathbf{m}})|\mathbf{X}_{\mathbf{m}}]\}
+\mbox{Var}\{\mbox{E}[M_p(\mathbf{X}_{\mathbf{m}})|\mathbf{X}_{\mathbf{m}}]\} 
\nonumber\\
&=& \mbox{E}[\sigma^2_{p}(\mathbf{X}_{\mathbf{m}})]
+\mbox{Var}[m_p(\mathbf{X}_{\mathbf{m}})].
\label{eq.varDecomp}
\end{eqnarray}}
The term $\sigma^2_M\equiv \mbox{E}[\sigma^2_{p}(\mathbf{X}_{\mathbf{m}})]$ is a measure of the simulation uncertainty: the expected metamodel variance weighted by the density of moment estimator $\mathbf{X}_{\mathbf{m}}$.  This weighting makes sense because the accuracy of the metamodel in regions with higher density is more important for the estimation of system mean performance. The term $\sigma^2_I\equiv 
\mbox{Var}[m_p(\mathbf{X}_{\mathbf{m}})]$ is a measure of model uncertainty when we replace the unknown true response surface $\mu(\cdot)$ with its best linear unbiased estimate $m_p(\cdot)$. 

If the simulation uncertainty disappears (i.e., $\sigma^2_p(\cdot)=0$), then $\sigma^2_M=0$, $\mbox{CI}_0$ and $\mbox{CI}_+$ coincide. 
On the other hand, as $m\rightarrow\infty$ (more and more real-world data), $\mathbf{X}_{\mathbf{m}}\stackrel{a.s.}\rightarrow\bx_c$ and since $m_p(\bx)$ is continuous we have $\sigma^2_I=0$; therefore, the width of $\mbox{CI}_0$ shrinks to zero as does coverage since there is remaining simulation uncertainty in general. However, because $\mbox{CI}_+$ accounts for simulation uncertainty it still provides asymptotically consistent coverage.  This effect is demonstrated by the empirical study in Section~\ref{sec:empirical}.

Our decomposition allows us to express the total variance in Equation~(\ref{eq.varDecomp}) as the sum of two variances measuring model and simulation uncertainties: $\sigma^2_T = \sigma^2_I+\sigma^2_M.$ In the metamodel-assisted bootstrapping framework, we can estimate each variance component as follows:
\begin{itemize}

\item Total variance:
$\widehat{\sigma}^2_T=\sum_{b=1}^B(M_b-\bar{M})^2/(B-1)$, where $\bar{M}=\sum_{b=1}^B M_b/B.$

\item Model variance:
$\widehat{\sigma}^2_I=\sum_{b=1}^B(\textcolor{black}{\mu}_b-\bar{{\mu}})^2/(B-1)$, where $\bar{{\mu}}=\sum_{b=1}^B\textcolor{black}{\mu}_b/B$.  

\item Simulation variance:
$\widehat{\sigma}^2_M= \sum_{b=1}^B\textcolor{black}{\sigma}_p^2 (\widehat{\mathbf{X}}^{(b)}_{\mathbf{m}})/B$.
\end{itemize}

The ratio $\widehat{\sigma}_I/\widehat{\sigma}_T$ provides an estimate of the relative contribution from model uncertainty on $\mbox{CI}_+$. If it is close to 1, the contribution from simulation uncertainty can be ignored.  Thus, this ratio can help a decision maker determine where to put more effort: If the model variance dominates, then get more real-world data (if possible). If the simulation variance dominates, then it can be reduced by more simulations, which can be a combination of additional design points and additional replications at existing design points. If neither dominates, then both activities are necessary to reduce $\mbox{CI}_+$ to a practically useful size.


\subsection{Variance Decomposition for Model Uncertainty Analysis}
\label{subSec:varDecModelUncertainty}

\begin{sloppypar}
The overall model variance $\sigma^2_I= \mbox{Var}[m_p({\mathbf{X}}_\mathbf{m})]$ is induced by the estimation uncertainty of 
correct moments $\mathbf{x}_c=(\mathbf{x}_{[1],c},\mathbf{x}_{[2],c},\ldots,\mathbf{x}_{[L],c})$ for process models $F^c=\{F_1^c,F_2^c,\ldots,F_L^c\}$. To efficiently identify the bottlenecks and reduce the impact of model uncertainty, we are interested in quantifying the contribution of moment estimation uncertainty of ${\mathbf{X}}_{\ell,m_\ell} = \mathbf{X}_\ell(\mathbf{Z}_{\ell,m_\ell})$ for each $\ell$-th model $F_\ell$. To approximate the estimation uncertainty of ${\mathbf{X}}_{\ell,m_\ell}$ with $\ell=1,2,\ldots,L$, the bootstrap resampled moments are drawn from the bootstrap distribution, $\widehat{\mathbf{X}}_{\ell,m_\ell}\sim F_{\mathbf{X}_{\ell,m_\ell}}(\cdot|\mathbf{z}_{\ell,m_\ell}^{(0)})$.
\end{sloppypar}

Motivated by the SV based sensitivity analysis (see for example \cite{
song2016shapley}),
the overall model variance $\sigma^2_I$ in (\ref{eq.varDecomp}) can be decomposed as the sum of contributions from each source of model uncertainty,
\begin{equation}
\begin{split}
\sigma^2_I  = \mbox{Var}[m_p({\mathbf{X}}_\mathbf{m})]  =\mbox{Var}[m_p({\mathbf{X}}_{1,m_1},
{\mathbf{X}}_{2,m_2},\ldots,
{\mathbf{X}}_{L,m_L})] 
= \sum_{\ell=1}^L s_{\ell},
\label{eq.SVvarDecomp}
\end{split}
\end{equation}
with $s_\ell$ quantifying the 
contribution from the $\ell$-th model uncertainty,
\begin{equation}\label{eq.SV}
s_{\ell} = \sum_{\mathcal{J}\subseteq \mathcal{L}/\{\ell\}} \dfrac{(L-|\mathcal{J}|-1)!|\mathcal{J}|!}{L!} \left[ c(\mathcal{J}\cup\{\ell\}) - c(\mathcal{J}) \right],
\end{equation}
where $\mathcal{L} = \{1, 2, \ldots, L\}$ denotes the index set of $L$ sources of model uncertainty and $|\cdot|$ indicates the set size. 
Here, 
for any subset $\mathcal{J}\subseteq \mathcal{L}$,
we use the total effect based cost function
$c(\mathcal{J})=\mbox{E}[\mbox{Var}[m_p(\mathbf{X}_{\mathbf{m}}) | \mathbf{X}_{-\mathcal{J}}]]$
measuring the expected remaining variance when all other model moments, denoted by $\mathbf{X}_{-\mathcal{J}}$, are conditionally fixed, where ${-\mathcal{J}}$ denotes the remaining subset $\mathcal{L}/\mathcal{J}$.

\begin{sloppypar}
The metamodel-assisted bootstrap resampling is used to estimate the contribution from each source of model uncertainty (Algorithm~\ref{algo:ACI}). 
Basically, for any model with the index $i\notin \mathcal{J}$ or $i\in \mathcal{L}/\mathcal{J}$, we take the sample moment $\mathbf{x}_{i,m_i}^{(0)}$ as true one. Denote these model moments by $\mathbf{x}_{-\mathcal{J}}^{(0)}$.
Then, for the model with index $j\in \mathcal{J}$, we draw with replacement to generate the bootstrap sample moments quantifying the corresponding model uncertainty, $\widehat{\mathbf{X}}_{j,m_j}^{(b)}\sim F_{\mathbf{X}_{j,m_j}}(\cdot|\mathbf{z}_{j,m_j}^{(0)})$ with $b=1,2,\ldots,B'$. We represent the combination of bootstrap moment samples for all model moments with index $j\in \mathcal{J}$ by $\widehat{\mathbf{X}}_{\mathcal{J}}^{(b)}$.
Thus, 
we estimate $c(\mathcal{J})$ by a Monte Carlo sampling approach,
\begin{equation}\label{eq.c_hat}
    \widehat{c}(\mathcal{J}) = \frac{1}{B'-1} \sum_{b=1}^{B'}
    \left[ \widehat{m}_p\left(\mathbf{x}_{-\mathcal{J}}^{(0)}, 
    \widehat{\mathbf{X}}_{\mathcal{J}}^{(b)} \right)
    -\bar{m}_{\mathcal{J}} \right]^2
\end{equation}
where $\bar{m}_{\mathcal{J}} = \sum_{b=1}^{B'} \widehat{m}_p\left(\mathbf{x}_{-\mathcal{J}}^{(0)}, 
\widehat{\mathbf{X}}_{\mathcal{J}}^{(b)} \right) / B'$. By plugging $\widehat{c}(\mathcal{J})$ into Equation~(\ref{eq.SV}), we can get the estimator $\widehat{s}_\ell$ quantifying the contribution from the $\ell$-th model uncertainty to $\mbox{Var}[m_p(\mathbf{X}_{\mathbf{m}})]$. An efficient approximation algorithm, using the randomly selected subset instead of all possible index sets permutations, can be used to reduce the computational burden; see \cite{song2016shapley}.
\end{sloppypar}

\subsection{Asymptotic Consistency Study of Variance Contribution Estimation}
\label{subsec:VarianceComponentAsymptoticConsistency}

We provide the asymptotic consistency study of variance contribution estimation from each source of uncertainty; see 
 Theorems~\ref{thm:decomposition1}, \ref{thm:decomposition2}, and \ref{thm:decomposition3}.

\begin{theorem} \label{thm:decomposition1}
Suppose that Assumptions~1--4 in Appendix \ref{sec:assumptions} 
hold.
Then conditional on $\bar{\mathbf{Y}}_\mathcal{D}$, the variance component estimators $\widehat{\sigma}^2_M, \widehat{\sigma}^2_I, \widehat{\sigma}^2_T$ are consistent as $m, B\rightarrow\infty$, where as $m \rightarrow \infty$ we have $m_\ell/m \rightarrow c_\ell$, $\ell=1,2,\ldots,L$, for a constant $c_\ell>0$. Specifically, 
\begin{itemize}

\item As $m\rightarrow\infty$, the model uncertainty disappears:
\[
\lim_{m\rightarrow\infty}\sigma^2_M=\sigma^2_{p}(\bx_c), 
\lim_{m\rightarrow\infty}\sigma^2_I=0 \mbox{ and
}\lim_{m\rightarrow\infty}\sigma^2_T=\sigma^2_{p}(\bx_c).
\]

\item As $m\rightarrow\infty$ and $B\rightarrow\infty$ in an iterated limit, the variance component estimators are consistent: 
\begin{eqnarray*}
\lim_{m\rightarrow\infty}\lim_{B\rightarrow\infty}\widehat{\sigma}_M^2 &=&\lim_{m\rightarrow\infty}\sigma_M^2=\sigma_p^2(\bx_c), \\
\lim_{m\rightarrow\infty}\lim_{B\rightarrow\infty}\widehat{\sigma}_I^2 &=&\lim_{m\rightarrow\infty}\sigma_I^2=0, \\
\lim_{m\rightarrow\infty}\lim_{B\rightarrow\infty}\widehat{\sigma}_T^2 &=&\lim_{m\rightarrow\infty}\sigma_T^2=\sigma^2_p(\bx_c), \\ 
\lim_{m\rightarrow\infty}\lim_{B\rightarrow\infty}\widehat{s}_\ell &=&\lim_{m\rightarrow\infty}s_{\ell}=0 \mbox{~ for~} \ell = 1,2,\ldots,L.
\end{eqnarray*}

\end{itemize}

\end{theorem}

Theorem~\ref{thm:decomposition1} demonstrates that the variance components estimators $\widehat{\sigma}^2_I$, $\widehat{\sigma}^2_M$, $\widehat{\sigma}^2_T$, and $\widehat{s}_\ell$ for $\ell=1,2,\ldots,L$ are consistent. However, we can see that the model uncertainty disappears as $m\rightarrow\infty$. 
In addition, we study the consistency of scaled versions of $\sigma^2_I$ and $\widehat{\sigma}^2_I$ in Theorem~\ref{thm:decomposition2}, showing that $m\sigma^2_I$ and $m\widehat{\sigma}_I^2$ converge to the same non-zero constant.

\begin{theorem} \label{thm:decomposition2}
Suppose Assumptions~1--6 in Appendix \ref{sec:assumptions} 
hold. Then we have $\lim_{m\rightarrow\infty}m\sigma_I^2=\lim_{m\rightarrow\infty} \lim_{B\rightarrow\infty} m\widehat{\sigma}^2_I=\sigma^2_\mu$ almost surely, where $\sigma^2_\mu$ is a positive constant.

\end{theorem}



\begin{sloppypar}
\begin{theorem}
\label{thm:decomposition3}
Suppose Assumptions~1--6 in Appendix \ref{sec:assumptions}
hold.
Then we have $\lim_{m\rightarrow\infty}ms_{\ell}=\lim_{m\rightarrow\infty} \lim_{B\rightarrow\infty} m\widehat{s}_{\ell}=\sigma^2_s$ almost surely, where $\sigma^2_s$ is a positive constant.
\end{theorem}
\end{sloppypar}

Theorems~\ref{thm:decomposition1}--\ref{thm:decomposition3} give the asymptotic properties of the variance component estimators,
guaranteeing: (1) $\widehat{\sigma}_I/\widehat{\sigma}_T$ is a consistent estimator for the relative contribution of model uncertainty to the overall estimation uncertainty; and (2) $\widehat{s}_\ell$ is a consistent estimator of the contribution from the $\ell$-th model uncertainty. The detailed proof is provided in Appendix \ref{app:consistencyVarCom}. We will empirically evaluate its finite-sample performance in Section~\ref{sec:empirical} \textcolor{black}{where we form the variance component estimators by inserting $(\widehat{\tau}^2, \widehat{\pmb{\theta}}, \widehat{C})$ for the unknown parameters $(\tau^2, \pmb{\theta}, C)$.}


\section{Empirical Study}
\label{sec:empirical}

We study the finite sample performance of the proposed metamodel-assisted uncertainty analysis framework and compare it with the direct bootstrap approach. We consider a biopharmaceutical manufacturing example 
in Sections~\ref{subSec:bioempirical_CI}. 
A cell culture process hybrid model for cell therapy manufacturing
is studied in Section~\ref{subsec:Cell_Exten}. Additionally, a queueing network example is provided in Appendix \ref{subsec:queueingNetworkExample}
. 
The proposed framework demonstrates good and robust performance under different experiment settings in terms of (1) the amount of real-world data $m$ which controls the level of model uncertainty; (2) the simulation budget $N$ which controls the simulation uncertainty; and (3) the number of design points $k$ for GP metamodel construction, with $N$, is used to control the metamodel uncertainty. 

The empirical results show that the proposed framework can provide better performance than the direct bootstrap approach. 
The new ACI $\mbox{CI}_+$ is robust to different levels of real-world data $m$, number of design points $k$, and simulation budget $N$ in terms of replications. 
When simulation uncertainty is significant, $\mbox{CI}_0$ tends to have undercoverage that becomes more serious as $m$ increases.
Since $\mbox{CI}_+$ accounts for both simulation and model uncertainties, it does not exhibit this degradation.
The ratio $\widehat{\sigma}_I/\widehat{\sigma}_T$ is a useful measure of the relative contribution of model uncertainty to overall statistical
uncertainty and the SV-based sensitivity analysis further quantifies the contribution from each source of model uncertainty. 

\subsection{A Biopharmaceutical Manufacturing Example}
\label{subSec:bioempirical_CI}

We consider the biomanufacturing example  illustrated in Figure~\ref{fig_api}; see the details in \cite{wang2019stochastic}.
We are interested in estimating  the expected productivity of an antigen protein drug, i.e., $\mu(\mathbf{x}_c)$.
The protein and impurity accumulations in the exponential-growth phase of fermentation process are modeled with the hybrid models, i.e., 
$
    X_t = X_0 \cdot e^{\gamma t} + \epsilon_P
$
and $
   I_t = I_0 \cdot e^{\gamma t} + \epsilon_I$
with $0 \leq t \leq T$, 
where 
$\gamma$ is the growth rate, $X_0$ and $I_0$ are the starting amounts of biomass and impurity. We consider the fixed harvest time $T = 54$ and the fixed initial impurity amount $I_0 = 14.64$.

The downstream purification process includes  centrifuge, chromatography, filtration, and quality control. Random proportions of protein and impurity are removed at each operation unit, except at the quality control step. 
\textbf{(1)~Centrifuge Step.} The protein and impurity levels before and after centrifuge are denoted by $(X_F,I_F)$ and $(X_C,I_C)$. We assume that this step does not change the protein level, i.e., $X_C \equiv X_F$ \cite{delahaye2015ultra}, and it removes a random proportion of impurity, i.e., $I_C = Q \cdot I_F$.
\textbf{(2)~Chromatography Step.} For chromatography, random removal proportions of protein and impurity, denoted by $Q_P$ and $Q_I$, 
follow uniform distributions \cite{martagan2017performance}.
The target protein and impurity levels before and after chromatography are denoted by $(X_C, I_C)$ and $(X_P, I_P)$, and we have $X_P = Q_P \cdot X_C$
and $I_P = Q_I \cdot I_C$. 
\textbf{(3)~Filtration Step.} Filtration works as a polishing procedure and it slightly reduces the impurity. Denote the protein and impurity levels before and after filtration with $(X_P, I_P)$ and $(X_{fr}, I_{fr})$. Thus, $I_{fr} = Q_{fr} \cdot I_P$. and $X_{fr} = X_P$. \textbf{(4)~ Quality Control Step.} During the quality control step, if the impurity percentage $\frac{I_{fr}}{X_{fr} + I_{fr}}$ is greater than the requirement, say $\omega = 25\%$, the corresponding batch is discarded. 
Therefore, the expected productivity of each batch is defined as:
\begin{equation}   \mu(\mathbf{x}_c) = 
\mbox{E}\left[X_{fr} \cdot {1}\left(
\frac{I_{fr}}{X_{fr} + I_{fr}} \leq \omega \right)
\right]. 
\nonumber 
\end{equation}

\begin{table*}[h!]
\small
\centering
\caption{The underlying true process model parameters.}
\label{tab:truepara}
\begin{tabular}{|c|cc|}
\hline
                       & \multicolumn{1}{c|}{Protein Concentration}                  & Impurity Concentration                 \\ \hline
Initial Biomass        & \multicolumn{1}{c|}{$X_0 \sim \mathcal{N}(15.98, 4.17^2)$}  & N.A.                                   \\ \hline
Growth Rate & \multicolumn{2}{c|}{ $\gamma \sim \mathcal{N}(0.0475,0.008^2)$}
\\ \hline
Residual               & \multicolumn{1}{c|}{$\epsilon_P \sim \mathcal{N}(0, 0.4918^2)$}             & $\epsilon_I \sim \mathcal{N}(0, 0.4918^2)$             \\ \hline
Centrifuge             & \multicolumn{1}{c|}{N.A.}                                   & $Q \sim \mbox{Unif}(0.4,0.5)$          \\ \hline
Chromatography         & \multicolumn{1}{c|}{$Q_p \sim \mbox{Unif}(0.4833, 0.5907)$} & $Q_I \sim \mbox{Unif}(0.1458, 0.1782)$ \\ \hline
Filtration             & \multicolumn{1}{c|}{N.A.}                                   & $Q_{fr} \sim \mbox{Unif}(0.99, 1)$     \\ \hline
\end{tabular}
\end{table*}

Thus, this biopharmaceutical manufacturing example has $L = 8$ process models:
(1) $F_1$ modeling the residual or measurement error $\epsilon_P$; (2) $F_2$ modeling the batch-to-batch variation of the growth rate $\gamma$; 
(3) $F_3$ modeling the variation of the initial biomass $X_0$; (4) $F_4$ modeling the residual $\epsilon_I$ of impurity and metabolic waste accumulation; (5) $F_5$ modeling the random impurity removal ratio $Q$ at centrifuge step; (6) $F_6$ and $F_7$ modeling the random removal ratios, $Q_p$ and $Q_I$, of protein and impurity at chromatography step; and (7) $F_8$ modeling the random impurity removal ratio $Q_{fr}$ at filtration step. All the underlying true model parameters are summarized in Table~\ref{tab:truepara}. 
In the empirical study, we assume that these parameters are unknown and they are estimated with finite observations with size $m$. 
Since we often have very limited biopharmaceutical manufacturing process data available in the real world, we focus on the cases with $m= 10,20,40$ and let $m_\ell = m$ for $\ell= 1,2,\dots,L$. 

We assess the performances of $\mbox{CI}_+$ and $\mbox{CI}_0$ especially under the situation when the system has large simulation uncertainty. Therefore, the run length for each replication is set as 2 after the warm up equal to 25 in terms of the number of batches. 
For the proposed metamodel-assisted uncertainty analysis framework, when we build the GP metamodel, we set 
the number of design points $k = 20,40,80$. The same number of replications is assigned to each design point, i.e., $n_j = n = N/k$ for $j = 1, 2,\ldots, k$. To precisely estimate the percentile interval quantifying the system mean performance estimation uncertainty, we set the number of bootstrap resampled moments $B = 1000$ \cite{barton_nelson_xie_2011}. We compare the performance of our proposed framework with \textit{direct bootstrap} under the same computational budget. 
In the direct bootstrap approach, we run simulations at each bootstrapped moments to estimate the system mean response and equally allocate the simulation budget. It means that the number of replications at each bootstrapped moment sample is $n^d = N/B$. 
To assess the coverage of CIs, we conduct a side experiment with $10^6$ run length and 40 replications to estimate the true mean response and obtain $\mu(\bx_c) = 116.759 \pm 0.006$.

\subsubsection{Biomanufacturing System Uncertainty Quantification}
\label{subsec:bioprocessUQ}

Tables~\ref{tab:bioN2000} and \ref{tab:bioN4000} 
show the mean and standard deviation (SD) results of width and coverage of 95\% CIs, 
quantifying the overall estimation uncertainty 
of the expected productivity, obtained by the proposed metamodel-assisted uncertainty analysis framework and the direct bootstrap approach, when the simulation computational budget is $N = 2000,4000$. We also record the ratio of model uncertainty to total variance $\widehat{\sigma}^2_I/\widehat{\sigma}^2_T$. 
All results are based on 500 macro-replications.
As $m$ increases, 
the contribution of model uncertainty, measured by $\widehat{\sigma}^2_I/\widehat{\sigma}^2_T$, decreases.
The coverage of $\mbox{CI}_+$ is constantly better and closer to the nominal value of 95\% compared with $\mbox{CI}_0$.
The direct bootstrap approach has substantial 
over coverage issue, 
which was described and explained in \cite{barton2007presenting}. 
Since each experiment can be expensive and the average value of each batch of bio-drugs excesses one million, this over coverage issue can lead to overly conservative decision making and dramatically impact the profit. Given the fixed computational budget, as the number of real-world data $m$ increases, the mean and SD of the interval widths decrease, and the coverage becomes closer to the nominal value. 
\textit{Overall, the proposed metamodel-assisted uncertainty analysis will provide better performance, 
especially under the situation with very limited amount of real-world data and high model uncertainty,} which often happens in the biopharmaceutical manufacturing industry.

\begin{table*}[h!]
\small
\centering
\caption{The CIs results 
(SD)
of the expected productivity and $\widehat{\sigma}^2_I/\widehat{\sigma}^2_T$
when $N=2000$. 
}
\label{tab:bioN2000}
\begin{tabular}{|c|ccc|c|}
\hline
\multirow{2}{*}{$m = 10$}             & \multicolumn{3}{c|}{Metamodel-Assisted Uncertainty Analysis}                                   & \multirow{2}{*}{\begin{tabular}[c]{@{}c@{}}Direct\\ Bootstrap\end{tabular}} \\ \cline{2-4}
                                        & \multicolumn{1}{c|}{$k$ = 20, $n$=100} & \multicolumn{1}{c|}{$k$ = 40, $n$=50} & $k$ = 80, $n$=25 &                                   \\ \hline
Coverage of   $\mbox{CI}_0$             & \multicolumn{1}{c|}{84.80\%}           & \multicolumn{1}{c|}{88.20\%}          & 89.40\%          & \multirow{2}{*}{99.60\%}          \\ \cline{1-4}
Coverage of   $\mbox{CI}_+$             & \multicolumn{1}{c|}{88.60\%}           & \multicolumn{1}{c|}{90.40\%}          & 92.00\%          &                                   \\ \hline
$\mbox{CI}_0$ Width                & \multicolumn{1}{c|}{89.60 (32.99)}             & \multicolumn{1}{c|}{99.19 (37.08)}            &  98.54 (33.62)           & \multirow{2}{*}{\begin{tabular}[c]{@{}c@{}}224.21\\ (81.25)\end{tabular}}            \\ \cline{1-4}
$\mbox{CI}_+$ Width                & \multicolumn{1}{c|}{103.21 (35.81)}             & \multicolumn{1}{c|}{109.60 (38.66)}            & 102.81 (35.09)           &                                   \\ \hline
$\widehat{\sigma}^2_I/\widehat{\sigma}^2_T$ & \multicolumn{1}{c|}{80.26\%}           & \multicolumn{1}{c|}{88.99\%}          & 89.51\%          & 61.10\%                           \\ \hline
\hline
\multirow{2}{*}{$m = 20$}             & \multicolumn{3}{c|}{Metamodel-Assisted Uncertainty Analysis}                                   & \multirow{2}{*}{\begin{tabular}[c]{@{}c@{}}Direct\\ Bootstrap\end{tabular}} \\ \cline{2-4}
                                        & \multicolumn{1}{c|}{$k$ = 20, $n$=100} & \multicolumn{1}{c|}{$k$ = 40, $n$=50} & $k$ = 80, $n$=25 &                                   \\ \hline
Coverage of    $\mbox{CI}_0$            & \multicolumn{1}{c|}{85.80\%}           & \multicolumn{1}{c|}{89.60\%}          & 89.60\%          & \multirow{2}{*}{100.00\%}          \\ \cline{1-4}
Coverage of   $\mbox{CI}_+$             & \multicolumn{1}{c|}{92.60\%}           & \multicolumn{1}{c|}{93.00\%}          & 92.60\%          &                                   \\ \hline
$\mbox{CI}_0$ Width              & \multicolumn{1}{c|}{64.09 (19.68)}             & \multicolumn{1}{c|}{66.98 (17.18)}            & 70.96 (17.84)            & \multirow{2}{*}{\begin{tabular}[c]{@{}c@{}}205.85 \\ (48.12)\end{tabular}}            \\ \cline{1-4}
$\mbox{CI}_+$ Width                & \multicolumn{1}{c|}{75.46 (20.93)}             & \multicolumn{1}{c|}{74.69 (17.29)}            & 79.01 (19.37)            &                                   \\ \hline
$\widehat{\sigma}^2_I/\widehat{\sigma}^2_T$ & \multicolumn{1}{c|}{76.55\%}           & \multicolumn{1}{c|}{84.15\%}          & 83.62\%          & 63.69\%                           \\ \hline
\hline
\multirow{2}{*}{$m = 40$}              & \multicolumn{3}{c|}{Metamodel-Assisted Uncertainty Analysis}                                   & \multirow{2}{*}{\begin{tabular}[c]{@{}c@{}}Direct\\ Bootstrap\end{tabular}} \\ \cline{2-4}
                                        & \multicolumn{1}{c|}{$k$ = 20, $n$=100} & \multicolumn{1}{c|}{$k$ = 40, $n$=50} & $k$ = 80, $n$=25 &                                   \\ \hline
Coverage of    $\mbox{CI}_0$            & \multicolumn{1}{c|}{84.20\%}           & \multicolumn{1}{c|}{91.00\%}          & 86.40\%          & \multirow{2}{*}{100.00\%}          \\ \cline{1-4}
Coverage of   $\mbox{CI}_+$             & \multicolumn{1}{c|}{93.80\%}           & \multicolumn{1}{c|}{95.40\%}          & 92.40\%          &                                   \\ \hline
$\mbox{CI}_0$ Width              & \multicolumn{1}{c|}{43.62 (11.38)}             & \multicolumn{1}{c|}{47.58 (10.26)}            & 48.09 (10.42)           & \multirow{2}{*}{\begin{tabular}[c]{@{}c@{}}196.75 \\ (32.74)\end{tabular}}            \\ \cline{1-4}
$\mbox{CI}_+$ Width                & \multicolumn{1}{c|}{55.23 (12.59)}             & \multicolumn{1}{c|}{56.79 (10.86)}            & 57.74 (11.73)           &                                   \\ \hline
$\widehat{\sigma}^2_I/\widehat{\sigma}^2_T$ & \multicolumn{1}{c|}{68.46\%}           & \multicolumn{1}{c|}{74.73\%}          & 73.96\%          & 65.03\%                           \\ \hline
\end{tabular}
\end{table*}

\begin{table*}[h!]
\small
\centering
\caption{The CIs results (SD) of the expected productivity and $\widehat{\sigma}^2_I/\widehat{\sigma}^2_T$
when $N=4000$.} 

\label{tab:bioN4000}
\begin{tabular}{|c|ccc|c|}
\hline
\multirow{2}{*}{$m = 10$}             & \multicolumn{3}{c|}{Metamodel-Assisted Uncertainty Analysis}                                   & \multirow{2}{*}{\begin{tabular}[c]{@{}c@{}}Direct\\ Bootstrap\end{tabular}} \\ \cline{2-4}
                                        & \multicolumn{1}{c|}{$k$ = 20, $n$=200} & \multicolumn{1}{c|}{$k$ = 40, $n$=100} & $k$ = 80, $n$=50 &                                   \\ \hline
Coverage of   $\mbox{CI}_0$             & \multicolumn{1}{c|}{86.80\%}           & \multicolumn{1}{c|}{89.40\%}          & 91.20\%          & \multirow{2}{*}{99.40\%}          \\ \cline{1-4}
Coverage of   $\mbox{CI}_+$             & \multicolumn{1}{c|}{91.20\%}           & \multicolumn{1}{c|}{90.60\%}          & 92.80\%          &                                   \\ \hline
$\mbox{CI}_0$ Width                & \multicolumn{1}{c|}{91.96 (33.84)}             & \multicolumn{1}{c|}{103.24 (37.66)}            &  99.45 (35.73)           & \multirow{2}{*}{\begin{tabular}[c]{@{}c@{}}178.19\\ (64.18)\end{tabular}}            \\ \cline{1-4}
$\mbox{CI}_+$ Width                & \multicolumn{1}{c|}{102.84 (34.90)}             & \multicolumn{1}{c|}{108.23 (38.96)}            & 103.46 (36.72)           &                                   \\ \hline
$\widehat{\sigma}^2_I/\widehat{\sigma}^2_T$ & \multicolumn{1}{c|}{83.52\%}           & \multicolumn{1}{c|}{92.78\%}          & 93.38\%          & 72.20\%                           \\ \hline
\hline
\multirow{2}{*}{$m = 20$}             & \multicolumn{3}{c|}{Metamodel-Assisted Uncertainty Analysis}                                   & \multirow{2}{*}{\begin{tabular}[c]{@{}c@{}}Direct\\ Bootstrap\end{tabular}} \\ \cline{2-4}
                                        & \multicolumn{1}{c|}{$k$ = 20, $n$=200} & \multicolumn{1}{c|}{$k$ = 40, $n$=100} & $k$ = 80, $n$=50 &                                   \\ \hline
Coverage of    $\mbox{CI}_0$            & \multicolumn{1}{c|}{88.80\%}           & \multicolumn{1}{c|}{91.20\%}          & 91.60\%          & \multirow{2}{*}{100.00\%}          \\ \cline{1-4}
Coverage of   $\mbox{CI}_+$             & \multicolumn{1}{c|}{92.40\%}           & \multicolumn{1}{c|}{93.00\%}          & 93.40\%          &                                   \\ \hline
$\mbox{CI}_0$ Width              & \multicolumn{1}{c|}{65.58 (21.10)}             & \multicolumn{1}{c|}{69.49 (16.92)}            & 73.37 (17.45)            & \multirow{2}{*}{\begin{tabular}[c]{@{}c@{}}156.94 \\ (39.69)\end{tabular}}            \\ \cline{1-4}
$\mbox{CI}_+$ Width                & \multicolumn{1}{c|}{75.87 (21.68)}             & \multicolumn{1}{c|}{74.64 (17.59)}            & 78.50 (18.43)            &                                   \\ \hline
$\widehat{\sigma}^2_I/\widehat{\sigma}^2_T$ & \multicolumn{1}{c|}{79.08\%}           & \multicolumn{1}{c|}{88.86\%}          & 89.13\%          & 75.56\%                           \\ \hline
\hline
\multirow{2}{*}{$m = 40$}              & \multicolumn{3}{c|}{Metamodel-Assisted Uncertainty Analysis}                                   & \multirow{2}{*}{\begin{tabular}[c]{@{}c@{}}Direct\\ Bootstrap\end{tabular}} \\ \cline{2-4}
                                        & \multicolumn{1}{c|}{$k$ = 20, $n$=200} & \multicolumn{1}{c|}{$k$ = 40, $n$=100} & $k$ = 80, $n$=50 &                                   \\ \hline
Coverage of    $\mbox{CI}_0$            & \multicolumn{1}{c|}{87.00\%}           & \multicolumn{1}{c|}{93.60\%}          & 91.60\%          & \multirow{2}{*}{100.00\%}          \\ \cline{1-4}
Coverage of   $\mbox{CI}_+$             & \multicolumn{1}{c|}{94.80\%}           & \multicolumn{1}{c|}{95.00\%}          & 94.80\%          &                                   \\ \hline
$\mbox{CI}_0$ Width              & \multicolumn{1}{c|}{45.69 (11.59)}             & \multicolumn{1}{c|}{49.66 (9.71)}            & 51.19 (10.27)           & \multirow{2}{*}{\begin{tabular}[c]{@{}c@{}}145.75 \\ (30.15)\end{tabular}}            \\ \cline{1-4}
$\mbox{CI}_+$ Width                & \multicolumn{1}{c|}{54.52 (12.42)}             & \multicolumn{1}{c|}{55.87 (10.04)}            & 57.18 (10.79)           &                                   \\ \hline
$\widehat{\sigma}^2_I/\widehat{\sigma}^2_T$ & \multicolumn{1}{c|}{74.93\%}           & \multicolumn{1}{c|}{82.37\%}          & 82.99\%          & 77.71\%                           \\ \hline
\end{tabular}
\end{table*}

\subsubsection{Biomanufacturing System Variance Decomposition}
\label{subsec:bioprocessshapley}

When the model uncertainty plays a dominate impact on the system performance estimation uncertainty, it is critical to identify the key source, which can be used to efficiently improve the simulation model. Based on the analytical study in Section~\ref{subSec:varDecModelUncertainty}, 
the means with 95\% CI of the relative contribution from each $\ell$-th model uncertainty, i.e., ($\widehat{s}_{\ell}/{\widehat{\sigma}^2_I}\times 100\%$), 
are recorded in Table~\ref{tab:ModelRiskPro}. 
The results are estimated based on 100 macro-replications.
We set the number of bootstrapped moments used for the variance estimation $B'= 2000$.  
Since the model uncertainty of protein generation process characterized by models for $\{\epsilon_P, \gamma, X_0\}$ dominates, we gradually increase $m_\ell$ with $\ell=1,2,3$ as $m'=10,20,40$, while fixing the number of real-world data for remaining models $m_\ell = 10$ for $\ell = 4,5, \ldots ,L$.  The order of importance, $\epsilon_P > \gamma > X_0$ is consistent across all sample sizes. Of the remaining variables, the removal proportion of protein at chromatography, $Q_P$, provides the largest proportion of contribution across all sample sizes and it increases dramatically as the sample size increases. 
As the sample size $m'$ increases, the relative contribution from model uncertainty of $\{X_0,\gamma,\epsilon_P\}$ reduces. The overall model uncertainty, measured by $\sigma_I$, also decreases with increasing sample size. 

This case study is motivated by a real animal bio-drug production. The quality requirement, i.e., $
\frac{I_{fr}}{X_{fr} + I_{fr}} \leq \omega$ with $\omega = 25\%$, is relatively easy to meet through downstream purification. 
Thus, the results in Table~\ref{tab:ModelRiskPro} indicate that the influence of the impurity pathway parameters is negligible.
This observation does not hold in general, especially for antigen proteins for human beings that typically have much more restrictive quality requirements (say $\omega = 1\%$).

\begin{table*}[hbt!]
\small
\centering
\caption{The relative contributions from each model uncertainty when $m'=10,20,40$.}
\label{tab:ModelRiskPro}
\begin{tabular}{|c|c|c|c|}
\hline
Process Model                & $m'$=10          & $m'$=20         & $m'$=40         \\ \hline
$\epsilon_P$                  & 44.51\% $\pm$ 5.05\%   & 40.32\% $\pm$ 4.74\%  & 37.73\% $\pm$ 4.53\% \\ \hline
$\gamma$               & 35.18\% $\pm$ 4.93\% & 31.89\% $\pm$ 4.18\% & 28.44\% $\pm$ 3.69\% \\ \hline
$X_0$               & 15.04\% $\pm$ 4.55\% & 14.34\% $\pm$ 4.28\% & 11.88\% $\pm$ 3.89\% 
\\ \hline \hline
$Q_P$                & 3.87\% $\pm$ 0.88\%   & 10.44\% $\pm$ 2.16\%  & 18.06\% $\pm$ 3.32\%  \\ \hline
$\epsilon_I$                & 0.83\% $\pm$ 1.11\%   & 1.90\% $\pm$ 1.81\%  & 2.28\% $\pm$ 2.02\%  \\ \hline
$Q$                & 0.18\% $\pm$ 0.35\%   & 0.33\% $\pm$ 0.53\%  & 0.70\% $\pm$ 0.89\%  \\ \hline
$Q_I$                & 0.29\% $\pm$ 0.23\%   & 0.35\% $\pm$ 0.37\%  & 0.45\% $\pm$ 0.54\%  \\ \hline
$Q_{fr}$             & 0.04\% $\pm$ 0.12\%   & 0.23\% $\pm$ 0.19\%  & 0.64\% $\pm$ 0.94\%  \\ \hline 
\hline
$\widehat{\sigma}_I$ & 25.43 $\pm$ 2.01 & 18.10 $\pm$ 1.21 & 13.51 $\pm$ 0.65 \\ \hline
\end{tabular}
\end{table*}

\subsection{Cell Culture Expansion Scheduling 
for Cell Therapy Manufacturing}
\label{subsec:Cell_Exten}


Here we use the erythroblast cell therapy manufacturing example presented in \cite{glen2018mechanistic} to assess the performance of proposed framework. The cell culture of erythroblast exhibits two phases: a relatively uninhibited growth phase followed by an inhibited phase. 
The hybrid model cell growth and inhibitor accumulation is
\begin{align}
    \rho_{t+1} &= \rho_t + \Delta t \cdot r^g \rho_t \Bigg (1 - \Big(1+e^{(k^s(k^c-I_t))} \Big) ^{-1} \Bigg ) + e^{\rho}_{t},\nonumber\\
    I_{t+1} &= I_t + \Delta t \cdot \Bigg (\frac{\rho_{t+1}-\rho_t}{\Delta t} - r^d I_t \Bigg) + e^{I}_{t}, \nonumber
\end{align}
where $\Delta t$ represents the time interval, $\rho_t$ and $I_t$ represent the cell density and the unobservable inhibitor concentration at the $t$-th time step
. The kinetic coefficients $r^g$, $k^s$, $k^c$ and $r^d$ 
denote the cell growth rate, inhibitor sensitivity, inhibitor threshold, and inhibitor decay. 
The residuals follow the normal distributions, i.e., $e_{t}^{\rho} \sim {N}(0,(v^{\rho})^{2})$ and $e_{t}^{I} \sim {N}(0,(v^{I})^{2})$.
There is raw material uncertainty for seed cell density, i.e., $\rho_0 \sim {N}(\mu_{\rho}, \sigma_{\rho}^2)$. The initial inhibitor concentration equals to 0 due to the fresh medium, i.e., $I_0 = 0$. Additionally, the investigation from \cite{glen2018mechanistic} shows that the growth rate has significant variability cross different donors.  Therefore, we incorporate batch-to-batch variation 
by considering the random effect on the growth rate, i.e., $r^g \sim \mathcal{N}(\mu^g,(\sigma^g)^2)$. 

\begin{sloppypar}
Thus, this erythroblast cell therapy manufacturing example has $L = 7$ process models: (1) $F_1$ for $\rho_0$; (2) $F_2$ for $e^\rho$; (3) $F_3$ for $e^I$; (4) $F_4$ for $r^g$; and 
(5--7) the degenerate distributions $F_5, F_6, F_7$ for bioprocess kinetic parameters ${k}^s, {k}^c,{r}^d$.
Set the underlying true parameters as $\{\mu_{\rho}, \sigma_{\rho}, v^{\rho}, v^{I}, \mu^g,\sigma^g\} =  \{3,0.03,0.01,0.01,0.037,0.008\}$ 
and $\{k^s, k^c,r^d\} = \{3.4,2.6,0.005\}$, which are validated by using the real-world data presented in \cite{glen2018mechanistic}.
In this empirical study, we assume that all these parameters are unknown and estimated with a finite amount of real-world data with size $m$. The cell density data are collected every 4 hours, i.e., $\Delta t=4$ hours. Thus, we have $m$ trajectory observations, i.e., $\pmb\tau^{(i)} \equiv (\rho_0^{(i)},\rho_1^{(i)},\ldots,\rho_{T}^{(i)})$ with $i = 1,2,\ldots,m$.
\end{sloppypar}

At any time $t$, if the batch-extension is performed, 
the original batch is scaled up to a $\lambda$ times larger cell culture vessel filling with fresh medium. That means the cell density $\rho$ and the concentration of inhibitor $I$ decrease to $1/\lambda$ of original values. In this example, suppose that the batch-extension is scheduled at the 24-th hour (corresponding to time step $t = \frac{24}{\Delta t} + 1 = 7$). Then, the original batch is scaled up to $\lambda=4$ fold. The cell culture process ends at $T=40$ hours (corresponding to time step $t = \frac{T}{\Delta t} + 1 = 11$).
Our goal is to estimate the expected productivity in terms of total biomass of target cells, i.e., $\mu(\mathbf{x}_c) = 
\mbox{E}[\rho_T \cdot \lambda]$.

We focus on the cases with $m=3,6,20$ and let $m_\ell = m$ for $\ell= 1,2,\dots,L$. The total simulation budget is set to be $N= 4000$ replications. 
We compare the performance of our proposed framework with direct bootstrap approach under the same computational budget. For the proposed metamodel-assisted uncertainty analysis framework, we set the number of design points $k = 20,40,80$. The same number of replications is assigned to each design point, i.e., $n_j = n = N/k$ for $j = 1, 2,\ldots, k$. The number of bootstrap resampled moments is set as $B = 1000$. In the direct bootstrap approach, the number of replications allocated at each bootstrapped moment sample is $n^d = N/B = 4$. To assess the coverage of CIs, we conduct a side experiment with $10^6$ batches and 20 replications to estimate the true mean response and obtain $\mu(\bx_c) = 17.32 \pm 0.004$.

Table~\ref{tab:CellextensionN4000} 
records the mean and standard deviation (SD) results of width and coverage of 95\% CIs, 
quantifying the overall estimation uncertainty 
of the expected productivity, obtained by the proposed metamodel-assisted uncertainty analysis framework and the direct bootstrap approach. We also record the ratio of model uncertainty to total variance $\widehat{\sigma}^2_I/\widehat{\sigma}^2_T$. 
All results are based on 500 macro-replications.
The coverage of $\mbox{CI}_+$ is much closer to the nominal value of 95\%, when compare with $\mbox{CI}_0$. 
The direct bootstrap again exhibits overcoverage and provides much wider confidence interval width means and standard deviations. Given the fixed computational budget, as the number of real-world data $m$ increases, the mean and SD of the interval widths decrease, and the coverage becomes closer to the nominal value. 

\begin{table*}[h!]
\footnotesize
\centering
\caption{The CIs results (SD) of the expected productivity and $\widehat{\sigma}^2_I/\widehat{\sigma}^2_T$
when $N=4000$.}
\label{tab:CellextensionN4000}
\begin{tabular}{|c|ccc|c|}
\hline
\multirow{2}{*}{$m$   = 3}              & \multicolumn{3}{c|}{Metamodel-Assisted Uncertainty Analysis}                                   & \multirow{2}{*}{\begin{tabular}[c]{@{}c@{}}Direct\\ Bootstrap\end{tabular}} \\ \cline{2-4}
                                        & \multicolumn{1}{c|}{$k$ = 20, $n$=200} & \multicolumn{1}{c|}{$k$ = 40, $n$=100} & $k$ = 80, $n$=50 &                                   \\ \hline
Coverage of   $\mbox{CI}_0$             & \multicolumn{1}{c|}{83.20\%}           & \multicolumn{1}{c|}{86.20\%}          & 84.40\%          & \multirow{2}{*}{99.80\%}          \\ \cline{1-4}
Coverage of   $\mbox{CI}_+$             & \multicolumn{1}{c|}{90.20\%}           & \multicolumn{1}{c|}{91.20\%}          & 90.80\%          &                                   \\ \hline
$\mbox{CI}_0$ Width              & \multicolumn{1}{c|}{4.67 (2.11)}             & \multicolumn{1}{c|}{4.23 (2.42)}            & 4.13 (2.45)          & \multirow{2}{*}{\begin{tabular}[c]{@{}c@{}}7.12 \\ (3.75)\end{tabular}}            \\ \cline{1-4}
$\mbox{CI}_+$ Width                & \multicolumn{1}{c|}{5.03 (2.75)}             & \multicolumn{1}{c|}{5.36 (2.76)}            & 5.21 (2.52)           &                                   \\ \hline
$\widehat{\sigma}^2_I/\widehat{\sigma}^2_T$ & \multicolumn{1}{c|}{86.17\%}           & \multicolumn{1}{c|}{90.23\%}          & 87.32\%          & 87.21\%                           \\ \hline
\hline
\multirow{2}{*}{$m$   = 6}             & \multicolumn{3}{c|}{Metamodel-Assisted Uncertainty Analysis}                                   & \multirow{2}{*}{\begin{tabular}[c]{@{}c@{}}Direct\\ Bootstrap\end{tabular}} \\ \cline{2-4}
                                        & \multicolumn{1}{c|}{$k$ = 20, $n$=100} & \multicolumn{1}{c|}{$k$ = 40, $n$=50} & $k$ = 80, $n$=25 &                                   \\ \hline
Coverage of   $\mbox{CI}_0$             & \multicolumn{1}{c|}{89.60\%}           & \multicolumn{1}{c|}{89.00\%}          & 89.20\%          & \multirow{2}{*}{100.00\%}          \\ \cline{1-4}
Coverage of   $\mbox{CI}_+$             & \multicolumn{1}{c|}{92.80\%}           & \multicolumn{1}{c|}{93.40\%}          & 91.60\%          &                                   \\ \hline
$\mbox{CI}_0$ Width   Mean              & \multicolumn{1}{c|}{2.99 (1.86)}             & \multicolumn{1}{c|}{3.14 (1.76)}            &  3.25  (1.78)        & \multirow{2}{*}{\begin{tabular}[c]{@{}c@{}}5.35 \\ (2.52) \end{tabular}}            \\ \cline{1-4}
$\mbox{CI}_+$ Width Mean                & \multicolumn{1}{c|}{3.34 (1.91)}             & \multicolumn{1}{c|}{3.42 (1.82)}            & 3.43 (1.84)           &                                   \\ \hline
$\widehat{\sigma}^2_I/\widehat{\sigma}^2_T$ & \multicolumn{1}{c|}{86.34\%}           & \multicolumn{1}{c|}{90.41\%}          & 91.02\%          & 74.83\%                           \\ \hline
\hline
\multirow{2}{*}{$m$ =   20}             & \multicolumn{3}{c|}{Metamodel-Assisted Uncertainty Analysis}                                   & \multirow{2}{*}{\begin{tabular}[c]{@{}c@{}}Direct\\ Bootstrap\end{tabular}} \\ \cline{2-4}
                                        & \multicolumn{1}{c|}{$k$ = 20, $n$=100} & \multicolumn{1}{c|}{$k$ = 40, $n$=50} & $k$ = 80, $n$=25 &                                   \\ \hline
Coverage of    $\mbox{CI}_0$            & \multicolumn{1}{c|}{93.40\%}           & \multicolumn{1}{c|}{94.00\%}          & 93.60\%          & \multirow{2}{*}{97.80\%}          \\ \cline{1-4}
Coverage of   $\mbox{CI}_+$             & \multicolumn{1}{c|}{95.40\%}           & \multicolumn{1}{c|}{95.00\%}          & 95.20\%          &                                   \\ \hline
$\mbox{CI}_0$ Width   Mean              & \multicolumn{1}{c|}{1.68 (1.05)}             & \multicolumn{1}{c|}{1.72 (1.09)}            & 1.74 (1.12)           & \multirow{2}{*}{\begin{tabular}[c]{@{}c@{}} 3.84\\ (1.72)\end{tabular}}            \\ \cline{1-4}
$\mbox{CI}_+$ Width Mean                & \multicolumn{1}{c|}{1.79 (1.10)}             & \multicolumn{1}{c|}{1.83 (1.12)}            & 1.86 (1.15)           &                                   \\ \hline
$\widehat{\sigma}^2_I/\widehat{\sigma}^2_T$ & \multicolumn{1}{c|}{81.20\%}           & \multicolumn{1}{c|}{85.16\%}          & 84.65\%          & 80.14\%                           \\ \hline
\end{tabular}
\end{table*}

\section{Conclusions}
\label{sec:conclusion}

To efficiently develop a simulation model 
to improve the assessment of the mean response for flexible and integrated biomanufacturing systems with modular design, we propose a metamodel-assisted bootstrapping 
uncertainty quantification and sensitivity analysis
framework. Process model uncertainty is approximated
by the bootstrap and an equation-based stochastic kriging metamodel is used to propagate the model uncertainty to the output mean. The
simulation uncertainty is derived using properties of stochastic kriging. This framework delivers an interval quantifying the system mean response estimation accuracy accounting for both simulation and model uncertainties. The asymptotic consistency of this interval is proved under the assumption that the true response surface is a realization of a Gaussian process and certain parameters are known.
Given very limited real-world observations and high stochastic uncertainty,  the model uncertainty often dominates, especially for personalized bio-drug manufacturing.
We provide a variance decomposition quantifying the relative contribution from each source of model uncertainty, as well as simulation uncertainty. 
While the asymptotic analysis shows correctness for the proposed framework, the empirical study on multiple biomanufacturing and service examples demonstrates that it also has good finite-sample performance.

\section*{Acknowledgments}

This paper is based upon work supported by the National Science Foundation under Grant No
.\ CMMI-0900354 and CMMI-1068473, National Institute of Standards and Technology (70NANB17H002), Department of Commerce. We also would like to thank the anonymous reviewers for their comments that have helped us improve the manuscript.
\bibliographystyle{plain}
\bibliography{paper2}

\begin{thebibliography}{10}

\bibitem{Adler_2010}
R.~J. Adler.
\newblock {\em The Geometry of Random Fields}.
\newblock SIAM, Philadelphia, PA, 2010.

\bibitem{ankenman_nelson_staum_2010}
B.~E. Ankenman, B.~L. Nelson, and J.~Staum.
\newblock Stochastic kriging for simulation metamodeling.
\newblock {\em Operations Research}, 58:371--382, 2010.

\bibitem{barton_nelson_xie_2011}
R.~R. Barton, B.~L. Nelson, and W.~Xie.
\newblock Quantifying input uncertainty via simulation confidence interval.
\newblock {\em Informs Journal on Computing}, 26:74--87, 2014.

\bibitem{barton2007presenting}
Russell~R Barton et~al.
\newblock Presenting a more complete characterization of uncertainty: Can it be
  done.
\newblock In {\em Proceedings of the 2007 INFORMS simulation society research
  workshop}, pages 26--60. INFORMS Simulation Society, 2007.

\bibitem{barton2001resampling}
Russell~R Barton and Lee~W Schruben.
\newblock Resampling methods for input modeling.
\newblock In {\em Proceeding of the 2001 Winter Simulation Conference (Cat. No.
  01CH37304)}, volume~1, pages 372--378. IEEE, 2001.

\bibitem{biller2011accounting}
Bahar Biller and Canan~G Corlu.
\newblock Accounting for parameter uncertainty in large-scale stochastic
  simulations with correlated inputs.
\newblock {\em Operations Research}, 59(3):661--673, 2011.

\bibitem{Billingsley_1995}
P.~Billingsley.
\newblock {\em Probability and Measure}.
\newblock Wiley-Interscience, New York, 1995.

\bibitem{borgonovo2016sensitivity}
Emanuele Borgonovo and Elmar Plischke.
\newblock Sensitivity analysis: a review of recent advances.
\newblock {\em European Journal of Operational Research}, 248(3):869--887,
  2016.

\bibitem{corlu2020stochastic}
Canan~G Corlu, Alp Akcay, and Wei Xie.
\newblock Stochastic simulation under input uncertainty: A review.
\newblock {\em Operations Research Perspectives}, page 100162, 2020.

\bibitem{delahaye2015ultra}
M~Delahaye, K~Lawrence, SJ~Ward, and M~Hoare.
\newblock An ultra scale-down analysis of the recovery by dead-end
  centrifugation of human cells for therapy.
\newblock {\em Biotechnology and Bioengineering}, 112(5):997--1011, 2015.

\bibitem{doran2013bioprocess}
Pauline~M Doran.
\newblock {\em Bioprocess Engineering Principles}.
\newblock Academic Press, 2012.

\bibitem{glen2018mechanistic}
Katie~E Glen, Elizabeth~A Cheeseman, Adrian~J Stacey, and Robert~J Thomas.
\newblock A mechanistic model of erythroblast growth inhibition providing a
  framework for optimisation of cell therapy manufacturing.
\newblock {\em Biochemical Engineering Journal}, 133:28--38, 2018.

\bibitem{hernandez2019predicting}
Tanja Hern{\'a}ndez~Rodr{\'\i}guez, Christoph Posch, Julia Schmutzhard, Josef
  Stettner, Claus Weihs, Ralf P{\"o}rtner, and Bj{\"o}rn Frahm.
\newblock Predicting industrial-scale cell culture seed trains--a bayesian
  framework for model fitting and parameter estimation, dealing with
  uncertainty in measurements and model parameters, applied to a nonlinear
  kinetic cell culture model, using an mcmc method.
\newblock {\em Biotechnology and Bioengineering}, 116(11):2944--2959, 2019.

\bibitem{Jones_1998}
D.~Jones, M.~Schonlau, and W~Welch.
\newblock Efficient global optimization of expensive black-box functions.
\newblock {\em Journal of Global Optimization}, 13:455--492, 1998.

\bibitem{kleijnen2017regression}
Jack~PC Kleijnen.
\newblock Regression and kriging metamodels with their experimental designs in
  simulation: a review.
\newblock {\em European Journal of Operational Research}, 256(1):1--16, 2017.

\bibitem{lam2018subsampling}
Henry Lam and Huajie Qian.
\newblock Subsampling variance for input uncertainty quantification.
\newblock In {\em 2018 Winter Simulation Conference (WSC)}, pages 1611--1622.
  IEEE, 2018.

\bibitem{lam2022subsampling}
Henry Lam and Huajie Qian.
\newblock Subsampling to enhance efficiency in input uncertainty
  quantification.
\newblock {\em Operations Research}, 70(3):1891--1913, 2022.

\bibitem{Lehmann_Casella_1998}
E.L. Lehmann and G.~Casella.
\newblock {\em Theory of Point Estimation}.
\newblock Springer-Verlag, New York, 1998.

\bibitem{Loeppky_2009}
J.~L. Loeppky, J.~Sachs, and W.~J Welch.
\newblock Choosing the sample size of a computer experiment: A practical guide.
\newblock {\em Technometrics}, 51:366--376, 2009.

\bibitem{martagan2017performance}
Tugce Martagan, Ananth Krishnamurthy, Peter~A Leland, and Christos~T
  Maravelias.
\newblock Performance guarantees and optimal purification decisions for
  engineered proteins.
\newblock {\em Operations Research}, 66(1):18--41, 2017.

\bibitem{mockus2015batch}
Linas Mockus, John~J Peterson, Jose~Miguel Lainez, and Gintaras~V Reklaitis.
\newblock Batch-to-batch variation: a key component for modeling chemical
  manufacturing processes.
\newblock {\em Organic Process Research \& Development}, 19(8):908--914, 2015.

\bibitem{moller2020model}
Johannes M{\"o}ller, Tanja~Hern{\'a}ndez Rodr{\'\i}guez, Jan M{\"u}ller, Lukas
  Arndt, Kim~B Kuchem{\"u}ller, Bj{\"o}rn Frahm, Regine Eibl, Dieter Eibl, and
  Ralf P{\"o}rtner.
\newblock Model uncertainty-based evaluation of process strategies during
  scale-up of biopharmaceutical processes.
\newblock {\em Computers \& Chemical Engineering}, 134:106693, 2020.

\bibitem{OBrien_2021}
Conor~M. O'Brien, Qi~Zhang, Prodromos Daoutidis, and Wei-Shou Hu.
\newblock A hybrid mechanistic-empirical model for in silico mammalian cell
  bioprocess simulation.
\newblock {\em Metabolic Engineering}, 66:31--40, 2021.

\bibitem{Picheny_2010}
V.~Picheny, D.~Ginsbourger, O.~Roustant, R.~T. Haftka, and N.~Kim.
\newblock Adaptive designs of experiments for accurate approximation of a
  target region.
\newblock {\em Journal of Mechanical Design}, 132:071008, 2010.

\bibitem{rodriguez2020digital}
Tanja~Hern{\'a}ndez Rodr{\'\i}guez and Bj{\"o}rn Frahm.
\newblock Digital seed train twins and statistical methods.
\newblock {\em Advances in Biochemical Engineering and Biotechnology},
  176:97--131, 2021.

\bibitem{Santer_2003}
T.~J. Santner, B.~J. Williams, and W.~I. Notz.
\newblock {\em The Design and Analysis of Computer Experiments}.
\newblock Springer, New York, 2003.

\bibitem{Serfling_2002}
R.~J. Serfling.
\newblock {\em Approximation Theorems of Mathematical Statistics}.
\newblock Wiley, New York, 2002.

\bibitem{Severini_2005}
T.A. Severini.
\newblock {\em Elements of Distribution Theory}.
\newblock Cambridge University Press, New York, 2005.

\bibitem{Shao_1995}
J.~Shao and D.~Tu.
\newblock {\em The Jackknife and Bootstrap}.
\newblock Springer, New York, 1995.

\bibitem{SongNelson_2013}
Eunhye Song and Barry~L Nelson.
\newblock A quicker assessment of input uncertainty.
\newblock In {\em 2013 Winter Simulations Conference (WSC)}, pages 474--485.
  IEEE, 2013.

\bibitem{song2016shapley}
Eunhye Song, Barry~L Nelson, and Jeremy Staum.
\newblock Shapley effects for global sensitivity analysis: Theory and
  computation.
\newblock {\em SIAM/ASA Journal on Uncertainty Quantification},
  4(1):1060--1083, 2016.

\bibitem{SunFarooq_2002}
H.~Sun and M.~Farooq.
\newblock Note on the generation of random points uniformly distributed in
  hyper-ellipsoids.
\newblock In {\em Proceedings of the Fifth International Conference on
  Information Fusion}, pages 489--496, 2002.

\bibitem{Vaart_1998}
A.~W. Van Der~Vaart.
\newblock {\em Asymptotic Statistics}.
\newblock Cambridge University Press, Cambridge, UK, 1998.

\bibitem{wang2019stochastic}
Bo~Wang, Wei Xie, Tugce Martagan, Alp Akcay, and Canan~G Corlu.
\newblock Stochastic simulation model development for biopharmaceutical
  production process risk analysis and stability control.
\newblock In {\em 2019 Winter Simulation Conference (WSC)}, pages 1989--2000.
  IEEE, 2019.

\bibitem{wang2021hyperparm}
Peng Wang, Lyudmila Mihaylova, Rohit Chakraborty, Said Munir, Martin Mayfield,
  Khan Alam, Muhammad~Fahim Khokhar, Zhengkai Zheng, Chengxi Jiang, and Hui
  Fang.
\newblock A gaussian process method with uncertainty quantification for air
  quality monitoring.
\newblock {\em Atmosphere}, 12(1344):18, 2021.

\bibitem{xie_nelson_staum_2010}
W.~Xie, B.~L. Nelson, and J.~Staum.
\newblock The influence of correlation functions on stochastic kriging
  metamodels.
\newblock In {\em 2010 Winter Simulation Conference (WSC)}, pages 1067--1078.
  IEEE, 2010.

\bibitem{xie2017factor}
Wei Xie, Cheng Li, and Pu~Zhang.
\newblock A factor-based bayesian framework for risk analysis in stochastic
  simulations.
\newblock {\em ACM Transactions on Modeling and Computer Simulation (TOMACS)},
  27(4):1--31, 2017.

\bibitem{xie2022interpretable}
Wei Xie, Bo~Wang, Cheng Li, Dongming Xie, and Jared Auclair.
\newblock Interpretable biomanufacturing process risk and sensitivity analyses
  for quality-by-design and stability control.
\newblock {\em Naval Research Logistics (NRL)}, 69(3):461--483, 2022.

\bibitem{xie2018metamodel}
Wei Xie, Bo~Wang, and Qiong Zhang.
\newblock Metamodel-assisted risk analysis for stochastic simulation with input
  uncertainty.
\newblock In {\em 2018 Winter Simulation Conference (WSC)}, pages 1766--1777.
  IEEE, 2018.

\bibitem{zhang2022pooled}
Qiong Zhang, Bo~Wang, and Wei Xie.
\newblock A pooled percentile estimator for parallel simulations.
\newblock {\em Journal of Simulation}, 16(1):73--83, 2022.

\bibitem{Zouaoui_Wilson_2003}
F.~Zouaoui and J.~R. Wilson.
\newblock Accounting for parameter uncertainty in simulation input modeling.
\newblock {\em IIE Transactions}, 35:781--792, 2003.

\bibitem{Zouaoui_Wilson_2004}
F.~Zouaoui and J.~R. Wilson.
\newblock Accounting for input-model and input-parameter uncertainties in
  simulation.
\newblock {\em IIE Transactions}, 36:1135--1151, 2004.

\end{thebibliography}

\appendix


\vspace{0.2in}

\noindent \textbf{\Large Appendix}

In this appendix we prove Theorems \ref{thm:consistency}, \ref{thm:decomposition1}--\ref{thm:decomposition3} and provide a brief description of the experiment design used to build stochastic kriging metamodels. We also use a queue network example to illustrate the proposed framework is general even though it is motivated by the critical needs from biopharmaceutical manufacturing industry.

To be self-contained, we first state some definitions, lemmas and theorems that are used in the proofs. Let $\stackrel{D}\rightarrow$ denote convergence in distribution.
\begin{itemize}

\item \textbf{Borel-Cantelli Lemma} \cite{Billingsley_1995}: For
events $A_1, A_2, \ldots$, if
$\sum_{n=1}^\infty \Pr(A_n)$ converges, then 
\[
\Pr\left(\limsup_n A_n\right)=0 
\]
where 
\[
\limsup_n A_n = \cap_{n=1}^\infty \cup_{k=n}^\infty A_k
\]
is the set of outcomes that occur infinitely many times.

\item \textbf{Lemma 2.11} \cite{Vaart_1998}: Suppose that $\mathbf{X}_n \stackrel{D}\rightarrow \mathbf{X}$ for a random vector $\textbf{X}$ with a continuous distribution function. Then the distribution function of $\textbf{X}_n$ converges uniformly to that of $\textbf{X}$: $\parallel F_{\textbf{X}_n}-F_\textbf{X}\parallel_{\infty}\rightarrow 0$, where $\parallel h\parallel_{\infty}$ is the sup-norm of $h$ on $\Re$, $\parallel h\parallel_{\infty}=\sup_t|h(t)|$.

\item \textbf{Portmanteau Lemma} \cite{Vaart_1998}: For any random vectors $\mathbf{X}_n$ and $\mathbf{X}$ the following statements are equivalent.
\begin{enumerate}

\item $\mathbf{X}_n\stackrel{D}\rightarrow \mathbf{X}.$ 

\item $\mbox{E}[f(\mathbf{X}_n)]\rightarrow\mbox{E}[f(\mathbf{X})]$ for all bounded, continuous functions $f$.

\end{enumerate}

\item \textbf{Theorem 2.3} \cite{Vaart_1998}:  Let $g: \Re^k \rightarrow \Re^m$ be continuous at every point in a set $\mathcal{C}$ such that $\Pr\{X \in \mathcal{C}\} = 1$. Then
\begin{enumerate}

\item If $X_n \stackrel{D}{\longrightarrow} X$ then
$g(X_n) \stackrel{D}{\longrightarrow} g(X)$.

\item If $X_n \stackrel{P}{\longrightarrow} X$ then
$g(X_n) \stackrel{P}{\longrightarrow} g(X)$.

\item If $X_n \stackrel{a.s.}{\longrightarrow} X$ then
$g(X_n) \stackrel{a.s.}{\longrightarrow} g(X)$.

\end{enumerate}
In the proofs when we refer to the ``continuous mapping theorem'' we
will mean Theorem 2.3.

\item \textbf{Glivenko-Cantelli Theorem} \cite{Vaart_1998}: If $X_1,X_2,\ldots,X_n$ are i.i.d. random variables with distribution function $F$ and $F_n$ is the empirical cdf of $X_1,X_2,\ldots,X_n$, then $\parallel F_n-F\parallel_{\infty}\stackrel{a.s.}{\longrightarrow}0$ as $n\rightarrow\infty$.

\item \textbf{Lemma 21.2} \cite{Vaart_1998}: For cdf $F$, define the inverse cdf to be
\[
F^{-1}(p) = \inf \{ t{:}\ F(t) \ge p\}.
\]
Then a sequence of cdfs $F_n(t) \rightarrow F(t)$ for every $t$ where
$F$ is continuous if and only if $F_n^{-1}(p) \rightarrow F^{-1}(p)$
for every $p$ where $F^{-1}$ is continuous.

\item \textbf{Theorem~13.1} \cite{Severini_2005}: Let
$\mathbf{X}_1,\mathbf{X}_2,\ldots$ denote a sequence of
$d$-dimensional random vectors such that, for some vector $\bmu$,
\[
\sqrt{n}(\mathbf{X}_n-\bmu)\stackrel{D}\rightarrow \mbox{N}(\mathbf{0}_{d\times
1},\Sigma) \mbox{  as  } n\rightarrow\infty , 
\]
where $\Sigma$ is a $d \times d$ positive definite matrix with $|\Sigma|<\infty$.
Let $g{:}\ \Re^d\rightarrow\Re^k$ denote a continuously differentiable
function and let $\nabla g(\mathbf{x})$ denote the $d\times k$
matrix of partial derivatives of $g$ with respect to $\mathbf{x}$.
Then
\[ 
\sqrt{n}(g(\mathbf{X}_n)-g(\bmu))\stackrel{D}\rightarrow
\mbox{N}(\mathbf{0}_{k\times 1},\nabla g(\bmu)^\top\Sigma
\nabla g(\bmu) )\mbox{  as  } n\rightarrow \infty.  
\] 

\begin{sloppypar}

\item \textbf{Theorem~3.8} \cite{Shao_1995}: Let $\mathbf{X}_1,\mathbf{X}_2,\ldots,\mathbf{X}_m$ denote $d$-dimensional i.i.d.\ random vectors and
$\bar{\mathbf{X}}_m=m^{-1}\sum_{i=1}^m\mathbf{X}_i$. Let $\bar{\mathbf{X}}_m^*=m^{-1}\sum_{i=1}\mathbf{X}^*_i$ where $\{\mathbf{X}_1^*, \mathbf{X}_2^*, \ldots,\mathbf{X}_m^*\}$ are randomly and independently drawn with replacement from $\{\mathbf{X}_1,\mathbf{X}_2, \ldots,\mathbf{X}_m\}$. Let $g{:}\ \Re^d\rightarrow\Re^k$ denote a continuously differentiable function and $\nabla g(\bx)$ denote the $d\times k$ matrix of partial derivatives of $g$ with respect to $\bx$. Let $T_m=g(\bar{\mathbf{X}}_m)$ and denote the bootstrap variance estimator for $T_m$ by $v_m^*=\mbox{Var}_*[g(\bar{\mathbf{X}}_m^*)]$.

Suppose that $\mbox{E}[ \mathbf{X}_1^\top \mathbf{X}_1 ]<\infty$ and $\nabla g(\bmu)\neq \mathbf{0}_{d\times k}$ where $\bmu=\mbox{E}[\mathbf{X}_1]$.  Suppose further that
\begin{equation}
\label{Condition}
\max_{i_1,\ldots,i_m}|T_m(\mathbf{X}_{i_1},\ldots,
\mathbf{X}_{i_m})-T_m|/\tau_m\stackrel{a.s.}
\rightarrow 0, 
\end{equation}
where the maximum is taken over all integers $i_1,\ldots,i_m$
satisfying $1\leq i_1\leq \cdots\leq i_m\leq m$, and $\{\tau_m\}$ is a
sequence of positive numbers satisfying $\lim\inf_m\tau_m>0$ and
$\tau_m=O(e^{m^q})$ with a $q\in(0,1/2).$ Then $v_m^*$ is strongly
consistent, i.e., $v_m^*/\sigma_m^2\stackrel{a.s.}\rightarrow 1$, where
$\sigma_m^2=m^{-1}\nabla g(\bmu)^\top\Sigma \nabla
g(\bmu)$ and $\Sigma=\mbox{Var}(\mathbf{X}_1).$ 

\end{sloppypar}

\begin{sloppypar}
\item \textbf{Theorem~1.1} (\cite{Lehmann_Casella_1998},
Chapter~6): Let $X_1,X_2,\ldots,X_m$ be i.i.d. with $\mbox{E}(X_1)=\mu$,
$\mbox{Var}(X_1)=\sigma^2$, and finite fourth moment, and suppose $h$
is a function of a real variable whose first four derivatives
$h^\prime(x),h^{\prime\prime}(x),h^{(3)}(x)$ and $h^{(4)}(x)$ exist for all $x\in I$, where
$I$ is an interval with $\Pr(X_1\in I)=1$. Furthermore, suppose
that $|h^{(4)}(x)|\leq M$ for all $x\in I$, for some $M<\infty$. Then
\[ 
\mbox{E}[h(\bar{X})]=h(\mu)+\frac{\sigma^2}{2m}h^{\prime\prime}(\mu)+\mathcal{R}_m .
\]
If, in addition, the fourth derivative of $h^2$ is also bounded, then
\[ 
\mbox{Var}[h(\bar{X})]=\frac{\sigma^2}{m}[h^\prime(\mu)]^2+\mathcal{R}_m. 
\]
In both cases the remainder $\mathcal{R}_m$ is $O(1/m^2)$.
\end{sloppypar}


\item \textbf{Multivariate Taylor Formula} (\cite{Serfling_2002}, 
page~44): Let the function $g$ defined on $\Re^d$ posses continuous partial derivatives of order $n$ at each point of an open set
$S\subset \Re^d$. Let $\bx\in S$. For each point $\mathbf{y}$,
$\mathbf{y}\neq \bx$, such that the line segment $L(\bx,\mathbf{y})$
joining $\bx$ and $\mathbf{y}$ lies in $S$, there exists a point
$\mathbf{z}$ in the interior of $L(\mathbf{x},\mathbf{y})$ such that
\begin{eqnarray}
\lefteqn{
g(\mathbf{y})=g(\bx)+ \left. \sum_{k=1}^{n-1}\frac{1}{k!}\sum_{i_1=1}^d\cdots\sum_{i_k=1}^d
\frac{\partial^kg(t_1,\ldots,t_d)}{\partial t_{i_1}\cdots \partial
t_{i_k}}
\right|_{\mathbf{t}=\bx}\cdot \prod_{j=1}^k(y_{i_j}-x_{i_j}) }
\nonumber \\
&& + \left.\frac{1}{n!}\sum_{i_1=1}^d\cdots\sum_{i_n=1}^d\frac{\partial^n
g(t_1,\ldots,t_d)}{\partial t_{i_1}\cdots \partial t_{i_n}}
\right|_{\mathbf{t}=\mathbf{z}}\cdot\prod_{j=1}^n(y_{i_j}-x_{i_j}). \nonumber
\end{eqnarray}

\end{itemize}

\section{Assumptions for Asymptotic Study}
\label{sec:assumptions}

Assumptions that are needed for the asymptotic analysis of interval $\mbox{CI}_+$ and variance decomposition are summarized below. Assumptions~1--2 give sufficient conditions for the almost sure (a.s.) consistency of bootstrap moment estimators $\widehat{\mathbf{X}}_{\mathbf{m}}\stackrel{a.s.}\rightarrow\bx_c$ as $m\rightarrow\infty$ (see Lemma~1 in \ref{app:consistencyCI+}). Under Assumption~3, a GP $M(\cdot)$ with a correlation function satisfying Condition~(\ref{eq.corrCond}) has continuous sample paths almost surely; see \cite{Adler_2010} Theorem 3.4.1. Condition~(\ref{eq.corrCond}) is satisfied by many correlation functions used in practice, and in particular any power exponential correlation function $r(\bx-\bx^\prime)= \exp(- \sum_{j=1}^d \theta_j|x_j-x_j^\prime|^p )$ with $0<p\leq 2$ and $\theta_j>0$ \cite{Santer_2003}. Assumption~4 indicates that 
process data are collected independently of the simulation model, and that our uncertainty about the mean response surface as represented by $M(\bx)$ is independent of the stochastic simulation noise (although both can depend on $\bx$). Assumptions 5--6 are for the asymptotic consistency study of variance component estimators $\widehat{\sigma}_I^2$ and $\widehat{s}_\ell^2$.

\noindent\textbf{Assumptions~($\star$):}
\begin{enumerate}

\item The $\ell$th model distribution is uniquely determined by its first $h_\ell$ moments and it has finite first $4h_\ell$ moments for $\ell=1,2,\ldots,L$.

\item We have i.i.d observations $Z_{\ell,1}^{(0)},Z_{\ell,2}^{(0)},\ldots,Z_{\ell,m_\ell}^{(0)}$ from the $\ell$th distribution for $\ell=1,2,\ldots,L$. As $m\rightarrow \infty$, we have $m_\ell/m \rightarrow c_\ell$, $\ell=1,2,\ldots,L$, for a constant $c_\ell>0$.

\item The $\epsilon_j(\mathbf{x})\stackrel{i.i.d.}\sim \mbox{N}(0,\sigma^2_{\epsilon}(\mathbf{x}))$ for any $\bx$, and $M(\mathbf{x})$ is a stationary, separable GP with a continuous correlation function satisfying 
\begin{equation}
1-r(\mathbf{x}-\mathbf{x}^\prime)\leq\frac{c}
{|\mbox{log}(\parallel\mathbf{x}-\mathbf{x}^\prime \parallel_2)|^{1+\gamma}}
\mbox{ for all } \parallel\mathbf{x}-\mathbf{x}^\prime \parallel_2\leq\delta 
\label{eq.corrCond}
\end{equation}
for some $c>0$, $\gamma>0$ and $\delta<1$, where $\parallel\mathbf{x}-\mathbf{x}^\prime \parallel_2 =\sqrt{\sum_{j=1}^d({x}_j-{x}^\prime_j)^2}.$ 
\item 
Process observations $Z_{\ell j}^{(0)}$, simulation noise $\epsilon_j(\bx)$ and GP $M(\mathbf{x})$ are mutually independent. The bootstrap process is independent of all of them.

\item [5.] The first three derivatives of the correlation function of the GP $M(\bx)$ exist and the third derivative is bounded.

\item [6.] $m_\ell/m \rightarrow 1$ for $\ell=1,2,\ldots,L$.
\end{enumerate}

\section{Asymptotic Consistency of $\mbox{CI}_+$}
\label{app:consistencyCI+}

To prove Theorem~1, we first establish three supporting lemmas.

\noindent\emph{Lemma} 1. Suppose that Assumptions~1--2 hold. Then
the bootstrap resampled moments converge almost surely to the true
moments $\widehat{\mathbf{X}}_m\stackrel{a.s.}\rightarrow
\mathbf{x}_c$ as $m\rightarrow\infty$.



\noindent \textbf{Proof:} Since all of the input processes are
independent, we establish the result for one input distribution $F^c$
without loss of generality. We prove the result for $\mathbf{x}_c$
being the generic $h$th-order moment, $\alpha_h \equiv \mbox{E}(Z^h) <
\infty$, for $Z\sim F^c$. 

The $h$th-order bootstrap resampled moment is
\begin{equation}
\label{eq:boot.moment}
\widehat{X}_{m}=\frac{1}{m}\sum_{j=1}^m (\textcolor{black}{Z^{(j;m)}})^h \mbox{ with }
\textcolor{black}{Z^{(j;m)}}\stackrel{i.i.d}\sim  \mathbf{Z}_{m}^{(0)}  
\end{equation}
where ``$\textcolor{black}{Z^{(j;m)}}\sim \mathbf{Z}_{m}^{(0)}$"
denotes the $j$th
independent sample with replacement from $\mathbf{Z}_{m}^{(0)}$. We
use the Chebychev Inequality and the Borel-Cantelli Lemma to prove the
result.

By the Chebychev Inequality, for every $\epsilon>0$, we have
\begin{equation}
\label{eq.ChebIneq}
 \mbox{Pr}\left\{|\widehat{X}_m-\alpha_h|>\epsilon \right\}
\leq
\frac{\mbox{E}\left[ (\widehat{X}_m-\alpha_h)^4 \right]}{\epsilon^4}. 
\end{equation}
Notice that
\begin{equation}
\mbox{E}\left[(\widehat{X}_m-\alpha_h)^4\right]
=\mbox{E}\left[\widehat{X}_m^4\right]-4\alpha_h\mbox{E}\left[\widehat{X}_m^3\right]
+6\alpha_h^2\mbox{E}\left[\widehat{X}_m^2\right]-4\alpha_h^3\mbox{E}\left[\widehat{X}_m\right]+\alpha_h^4.
\label{eq.fourM}
\end{equation}
We will analyze each term in Equation~(\ref{eq.fourM}). First, we show
that any $i$th bootstrap resampled moment, denoted as
$\widehat{\alpha}_i$, is unbiased,
\begin{eqnarray}
\label{eq.unbiasedM}
\lefteqn{ \mbox{E}\left[ \widehat{\alpha}_i \right]\equiv\mbox{E}\left[\frac{1}{m}\sum_{j=1}^m(\textcolor{black}{Z^{(j;m)}})^i
\right]} \\ 
&=& \mbox{E}\left[\mbox{E}\left[(\textcolor{black}{Z^{(j;m)}})^i|Z_1^{(0)},\ldots,
Z_m^{(0)}\right]\right] \nonumber \\ 
&=& \mbox{E}\bigg[\frac{1}{m}\sum_{j=1}^m (Z_j^{(0)})^i\bigg]   \nonumber \\ \nonumber
&=& \alpha_i.  \nonumber
\end{eqnarray}
Thus, $\mbox{E}[\widehat{X}_m]=\alpha_h$. Notice that
\begin{eqnarray}
\lefteqn{\mbox{E}\left[ \widehat{X}_m^2 \right]= \mbox{E}\left[\left(\frac{1}{m}\sum_{j=1}^m(\textcolor{black}{Z^{(j;m)}})^h\right)^2\right] } \nonumber \\ \nonumber
&=& \frac{1}{m^2}\mbox{E}\left[\left(\sum_{j=1}^m(\textcolor{black}{Z^{(j;m)}})^h\right)^2\right] \nonumber \\ 
&=& \frac{1}{m^2}\mbox{E}\left[\sum_{j=1}^m(\textcolor{black}{Z^{(j;m)}})^{2h}
+\sum_{i\neq j}(\textcolor{black}{Z^{(i;m)}})^h(\textcolor{black}{Z^{(j;m)}})^h\right]  \nonumber \\
&=&  \frac{1}{m^2}\left(m\alpha_{2h} +m(m-1)\right.
\nonumber \\
&& \left.  \cdot \mbox{E}\left[\mbox{E}[(\textcolor{black}{Z^{(i;m)}})^h | Z_1^{(0)},\ldots,Z_m^{(0)}]\cdot \mbox{E}[(\textcolor{black}{Z^{(j;m)}})^h |Z_1^{(0)},\ldots,Z_m^{(0)}]\right] \right) \nonumber \\ 
&=&  \frac{1}{m^2}\left(m\alpha_{2h}+
m(m-1)
\mbox{E}\bigg[\bigg(\frac{1}{m}\sum_{i=1}^m (Z_i^{(0)})^h\bigg)^2\bigg]\right)  \nonumber\\ 
&=&   \frac{1}{m^2}\left(m\alpha_{2h}+
\frac{m(m-1)}{m^2} \mbox{E}\bigg[\sum_{i=1}^m(Z_i^{(0)})^{2h} + \sum_{i\neq j}(Z_i^{(0)})^h(Z_j^{(0)})^h\bigg]\right)  \nonumber\\ 
&=&  \frac{1}{m^2}\left(m\alpha_{2h}+
\frac{m(m-1)}{m^2}(m\alpha_{2h}+m(m-1)\alpha_h^2) \right)  \nonumber\\
&=& \frac{1}{m^2}[(2m-1)\alpha_{2h}+(m-1)^2\alpha_h^2] \nonumber\\
&=&\frac{2}{m}\alpha_{2h}+\left(1-\frac{2}{m}\right)\alpha_h^2+O(m^{-2}) \nonumber
\end{eqnarray}
where $O(m^{-2})$ means terms at most order $1/m^2$. 
Similar derivations show that 
\begin{eqnarray}
\lefteqn{\mbox{E}\left[\widehat{X}_m^3\right]
=\frac{1}{m^4}\Big([m(4m-3)+(m-1)(m-2)]\alpha_{3h}} \nonumber\\ 
&& +\ [3m(m-1)^2+3(m-1)^2(m-2)]\alpha_h\alpha_{2h}
+(m-1)^2(m-2)^2\alpha_h^3\Big)  \nonumber\\ 
&=& \frac{6}{m}\alpha_h\alpha_{2h}+\left(1-\frac{6}{m}\right)\alpha_h^3+O(m^{-2}) \nonumber 
\end{eqnarray}
and 
\[
\mbox{E}\left[\widehat{X}_m^4\right]=\frac{12}{m}\alpha_h^2\alpha_{2h}
+\left(1-\frac{12}{m}\right)\alpha_h^4+ O(m^{-2}).
\]
Thus,
\begin{eqnarray}
\label{eq.FourCentM}
\lefteqn{
\mbox{E}\left[(\widehat{X}_m-\alpha_h)^4\right] 
= \mbox{E}[\widehat{X}_m^4]-4\alpha_h\mbox{E}[\widehat{X}_m^3]
+6\alpha_h^2\mbox{E}[\widehat{X}_m^2]-4\alpha_h^3\mbox{E}[\widehat{X}_m]
+\alpha_h^4} \\
&=& 
\frac{12}{m}\alpha_h^2\alpha_{2h}
+\left(1-\frac{12}{m}\right)\alpha_h^4
-4\alpha_h\left[\frac{6}{m}\alpha_h\alpha_{2h}
+\left(1-\frac{6}{m}\right)\alpha_h^3\right]
\nonumber\\ 
&& +\ 6\alpha_h^2\left[\frac{2}{m}\alpha_{2h}+\left(1-\frac{2}{m}\right)\alpha_h^2\right]
-3\alpha_h^4+O(m^{-2}) \nonumber\\ 
&=& 0+ O(m^{-2}) \nonumber
\end{eqnarray}
because all of the $O(m^{-1})$ terms cancel.
Therefore, combining Equations~(\ref{eq.ChebIneq}), (\ref{eq.fourM})
and (\ref{eq.FourCentM}), we have
\[ \sum_{m=1}^{\infty}\mbox{Pr}\{|\widehat{X}_m-\alpha_h|>\epsilon\}\leq 
\sum_{m=1}^{\infty}\frac{c}{m^2\epsilon^4} < \infty\] where $c$ is
some finite constant.  Thus, if $\alpha_{4h} <\infty$, then
$\widehat{X}_m\stackrel{a.s.}\rightarrow \alpha_h$ by the
first Borel-Cantelli Lemma in Section~4 of \cite{Billingsley_1995}. 

Since Assumption~2 guarantees $m_\ell\rightarrow\infty$ for each
moment associated with the $\ell$th model, we can generalize
the almost sure convergence to a vector of moments by applying the
converging together lemma. Therefore, we have
$\widehat{\mathbf{X}}_{\mathbf{m}}\stackrel{a.s.}\rightarrow
\mathbf{x}_c$. \blot

\vspace{12pt}

\noindent \textbf{Remark:} The independent variables in our stochastic
kriging metamodel consist of central moments and standardized central
moments.  Since standardized moments are continuous functions of raw
moments, we can use the continuous mapping theorem to obtain
corresponding almost sure convergence of the standardized moments.

\vspace{12pt}

\begin{sloppypar} 
Given a fixed and finite number of design points
$\bx_1,\bx_2,\ldots,\bx_k$, let
$\mathbf{M}=(M(\bx_1),M(\bx_2),\ldots,M(\bx_k))^\top$.  The simulation
error at design point $\bx_i$ is $\epsilon(\bx_i)$, so let
$\bar{\epsilon}(\bx_i)=\sum_{j=1}^{n_i}\epsilon(\bx_i)/n_i$ for
$i=1,2,\ldots,k$ denote the average. Therefore, the sample means of
simulation outputs at all design points can be represented as
$\bar{\mathbf{Y}}_\mathcal{D}=\mathbf{M}+\bar{\boldmath{\epsilon}}$,
where $\bar{\boldmath{\epsilon}}=(\bar{\epsilon}(\bx_1),
\bar{\epsilon}(\bx_2),\ldots,\bar{\epsilon}(\bx_k))^\top$. Finally,
let $M_p(\cdot)$ be a GP having the conditional distribution of
$M(\cdot)$ given $\bar{\mathbf{Y}}_\mathcal{D}$. 
\end{sloppypar}

\vspace{12pt}

\noindent\emph{Lemma} 2. Suppose Assumptions~3--4 hold. Then
$M_p(\cdot)$ has continuous sample paths almost surely.

\noindent \textbf{Proof:} Let $(\Omega_M, P_M)$ be the underlying
probability space for the GP $M(\cdot)$, and $(\Omega_\epsilon, P_\epsilon)$
be the underlying probability space for $\bar{\boldmath{\epsilon}}$. Notice
that $(\Omega_\epsilon, P_\epsilon)$ depends on the particular design
points $\bx_1, \bx_k, \ldots, \bx_k$ and corresponding numbers of
replications $n_1, n_2, \ldots, n_k$ which we consider fixed and
given, while $(\Omega_M, P_M)$ does not.

\begin{sloppypar}
Let $\omega_M \in \Omega_M$ be an elementary outcome and $M(\cdot,
\omega_M)$ the resulting random function. For notational convenience,
let $\mathbf{M}(\omega_M) = \left( M(\bx_1, \omega_M), M(\bx_2,
\omega_M), \ldots, M(\bx_k, \omega_M) \right)^\top$ the random
function evaluated at $\bx_1, \bx_2, \ldots, \bx_k$. Similarly,
$\bar{\boldmath{\epsilon}} =
\bar{\boldmath{\epsilon}}(\omega_\epsilon)$ for elementary outcome
$\omega_\epsilon \in \Omega_\epsilon$. Notice that under Assumption~3,
$\bar{\boldmath{\epsilon}}(\omega_\epsilon)$ has a multivariate normal
distribution.
\end{sloppypar}

Theorem 3.4.1 of \cite{Adler_2010} asserts that there is a
$P_M$-measurable set $\Omega_M^c \subset \Omega_M$ such that $\Pr\{
\omega_M \in \Omega_M^c\} = P_M(\Omega_M^c) = 1$, and for every
$\omega_M \in \Omega_M^c$ the function $M(\cdot, \omega_M)$ is
continuous.


The random variable $\bar{\bY}_\mathcal{D}$ maps $\Omega = \Omega_M
\times \Omega_\epsilon \rightarrow \Re^k$ as
$\bar{\bY}_\mathcal{D}(\omega) = \mathbf{M}(\omega_M) +
\bar{\mathbf{\epsilon}}(\omega_\epsilon)$ for $\omega = (\omega_M,
\omega_\epsilon) \in \Omega$ with probability measure $P = P_M \cdot
P_\epsilon$ since they are independent.  Our goal is to prove that
\begin{equation}
\label{eq:Prob0}
\Pr\{\omega_M \in \bar{\Omega}_M^c | \bar{\bY}_{D}\} = 0
\end{equation}
almost surely.

We know that $0 \le \Pr\{\omega_M \in \bar{\Omega}_M^c | \bar{\bY}_{D}\} \le 1$
with probability $1$. But also
\[
0 = \Pr\{\omega_M \in \bar{\Omega}_M^c \} = 
\E\left[\Pr\{\omega_M \in \bar{\Omega}_M^c | \bar{\bY}_{D}\} \right] .
\]
Therefore, (\ref{eq:Prob0}) must hold. \blot

\vspace{12pt}


\noindent\emph{Lemma} 3. Suppose that Assumptions~1--4 hold.  Then
$M_p(\widehat{\mathbf{X}}_{\mathbf{m}})\stackrel{a.s.} \rightarrow
M_p(\mathbf{x}_c)$ as $m\rightarrow\infty$.

\noindent \textbf{Proof:} Under Assumption~3, the GP $M(\cdot)$ has
continuous sample paths almost surely; applying Lemma~2, $M_p(\cdot)$
also has continuous sample paths almost surely.  Under
Assumptions~1--2,
$\widehat{\mathbf{X}}_{\mathbf{m}}\stackrel{a.s.}\rightarrow\mathbf{x}_c$
as $m\rightarrow \infty$ by Lemma~1. And $M_p(\cdot)$
and $\widehat{\mathbf{X}}_{\mathbf{m}}$ are independent. The result
follows by applying the continuous mapping theorem. \blot

\vspace{12pt}


\noindent\textbf{Theorem~3.1.} Suppose that Assumptions 1--4 hold.  Then the interval $[M_{(\lceil
B\frac{\alpha}{2} \rceil)},M_{(\lceil B(1-\frac{\alpha}{2})\rceil)}]$
is asymptotically consistent, meaning
\begin{equation}
\lim_{m\rightarrow\infty} \lim_{B\rightarrow \infty} 
\Pr\{M_{(\lceil B\alpha/2 \rceil)} \leq M_p(\mathbf{x}_c) \leq M_{(\lceil B(1-\alpha/2)\rceil)} \} = 1-\alpha.
\label{eq.11}
\end{equation} 

\noindent \textbf{Proof:} Define $K_{\mathbf{m}}(t)\equiv
\Pr\left\{ M_p(\widehat{\mathbf{X}}_{\mathbf{m}})\leq t \right\}. $ Notice
that the distribution $K_{\mathbf{m}}(t)$ depends on both the
distributions of $ M_p(\cdot)$ and
$\widehat{\mathbf{X}}_{\mathbf{m}}$. Specifically,
\begin{eqnarray*}
K_{\mathbf{m}}(t) &=& 
	\int \Pr\left\{M_p(\bx) \le t | \widehat{\bX}_\bm = \bx\right\}
	d\widehat{F}_{\bX_\bm}(\bx |\bz_\bm^{(0)}) \\
&=&
\int \Phi\left(
\frac{t - m_p(\bx)}{\sigma_p(\bx)}
\right)
	d\widehat{F}_{\bX_\bm}(\bx |\bz_\bm^{(0)}).
\end{eqnarray*}
Thus, $K_{\mathbf{m}}(t)$ is a continuous distribution almost surely.
Let $\widehat{K}_\bm$ be the empirical cdf of $M_1, M_2, \ldots, M_B$,
which are i.i.d.\ from $K_{\mathbf{m}}(t)$.  Notice that
$\widehat{K}_\bm^{-1}(\gamma) = M_{(\lceil B\gamma\rceil)}$ for
$\gamma = \alpha/2$ and $1 - \alpha/2$.

By the Glivenko-Cantelli Theorem \cite{Vaart_1998}, $||K_\bm -
\widehat{K}_\bm||_\infty \stackrel{a.s.}{\longrightarrow} 0$ as $B
\rightarrow \infty$. 
Therefore, by Lemma~21.2 of \cite{Vaart_1998}, 
\[
| M_{(\lceil B\gamma\rceil)} - K_\bm^{-1}(\gamma) |
\stackrel{a.s.}{\longrightarrow} 0
\]
as $B \rightarrow \infty$ for $\gamma = \alpha/2, 1 - \alpha/2$. As a result,
\begin{eqnarray}
\lefteqn{ \lim_{B\rightarrow \infty} 
\Pr\{M_{(\lceil B\alpha/2 \rceil)} \leq M_p(\mathbf{x}_c) \leq
M_{(\lceil B(1-\alpha/2)\rceil)} \} } 
\nonumber \\
&= &
\Pr\{K_\bm^{-1}(\alpha/2) \leq M_p(\mathbf{x}_c) \leq 
K_\bm^{-1}(1-\alpha/2) \}.
\nonumber 
\end{eqnarray}
Therefore, Equation~(\ref{eq.11}) becomes
\begin{equation}
\lim_{m\rightarrow\infty}\Pr\{K_{\mathbf{m}}^{-1}(\alpha/2)\leq
M_p(\mathbf{x}_c)  \leq K_{\mathbf{m}}^{-1}(1-\alpha/2)\}=1-\alpha.
\label{eq.2}
\end{equation}

To show Equation~(\ref{eq.2}), we only need to show that
\begin{equation}
\lim_{m\rightarrow \infty} \Pr\{K_{\mathbf{m}}^{-1}(\alpha/2) > M_p(\mathbf{x}_c)\}=\alpha/2 \nonumber
\label{eq.3}
\end{equation}
because the proof of the upper bound is similar.


Since, conditional on $\bar{\bY}_\mathcal{D}$, $M_p(\mathbf{x}_c)\sim
\mbox{N}(m_p(\mathbf{x}_c),\sigma_p^2(\mathbf{x}_c))$, the cdf $H(t)\equiv
\Pr\{M_p(\mathbf{x}_c)\leq t\}$ is continuous. By Lemma~3 and
Lemma~2.11 in \cite{Vaart_1998},
\begin{eqnarray}
\lefteqn{\sup_t |\Pr\{M_p(\widehat{\mathbf{X}}_{\mathbf{m}})\leq t \}-\Pr\{M_p(\mathbf{x}_c)\leq t\} |} \nonumber\\ 
&=&\parallel K_{\mathbf{m}}-H \parallel_{\infty}  \rightarrow 0 \mbox{ as } m\rightarrow \infty. \nonumber
\end{eqnarray}
 
Therefore, 
\begin{eqnarray}
\lefteqn{ \Pr\{K_{\mathbf{m}}^{-1}(\alpha/2)>M_p(\mathbf{x}_c)  \}}  \nonumber\\ 
&=& \Pr\{\alpha/2 \geq K_{\mathbf{m}}(M_p(\mathbf{x}_c))  \}  \nonumber\\ 
\label{eq.temp1}
&=& \Pr\{\alpha/2 \geq H(M_p(\mathbf{x}_c))\} + \textcolor{black}{o(1)}  \\ 
&=& \Pr\{M_p(\mathbf{x}_c)\leq H^{-1}(\alpha/2) \} +\textcolor{black}{o(1)} \nonumber\\ 
&=& \alpha/2 + \textcolor{black}{o(1)}. \nonumber
\end{eqnarray}
Equation~(\ref{eq.temp1}) is obtained because 
\[ |K_{\mathbf{m}}(M_p(\mathbf{x}_c))-H(M_p(\mathbf{x}_c))|\leq \parallel K_{\mathbf{m}}-H\parallel_{\infty}\rightarrow 0 \mbox{  as $m\rightarrow \infty$}. \]
Thus, we have
\begin{eqnarray}
\lefteqn{ \lim_{m\rightarrow\infty} \Pr\{ K_{\mathbf{m}}^{-1}(\alpha/2) \leq M_p(\mathbf{x}_c) \leq K_{\mathbf{m}}^{-1}(1-\alpha/2)\} }   \nonumber\\ 
&=& \lim_{m\rightarrow\infty} \Pr \{M_p(\mathbf{x}_c)\leq K_{\mathbf{m}}^{-1}(1-\alpha/2)\}-\lim_{m\rightarrow\infty} \Pr\{M_p(\mathbf{x}_c)<K_{\mathbf{m}}^{-1}(\alpha/2)\}  \nonumber\\ 
&=& (1-\alpha/2)-\alpha/2  \nonumber\\ 
&=& 1-\alpha. \nonumber
\end{eqnarray} \blot

\vspace{12pt}


\section{Asymptotic Analysis of Variance Contribution Estimation}
\label{app:consistencyVarCom}

\noindent\textbf{Theorem~4.1.} Suppose that Assumptions~1--4 hold. Then
the variance component estimators $\widehat{\sigma}^2_M,
\widehat{\sigma}^2_I, \widehat{\sigma}^2_T$, and $\widehat{s}_\ell$ for $\ell=1,2,\ldots,L$ are consistent as 
$m, B\rightarrow\infty$. 



\noindent \textbf{Proof:} When a GP $M(\cdot)$ has a continuous
correlation function with all parameters finite, the SK predictor
\begin{equation}
m_{p}(\bx)=\widehat{\beta}_0
+\tau^2R(\bx)^\top[\Sigma+C]^{-1}(\bar{\bY}_\mathcal{D}-\widehat{\beta}_0\cdot 1_{k\times 1}),
\end{equation}
and corresponding variance
\begin{equation}
\sigma_{p}^2(\bx) = \tau^2-\tau^4R(\bx)^\top[\Sigma+C]^{-1}R(\bx) 
 +\mathbf{\eta}^\top[1_{k\times 1}^\top(\Sigma+C)^{-1}1_{k\times 1}]^{-1}\mathbf{\eta}  \nonumber
\end{equation}
where $R(\bx)^\top=(r(\bx-\bx_1),r(\bx-\bx_2),\ldots,r(\bx-\bx_k))$ and $\eta=1-1_{k\times 1}^\top(\Sigma+C)^{-1}\tau^2R(\bx)$, are continuous and bounded functions of $\bx$. 

By the Strong Law of Large Numbers, the raw moment estimator
$\mathbf{X}_{\mathbf{m}}\stackrel{a.s.}\rightarrow\bx_c$ as
$m\rightarrow \infty$ under Assumptions~1--2. This almost sure
convergence can be extended to central moments and standardized
central moments by the continuous mapping theorem. By applying
the Portmanteau Lemma in \cite{Vaart_1998}, we have 
\[\lim_{m\rightarrow\infty}\sigma_M^2
=\lim_{m\rightarrow\infty} \int \sigma^2_p(\mathbf{x})\, dF^c_{\mathbf{X}_{\mathbf{m}}}(\mathbf{x})
=\lim_{m\rightarrow\infty}\mbox{E}[\sigma^2_p(\mathbf{X}_{\mathbf{m}})]
=\sigma_p^2(\bx_c),  \]
and
\begin{eqnarray}
\lefteqn{\lim_{m\rightarrow\infty}\sigma_I^2
=\lim_{m\rightarrow\infty}\int(m_p(\mathbf{x})-\mu_0)^2\, dF^c_{\mathbf{X}_{\mathbf{m}}}(\mathbf{x})} \nonumber\\ 
&=& \lim_{m\rightarrow\infty}\mbox{E}\left[\big(m_p(\mathbf{X}_{\mathbf{m}})
-\mbox{E}[m_p(\mathbf{X}_{\mathbf{m}})]\big)^2\right] \nonumber\\ 
&=& \big(m_p(\bx_c)-m_p(\bx_c)\big)^2=0  \nonumber
\end{eqnarray} 
where $\mu_0=\int\int \nu \, dF(\nu|\mathbf{x})\,
dF^c_{\mathbf{X}_{\mathbf{m}}}(\mathbf{x})=\int m_p(\mathbf{x})\,
dF^c_{\mathbf{X}_{\mathbf{m}}}(\mathbf{x})$.

Recall that $F(\nu|\bx)$ is a normal distribution
$\mbox{N}(m_p(\bx),\sigma_p^2(\bx))$. Let $g(\bx)\equiv
\int(\nu-\mu_0)^2\, dF(\nu|\bx)$. 
Then 
\begin{eqnarray}
\lefteqn{\lim_{m\rightarrow\infty}\sigma_T^2 = 
\lim_{m\rightarrow\infty}
 \int\int (\nu-\mu_0)^2\, dF(\nu|\mathbf{x})\, dF^c_{\mathbf{X}_{\mathbf{m}}}(\mathbf{x}) } 
 \nonumber\\[6pt]
&=&
\lim_{m\rightarrow\infty}
 \int g(\bx)\, dF^c_{\mathbf{X}_{\mathbf{m}}}(\mathbf{x}).  \nonumber
\end{eqnarray}
However,
\begin{eqnarray*}
g(\bx) &=& 
\int (\nu - m_p(\bx) + m_p(\bx) - \mu_0 )^2\, dF(\nu | \mathbf{x}) \\
&=&
\int (\nu - m_p(\bx))^2\, dF(\nu |\mathbf{x})
 + (m_p(\bx) - \mu_0 )^2  
\\
&& 
+ (m_p(\bx) - \mu_0 ) \int (\nu - m_p(\bx))\, dF(\nu | \mathbf{x}) \\
&=&
\sigma_p^2(\bx) + (m_p(\bx) - \mu_0 )^2   + 0.
\end{eqnarray*}
Since $m_p(\bx)$ and $\sigma_p^2(\bx)$ are continuous and bounded
functions, so is $g(\bx)$. Therefore,  
\[
\lim_{m\rightarrow\infty}\sigma_T^2 = 
 \lim_{m\rightarrow\infty}\mbox{E}[g(\mathbf{X}_{\mathbf{m}})]
= g(\bx_c) =\sigma^2_p(\bx_c)
\]
by applying the Portmanteau Lemma.

Next, we will show consistency of the variance estimators. By Lemma~1, $\widehat{\mathbf{X}}_{\mathbf{m}}\stackrel{a.s.}\rightarrow \bx_c.$
For the metamodel uncertainty estimator, we have
\begin{eqnarray}
\lefteqn{\lim_{m\rightarrow\infty}\lim_{B\rightarrow\infty}\widehat{\sigma}^2_M
=\lim_{m\rightarrow\infty}\lim_{B\rightarrow\infty}\frac{1}{B}
\sum_{b=1}^B\sigma^2_p
(\widehat{\mathbf{X}}_{\mathbf{m}}^{(b)}) }
\nonumber\\ 
&=& \lim_{m\rightarrow\infty}\mbox{E}\Big[\sigma_p^2
(\widehat{\mathbf{X}}_{\mathbf{m}})|\mathbf{Z}_{\mathbf{m}}^{(0)}\Big] 
\nonumber\\ 
&=&\sigma_p^2(\bx_c). \nonumber
\end{eqnarray}
The last step follows by applying the Portmanteau Lemma.

For the input uncertainty estimator, we have
\begin{eqnarray}
\lefteqn{
\lim_{m\rightarrow\infty}\lim_{B\rightarrow\infty}\widehat{\sigma}^2_I
=\lim_{m\rightarrow\infty}\lim_{B\rightarrow\infty}
\frac{B}{B-1}\left[\frac{1}{B}\sum_{b=1}^B m_p^2(
\widehat{\mathbf{X}}_{\mathbf{m}}^{(b)})-\bar{{\mu}}^2\right] } 
\nonumber\\ 
&=& \lim_{m\rightarrow\infty}\bigg(
\mbox{E}\Big[m_p^2(\widehat{\mathbf{X}}_{\mathbf{m}})|\mathbf{Z}_{\mathbf{m}}^{(0)}\Big]
-\mbox{E}^2\Big[m_p(\widehat{\mathbf{X}}_{\mathbf{m}})
|\mathbf{Z}_{\mathbf{m}}^{(0)}\Big] \bigg) \nonumber\\ 
&=& m_p^2(\bx_c)-m_p^2(\bx_c)=0. \nonumber
\end{eqnarray}
The last step follows by applying Lemma~1 and the Portmanteau Lemma.

For the total variance estimator, we have
\begin{eqnarray}
\lefteqn{\lim_{m\rightarrow\infty}\lim_{B\rightarrow\infty}
\widehat{\sigma}^2_T = \lim_{m\rightarrow\infty}\lim_{B\rightarrow\infty} \frac{B}{B-1}\left(\frac{1}{B}\sum_{b=1}^B M_b^2-\bar{M}^2\right)} \nonumber\\ 
&=&\lim_{m\rightarrow\infty}\mbox{E}\big[M_p^2(\widehat{\mathbf{X}}_{\mathbf{m}})
|\mathbf{Z}_{\mathbf{m}}^{(0)}\big] - \lim_{m\rightarrow\infty}\mbox{E}^2\big[M_p
(\widehat{\mathbf{X}}_{\mathbf{m}})|\mathbf{Z}_{\mathbf{m}}^{(0)}\big] 
\nonumber \\ 
\label{eq.Th2Mid1}
&=& \mbox{E}[M_p^2(\bx_c)]-\mbox{E}^2[M_p(\bx_c)]  \\
&=& m_p^2(\bx_c)+\sigma_p^2(\bx_c)-m_p^2(\bx_c) \nonumber\\ 
&=& \sigma^2_p(\bx_c). \nonumber
\end{eqnarray}
By Lemma~3,
$M_p(\widehat{\mathbf{X}}_{\mathbf{m}})\stackrel{a.s.}\rightarrow
M_p(\bx_c) \mbox{ as } m\rightarrow\infty$. Then
Step~(\ref{eq.Th2Mid1}) follows by applying Portmanteau Lemma. 

To show the consistency of $\widehat{s}_\ell$ for $\ell=1,2,\ldots,L$, we first study the cost function $\widehat{c}(\mathcal{J})$ and show it converges for any set $\mathcal{J}$,
\begin{eqnarray}
\lefteqn{
\lim_{m\rightarrow\infty}\lim_{B\rightarrow\infty}\widehat{c}(\mathcal{J})
=\lim_{m\rightarrow\infty}\lim_{B\rightarrow\infty}
\frac{1}{B-1} \sum_{b=1}^B
    \left[ m_p\left(\mathbf{x}_{-\mathcal{J}}^{(0)}, 
    \widehat{\mathbf{X}}_{\mathcal{J}}^{(b)} \right)
    -\bar{m}_{\mathcal{J}} \right]^2 } 
\nonumber\\ 
&=& \lim_{m\rightarrow\infty}\lim_{B\rightarrow\infty}
\frac{B}{B-1}\left[\frac{1}{B}\sum_{b=1}^B  m_p^2\left(\mathbf{x}_{-\mathcal{J}}^{(0)}, 
    \widehat{\mathbf{X}}_{\mathcal{J}}^{(b)} \right)
    -\bar{m}_{\mathcal{J}}^2\right] 
\nonumber\\ 
&=& \lim_{m\rightarrow\infty}\bigg(
\mbox{E}\Big[m_p^2\left(\mathbf{x}_{-\mathcal{J}}^{(0)}, 
    \widehat{\mathbf{X}}_{\mathcal{J}} \right)|\mathbf{Z}_{\mathbf{m}}^{(0)}\Big]
-\mbox{E}^2\Big[m_p\left(\mathbf{x}_{-\mathcal{J}}^{(0)}, 
    \widehat{\mathbf{X}}_{\mathcal{J}} \right)
|\mathbf{Z}_{\mathbf{m}}^{(0)}\Big] \bigg) \nonumber\\ 
&\stackrel{(*)}=& m_p^2(\bx_c)-m_p^2(\bx_c)=0. 
\label{eq.mid111}
\end{eqnarray}
Step (*) follows by applying Lemma~1 and the Portmanteau Lemma. Then, for the Shapley Value based variance estimator, we can show 
\begin{eqnarray}
\lefteqn{
\lim_{m\rightarrow\infty}\lim_{B\rightarrow\infty}\widehat{s}_{\ell} 
= \lim_{m\rightarrow\infty}\lim_{B\rightarrow\infty}\sum_{\mathcal{J}\subset \mathcal{L}/\{\ell\}} \dfrac{(L-|\mathcal{J}|-1)!|\mathcal{J}|!}{L!} \left[ \widehat{c}(\mathcal{J}\cup\{\ell\}) - \widehat{c}(\mathcal{J}) \right] }
\nonumber\\ 
\label{eq.Th2Mid2}
&=& \sum_{\mathcal{J}\subset \mathcal{L}/\{\ell\}} \dfrac{(L-|\mathcal{J}|-1)!|\mathcal{J}|!}{L!} \left[ \lim_{m\rightarrow\infty}\lim_{B\rightarrow\infty}\widehat{c}(\mathcal{J}\cup\{\ell\}) - \lim_{m\rightarrow\infty}\lim_{B\rightarrow\infty}\widehat{c}(\mathcal{J}) \right]
\nonumber 
\\ 
&=& 0. \nonumber
\end{eqnarray}
For the finite number of set $\mathcal{J}$, the last step follows by applying (\ref{eq.mid111}).
\blot

\vspace{12pt}


\noindent\textbf{Theorem~4.2.}
Suppose that Assumptions~1--4 and the
following additional assumptions hold:
\setcounter{enumi}{4}
\begin{enumerate}

\item [5.]The first three derivatives of the correlation function of
the GP $M(\bx)$ exist and the third derivative is bounded; and

\item [6.] $m_\ell/m \rightarrow 1$ for $\ell=1,2,\ldots,L$.

\end{enumerate}
Then $\lim_{m\rightarrow\infty}m\sigma_I^2=\lim_{m\rightarrow\infty}
\lim_{B\rightarrow\infty} m\widehat{\sigma}^2_I=\sigma^2_\mu$ almost
surely, where
$\sigma^2_\mu$ is a positive constant.

\noindent \textbf{Proof:} 
Under Assumptions~1--2, and applying the multivariate central limit
theorem, we have as $m\rightarrow\infty$,
\[
\sqrt{m}(\mathbf{X}_m-\mathbf{x}_c)\stackrel{D}\rightarrow\mbox{N}(\mathbf{0}_{d\times 1},\Lambda)
\]
where $\Lambda$ denotes the $d\times d$ positive definite  asymptotic
variance-covariance matrix of $\bX_{\mathbf{m}}$.

When a GP
$M(\bx)$ has a continuous correlation function with all parameters
finite, the SK predictor 
\begin{equation}
m_{p}(\bx)=\widehat{\beta}_0
+\tau^2R(\bx)^\top[\Sigma+C]^{-1}
(\bar{\bY}_\mathcal{D}-\widehat{\beta}_0\cdot 1_{k\times 1}),
\label{eq.predictor}
\end{equation} 
given the simulation sample mean $\bar{\bY}_\mathcal{D}$, is continuous and bounded. 
Under Assumption~5, the gradient $\nabla m_p(\bx)$ exists and
is
continuous. We will show that $\nabla m_p(\bx)\neq
\mathbf{0}_{d\times 1}$ almost surely. By taking the derivative of
$m_p(\bx)$ in Equation~(\ref{eq.predictor}), we have
\begin{equation}
 \frac{\partial m_{p}(\bx)}{\partial x_j}=\underbrace{
\frac{\partial{R}(\bx)^\top}{\partial
x_j}\tau^2[\Sigma+C]^{-1}}_\mathbf{A}(\bar{\bY}_\mathcal{D}-\widehat{\beta}_0\cdot
1_{k\times 1}).
\end{equation}
Since $\partial{R}(\bx)^\top/\partial x_j=
(\partial{r}(\bx-\bx_1)/\partial x_j,\partial{r}(\bx-\bx_2)/\partial
x_j,\ldots, \partial{r}(\bx-\bx_k)/\partial x_j)\neq
\mathbf{0}_{1\times k}$ and $\tau^2[\Sigma+C]^{-1}$ is positive
definite, $\mathbf{A}$ is a non-zero constant vector. Under
Assumption~3, $\mathbf{A}(\bar{\bY}_\mathcal{D}-\widehat{\beta}_0\cdot
1_{k\times 1})$ is a normal random variable that is equal to 0 with
probability 0. Thus, $\nabla m_p(\bx)\neq \mathbf{0}_{d\times
1}$ almost surely.  Applying Theorem~13.1 in \cite{Severini_2005}, we have
\[\sqrt{m}(m_p(\mathbf{X}_{{m}})-m_p(\mathbf{x}_c))\stackrel{D}
\rightarrow\mbox{N}(0,\sigma^2_{\mu}) \]
where $\sigma_{\mu}^2=\nabla
m_p(\bx_c)^\top\Lambda\nabla m_p(\bx_c)>0$. This establishes
the constant. 

Since $m_p(\cdot)$ is continuous and bounded, there always exists a
finite $M_1>0$ such that $|m_p(\bx)|<M_1$ for all $\bx\in\Re^d$.
Therefore, $\max_{\bx\in\Re^d}|m_p(\bx)-m_p(\bX_{\mathbf{m}})|<2M_1$.
Let $\tau_m=\mbox{e} ^{m^{1/4}}$. Since $2M_1/\tau_m\rightarrow 0$ as
$m\rightarrow\infty$, Condition~(\ref{Condition}) of Theorem~3.8 of
\cite{Shao_1995} holds. Thus, the bootstrap variance estimator is
strongly consistent:
$\lim_{m\rightarrow\infty}\lim_{B\rightarrow\infty}
m\widehat{\sigma}_{I}^2=\sigma_{\mu}^2$ almost surely. 

Next, we will show
$\lim_{m\rightarrow\infty}m\sigma_I^2=\sigma_{\mu}^2$ by proving a
multi-variate version of Theorem~1.1 in \cite{Lehmann_Casella_1998}, 
Chapter~6. Let $L(\bX_{\mathbf{m}},\bx_c)$ denote the line segment
joining $\bX_{\mathbf{m}}$ and $\bx_c$. By the Multivariate Taylor
Formula \cite{Serfling_2002}, 
\begin{eqnarray}
m_p(\bX_{\mathbf{m}}) &= & m_p(\bx_c)+\nabla 
m_p(\bx_c)^\top
(\bX_{\mathbf{m}}-\bx_c)
\nonumber \\
&& +\frac{1}{2}(\bX_{\mathbf{m}}-\bx_c)^\top 
\nabla
^2m_p(\bx_c)(\bX_{\mathbf{m}}-\bx_c)+\mathcal{R}(\bX_{\mathbf{m}},\bx_c).
\nonumber 
\end{eqnarray}
The remainder term 
\[
\mathcal{R}(\bX_{\mathbf{m}},\bx_c)=
\frac{1}{3!}\sum_{i_1=1}^d\sum_{i_2=1}^d\sum_{i_3=1}^d
\left. \frac{\partial^3 
m_p(x_1,\ldots,x_d)}{\partial x_{i_1}\partial x_{i_2}
\partial
x_{i_3}} \right|_{\bx=\bz}\prod_{j=1}^3(X_{\mathbf{m},i_j}-x_{c,i_j})  
\]
where $\bz$ denotes a value in the interior of
$L(\bX_{\mathbf{m}},\bx_c)$, and $X_{\mathbf{m},i}$ and $x_{c,i}$
denote the $i$th components of the vectors $\bX_{\mathbf{m}}$ and
$\bx_c$.  By taking the expectation over both sides, we have 
\begin{equation}
\label{eq.Theorem3_1}
\mbox{E}[m_p(\bX_{\mathbf{m}})]=m_p(\bx_c)+\frac{1}{2}\mbox{E}
\left[
(\bX_{\mathbf{m}}-\bx_c)^\top\nabla^2
m_p(\bx_c)(\bX_{\mathbf{m}}-\bx_c) \right]+ 
\mbox{E}[\mathcal{R}(\mathbf{X}_{\mathbf{m}},\bx_c)] 
\end{equation}
where $\nabla^2$ is the Hessian operator.

We will show that the second and third terms on the RHS of
Equation~(\ref{eq.Theorem3_1}) are $O(m^{-1})$ and $O(m^{-2})$,
respectively, under Assumption~5.  Since all of the input processes
are independent, we establish the result for one input distribution
$F^c$ without loss of generality. 

We prove the result for $\bx_c$ being the generic $h$th-order moment,
$x_{c,h}=\mbox{E}(Z_1^h)<\infty$, and $X_{m,h}=
m^{-1}\sum_{j=1}^mZ_j^h$ for $Z_j\stackrel{\mbox{iid}}\sim F^c$.  

Let $C_{ij}\equiv\frac{1}{2}[\nabla^2 m_p(\bx_c)]_{i,j}$. We
first consider components of the second term on the RHS
of Equation~(\ref{eq.Theorem3_1}).
\begin{eqnarray}
\lefteqn{ \mbox{E}[C_{ij}(X_{m,i}-x_{c,i})(X_{m,j}-x_{c,j})] }
\nonumber \\
&=& C_{ij}\mbox{E}\left[\frac{1}{m}\sum_{k_1=1}^m(Z_{k_1}^i-x_{c,i})\cdot\frac{1}{m}
\sum_{k_2=1}^m(Z_{k_2}^j-x_{c,j})\right]  \nonumber \\
&=& \frac{C_{ij}}{m^2}\mbox{E}\left[\sum_{k=1}^m(Z_k^i-x_{c,i})(Z_k^j-x_{c,j})
+\sum_{k_1\neq k_2}(Z_{k_1}^i-x_{c,i})(Z_{k_2}^j-x_{c,j})\right] \nonumber  \\
&=& \frac{C_{ij}}{m^2}\mbox{E}\left[\sum_{k=1}^m(Z_k^i-x_{c,i})(Z_k^j-x_{c,j})
+0\right]=O\left(\frac{1}{m}\right) \nonumber. 
\end{eqnarray}
The last two steps follow because the $Z_i$ are i.i.d.\ and Assumption~5
holds.  Thus, the second term on the RHS of
Equation~(\ref{eq.Theorem3_1}) is
\begin{eqnarray}
\lefteqn{ \frac{1}{2}\mbox{E}
[(\bX_{\mathbf{m}}-\bx_c)^\top\nabla^2
m_p(\bx_c)(\bX_{\mathbf{m}}-\bx_c)] } 
\nonumber \\
&= &\sum_{i=1}^d\sum_{j=1}^d
\mbox{E}[C_{ij}(X_{m,i}-x_{c,i})(X_{m,j}-x_{c,j})] =
O\left(\frac{1}{m}\right). 
\nonumber 
\end{eqnarray}
Similarly, for the components of the third term of the RHS of
Equation~(\ref{eq.Theorem3_1}), we have 
\begin{eqnarray}
\lefteqn{ D_{ijk}\mbox{E}[(X_{m,i}-x_{c,i})(X_{m,j}-x_{c,j})(X_{m,k}-x_{c,k})] } \nonumber \\
&=& D_{ijk}\mbox{E}\left[\frac{1}{m^3}\sum_{k_1=1}^m(Z_{k_1}^i-x_{c,i})
\cdot\sum_{k_2=1}^m(Z_{k_2}^j-x_{c,j})\cdot\sum_{k_3=1}^m(Z_{k_3}^k-x_{c,k})\right] \nonumber \\
&=& \frac{D_{ijk}}{m^3}\mbox{E}\left[\sum_{k_1=1}^m(Z_{k_1}^i-x_{c,i})(Z_{k_1}^j-x_{c,j})
(Z_{k_1}^k-x_{c,j})+0\right] = O\left(\frac{1}{m^2}\right). \nonumber 
\end{eqnarray}
where
\[
D_{ijk}\equiv \left. \frac{1}{3!}\frac{\partial^3
m_p(x_1,\ldots,x_d)}{\partial x_{i}\partial x_{j}\partial
x_{k}} \right|_{\bx=\bz}.
\]
Again, the last two steps follow because the $Z_i$ are
i.i.d.\ and Assumption~5 holds.  Thus, the third term in
Equation~(\ref{eq.Theorem3_1}) is
\begin{eqnarray}
\lefteqn{ \mbox{E}[\mathcal{R}(\mathbf{X}_{\mathbf{m}},\bx_c)] }
\nonumber \\
&=& 
\sum_{i=1}^d\sum_{j=1}^d\sum_{k=1}^d
D_{ijk}\mbox{E}[(X_{m,i}-x_{c,i})(X_{m,j}-x_{c,j})(X_{m,k}-x_{c,k})]
=O\left(\frac{1}{m^2}\right). 
\nonumber 
\end{eqnarray}

Squaring both sides of Equation~(\ref{eq.Theorem3_1}), we have
\begin{equation}
[\mbox{E}(m_p(\bX_{\mathbf{m}}))]^2=m_p^2(\bx_c)+m_p(\bx_c)
\mbox{E}[(\bX_{\mathbf{m}}-\bx_c)^\top\nabla^2 m_p(\bx_c)
(\bX_{\mathbf{m}}-\bx_c)]
+O\left(\frac{1}{m^2}\right) .
\end{equation}

By repeating the same derivation that results in
Equation~(\ref{eq.Theorem3_1}) but using $m_p^2(\cdot)$ instead of
$m_p(\cdot)$, we obtain 
\begin{eqnarray}
\lefteqn{\mbox{E}[m_p^2(\bX_{\mathbf{m}})]=m_p^2(\bx_c)+\frac{1}{2}
\mbox{E}[(\bX_{\mathbf{m}}-\bx_c)^\top\nabla^2m_p^2(\bx_c)(\bX_{\mathbf{m}}-\bx_c)]
+O\left(\frac{1}{m^2}\right) }\nonumber \\
\label{eq.Theorem3_2}
&=& m_p^2(\bx_c)+\mbox{E}\Big[(\bX_{\mathbf{m}}-\bx_c)^\top\nabla m_p(\bx_c)\nabla m_p(\bx_c)^\top (\bX_{\mathbf{m}}-\bx_c) \nonumber \\
&& +(\bX_{\mathbf{m}}-\bx_c)^\top m_p(\bx_c)\nabla ^2m_p(\bx_c)(\bX_{\mathbf{m}}-\bx_c)\Big]+O\left(\frac{1}{m^2}\right). 
\end{eqnarray}
Then,
\begin{eqnarray}
\lefteqn{\mbox{Var}[m_p(\bX_{\mathbf{m}})]=\mbox{E}[m_p^2(\bX_{\mathbf{m}})]
-\Big(\mbox{E}[m_p(\bX_{\mathbf{m}})]\Big)^2 }\nonumber \\
&=&\mbox{E}[(\bX_{\mathbf{m}}-\bx_c)^\top\nabla m_p(\bx_c)
\nabla m_p(\bx_c)^\top(\bX_{\mathbf{m}}-\bx_c)] + O\left(\frac{1}{m^2}\right) \nonumber \\
\label{eq.Th3_mid1}
&=& \mbox{E}[\nabla m_p(\bx_c)^\top(\bX_{\mathbf{m}}-\bx_c)
(\bX_{\mathbf{m}}-\bx_c)^\top\nabla m_p(\bx_c)]+O\left(\frac{1}{m^2}\right) \\
&=& \frac{1}{m}\nabla m_p(\bx_c)^\top \Lambda\nabla m_p(\bx_c)+O\left(\frac{1}{m^2}\right). \nonumber
\end{eqnarray}
Step~(\ref{eq.Th3_mid1}) follows because $\nabla m_p(\bx_c)^\top(\bX_{\mathbf{m}}-\bx_c)$ is a scalar.
Thus, we have $\lim_{m\rightarrow\infty}m\sigma_I^2=\sigma_{\mu}^2$. \blot


\noindent\textbf{Theorem~4.3.}
Suppose that Assumptions~1--6 hold. 
Then $\lim_{m\rightarrow\infty}ms_{\ell}=\lim_{m\rightarrow\infty}
\lim_{B\rightarrow\infty} m\widehat{s}_{\ell}=\sigma^2_s$ almost
surely, where
$\sigma^2_s$ is a positive constant.

\noindent \textbf{Proof:} 

Following the continuous mapping theorem, we have:
\begin{eqnarray}
\lefteqn{
\lim_{m\rightarrow\infty}\lim_{B\rightarrow\infty}m\widehat{s}_{\ell} 
= \lim_{m\rightarrow\infty}\lim_{B\rightarrow\infty}m\sum_{\mathcal{J}\subset \mathcal{L}/\{\ell\}} \dfrac{(L-|\mathcal{J}|-1)!|\mathcal{J}|!}{L!} \left[ \widehat{c}(\mathcal{J}\cup\{\ell\}) - \widehat{c}(\mathcal{J}) \right] }
\nonumber\\ 
&=& \sum_{\mathcal{J}\subset \mathcal{L}/\{\ell\}} \dfrac{(L-|\mathcal{J}|-1)!|\mathcal{J}|!}{L!} \left[
\underbrace{\lim_{m\rightarrow\infty}\lim_{B\rightarrow\infty}m\widehat{c}(\mathcal{J}\cup\{\ell\})}_\mathbf{\circled{1}} - \underbrace{\lim_{m\rightarrow\infty}\lim_{B\rightarrow\infty}m\widehat{c}(\mathcal{J}) }_\mathbf{\circled{2}}
\right]
\nonumber 
\end{eqnarray}

Therefore, to show the scaled consistency of $\widehat{s}_\ell$ for $\ell = 1, 2, \dots, L$, we need to study the scaled consistency of cost function \circled{1}~: $\widehat{c}(\mathcal{J}\cup\{\ell\})$ and \circled{2}~: $\widehat{c}(\mathcal{J})$.

For $\widehat{c}(\mathcal{J})$, based on the multivariate central limit theorem, we have as $m\rightarrow\infty$,
\[
\sqrt{m}(\mathbf{X}_\mathcal{J}^*-\mathbf{x}_c)\stackrel{D}\rightarrow\mbox{N}(\mathbf{0}_{d\times 1},\Lambda^*)
\]
where $\bX^*_\mathcal{J} = [\bX_{\mathcal{J}}^\top,\bx_{c,-\mathcal{J}}^\top]^\top $, $\Lambda^* = \begin{bmatrix} \Lambda_\mathcal{J} & 0 \\ 0 & 0 \end{bmatrix}$ and
$\Lambda_\mathcal{J}$ denotes sub-matrix of $\Lambda$ with respect to subset $\mathbf{X}_\mathcal{J}$. 


Since the condition of Theorem~13.1 in \cite{Severini_2005} still holds, we have
\[\sqrt{m}(m_p(
\bX^*_\mathcal{J}
)-m_p(\mathbf{x}_c))\stackrel{D}
\rightarrow\mbox{N}(0,\sigma^2_\mathcal{J}) \]
where $\sigma_{\mathcal{J}}^2=\nabla_\mathcal{J}
m_p(\bx_{c})^\top\Lambda_\mathcal{J}\nabla_\mathcal{J} m_p(\bx_{c})>0$, $\nabla_\mathcal{J}$ is gradient with respect to subset $\mathbf{X}_\mathcal{J}$ 
This establishes the constant. 
Moreover, the Condition~(\ref{Condition}) of Theorem~3.8 of
\cite{Shao_1995} holds. Thus, the cost function estimator $\widehat{c}(\mathcal{J})$ is
strongly consistent:
$\lim_{m\rightarrow\infty}\lim_{B\rightarrow\infty}
m\widehat{c}(\mathcal{J})=\sigma_{\mathcal{J}}^2$ almost surely. 

Similarly, we can prove the cost function estimator $\widehat{c}(\mathcal{J}\cup\{\ell\})$ is also
strongly consistent:
$\lim_{m\rightarrow\infty}\lim_{B\rightarrow\infty}
m\widehat{c}(\mathcal{J}\cup\{\ell\})=\sigma_{\mathcal{J}\cup\{\ell\}}^2$ almost surely, where $\sigma_{\mathcal{J}\cup\{\ell\}}^2=\nabla_{\mathcal{J}\cup\{\ell\}}
m_p(\bx_{c})^\top\Lambda_{\mathcal{J}\cup\{\ell\}}\nabla_{\mathcal{J}\cup\{\ell\}} m_p(\bx_{c})>0$, $\nabla_{\mathcal{J}\cup\{\ell\}}$ is gradient with respect to $\mathbf{X}_{\mathcal{J}\cup\{\ell\}}$, and $\Lambda_{\mathcal{J}\cup\{\ell\}}$ is sub-matrix of $\Lambda$ with respect to $\mathbf{X}_{\mathcal{J}\cup\{\ell\}}$. 

Consequently, we have Sharpley Value estimator $\widehat{s}_{\ell}$ is strongly consistent: \begin{equation*}
\lim_{m\rightarrow\infty}\lim_{B\rightarrow\infty}
m\widehat{s}_{\ell} = \sigma_s^2, ~~\text{with} ~~\sigma_s^2 = \sum_{\mathcal{J}\subset \mathcal{L}/\{\ell\}} \dfrac{(L-|\mathcal{J}|-1)!|\mathcal{J}|!}{L!} \left[
\sigma_{\mathcal{J}\cup\{\ell\}}^2 - 
\sigma_{\mathcal{J}}^2 \right]
\end{equation*} almost surely.

Next, we will show:
\begin{align*}
    \lim_{m\rightarrow\infty}ms_{\ell}&=\sigma_s^2 
\end{align*}
First, we need to show $\lim_{m\rightarrow\infty}mc(\mathcal{J}) =\sigma_{\mathcal{J}}^2$.
Let $L(\bX^*_\mathcal{J},\bx_c)$ denote the line segment
joining $\bX^*_\mathcal{J}$ and $\bx_c$. According to the Multivariate Taylor Formula \cite{Serfling_2002}, 
\begin{eqnarray}
\lefteqn{ m_p(\bX^*_\mathcal{J}) = m_p(\bx_c)+\nabla
m_p(\bx_{c})^\top
(\bX^*_\mathcal{J}-\bx_c) }
\nonumber \\
&& +\frac{1}{2}(\bX^*_\mathcal{J}-\bx_c)^\top 
\nabla
^2m_p(\bx_{c})(\bX^*_\mathcal{J}-\bx_c)+\mathcal{R}(\bX^*_\mathcal{J},\bx_c).
\nonumber 
\end{eqnarray}
And the remainder term 
\[
\mathcal{R}(\bX^*_\mathcal{J},\bx_c)=
\frac{1}{3!}\sum_{i_1=1}^{d_\mathcal{J}}\sum_{i_2=1}^{d_\mathcal{J}}\sum_{i_3=1}^{d_\mathcal{J}}
\left. \frac{\partial^3 
m_p(x_1,\ldots,x_d)}{\partial x_{i_1}\partial x_{i_2}
\partial
x_{i_3}} \right|_{\bx=\bz}\prod_{j=1}^3(X_{\mathcal{J},i_j}^*-x_{c,i_j})  
\]
where $d_\mathcal{J}=\sum_{\ell \in \mathcal{J}} h_\ell$, $\bz$ denotes a value in the interior of
$L(\bX^*_\mathcal{J},\bx_c)$, and $X_{\mathcal{J},i}$ and $x_{c,i}$
denote the $i$th components of the vectors $\bX^*_\mathcal{J}$ and
$\bx_c$.  

Following the same procedure of 
 \textbf{Theorem~4.2.}'s proof , we can obtain
\begin{equation*}
[\mbox{E}(m_p(\bX^*_\mathcal{J})]^2=m_p^2(\bx_c)+m_p(\bx_{c})
\mbox{E}[(\bX^*_\mathcal{J}-\bx_c)^\top\nabla^2 m_p(\bx_{c})
(\bX^*_\mathcal{J}-\bx_c)]
+O\left(\frac{1}{m^2}\right) .
\end{equation*}
and  
\begin{align*}
\mbox{E}[m_p^2(\bX^*_\mathcal{J})]
&= m_p^2(\bx_{c})+\mbox{E}\Big[(\bX^*_\mathcal{J}-\bx_c)^\top\nabla m_p(\bx_{c})\nabla m_p(\bx_{c})^\top (\bX^*_\mathcal{J}-\bx_c) \\
&~~ +(\bX^*_\mathcal{J}-\bx_c)^\top m_p(\bx_{c})\nabla ^2m_p(\bx_{c})(\bX^*_\mathcal{J}-\bx_c)\Big]+O\left(\frac{1}{m^2}\right). 
\end{align*}
Then,
\begin{eqnarray}
\lefteqn{\mbox{Var}[m_p(\bX^*_\mathcal{J})]=\mbox{E}[m_p^2(\bX^*_\mathcal{J})]
-\Big(\mbox{E}[m_p(\bX^*_\mathcal{J})]\Big)^2 }\nonumber \\
&=& \frac{1}{m}\nabla m_p(\bx_{c})^\top \Lambda^*\nabla m_p(\bx_{c})+O\left(\frac{1}{m^2}\right). \nonumber \\
&=& \frac{1}{m}\nabla_\mathcal{J} m_p(\bx_{c})^\top \Lambda_\mathcal{J}\nabla_\mathcal{J} m_p(\bx_{c})+O\left(\frac{1}{m^2}\right). \nonumber
\end{eqnarray}
Therefore, we have $\lim_{m\rightarrow\infty}mc(\mathcal{J})=\sigma_\mathcal{J}^2$. Similarly, we obtain $\lim_{m\rightarrow\infty}mc(\mathcal{J}\cup\{\ell\})=\sigma_{\mathcal{J}\cup\{\ell\}}^2$. Finally, by applying continuous mapping theorem, we have 
$\lim_{m\rightarrow\infty}ms_{\ell}=\sigma_s^2$
\blot

\vspace{12pt}

\noindent \textbf{Remark:} The independent variables in our stochastic
kriging metamodel consist of central moments and standardized central
moments, rather than raw moments. However, Theorem~3 can easily be
extended to central and standardized central moments as follows.

Since standardized moments are continuous functions of raw moments,
denoted generically as $g(\cdot)$, we can consider the composite
function $(m_p\circ g)(\cdot)$ and follow steps analogous to those in
the proof of Theorem~3.  Up to the third derivatives we have
\begin{eqnarray*}
(m_p\circ g)^\prime(t) &=& m_p^\prime(g(t))g^\prime(t) \\
(m_p\circ g)^{\prime\prime}(t) &=&
m_p^{\prime\prime}(g(t))[g^\prime(t)]^2+m_p^\prime(g(t))
g^{\prime\prime}(t) \\
(m_p\circ g)^{(3)}(t) &=& 
m_p^{(3)}(g(t))[g^\prime(t)]^3+2m_p^{\prime\prime}(g(t))g^\prime(t)g^{\prime\prime}(t) \\ &&
+m_p^{\prime\prime}(g(t))g^\prime(t)g^{\prime\prime}(t) + m_p^\prime(g(t))g^{(3)}(t). 
\end{eqnarray*}
Let $u$ denote the
mean, $u_i^\prime$ denote the $i$th order raw moment and $u_i$ denote the
$i$th order central moment. Then the first three central moments can
be expressed as functions of raw moments as follows:
\begin{eqnarray*}
u_1 &=& u, \\
u_2 &=& u_2^\prime-u^2 \\
u_3 &=& u_3^\prime-3uu_2^\prime+2u^3.
\end{eqnarray*}
The first three standardized central moments are $u_1,\sqrt{u_2}$ and
$u_3/u_2^{3/2}$. For a non-degenerate distribution, the second central
moment is positive and bounded away from $0$. Thus, the first three
derivatives $g^\prime,g^{\prime\prime},g^{(3)}$ exist and are finite.



\section{Experiment Design}
\label{appendix:ExperimentDesign}

To fit SK metamodels we recommend the experiment design developed in
\cite{barton_nelson_xie_2011} which demonstrated robust performance
over a number of test examples. In this section, we briefly review the
basic methodology; for detailed information please refer to
\cite{barton_nelson_xie_2011}.  

The experiment design is not specified a priori; instead the design
space, denoted by $\mathcal{D}$, depends on the real-world data
$\mathbf{z}_{\mathbf{m}}^{(0)}$ that will eventually be resampled. In
this way the design is adaptive.

At a high level, this is the approach: Generate a large number of
bootstrap samples from the real-world data
$\mathbf{z}_{\mathbf{m}}^{(0)}$ and compute the corresponding sample
moments. Find a regular region that encompasses a large fraction of
this sample; this will be the design space. Generate additional
bootstrap samples to test that the regular region does indeed cover
the desired fraction of the feasible space of sample moments, and
refine if necessary. Once satisfied, embed a space-filling design in
the regular region. These design points correspond to input
distribution moments at which to run simulation experiments to fit the
SK metamodel. We provide some more details below. 

Suppose we are interested in a $(1-\alpha)100\%$ CI; we set
$\alpha=0.05$ in our empirical study. We want the experiment
design to lead to a metamodel that is accurate for moments $\bx$ that are
the most likely bootstrap moment vectors generated from
$\mathbf{z}_{\mathbf{m}}^{(0)}$; by ``likely'' we mean, for instance,
covering $q=99\%>(1-\alpha)100\%=95\%$ of the feasible bootstrap
moments. 

To this end we find an ellipsoid that will contain an
independent bootstrap moment vector obtained by random sampling from
$\mathbf{z}_{\mathbf{m}}^{(0)}$ with probability at least $q$. We then
generate a space-filling experiment design inside this ellipsoid.  The
 procedure for constructing the design is as follows:
\begin{enumerate}

\item Generate $B_0$ bootstrap resamples from
$\mathbf{z}_{\mathbf{m}}^{(0)}$ and compute the corresponding sample
moments to generate a set of sample moments
$D_T=\{\widehat{\mathbf{X}}_{\mathbf{m}}^{(b)}, b=1,2,\ldots,B_0\}$. 

\item Find the smallest ellipsoid $E$ such that it contains the fraction $q$
of the data in $D_T$ when the ellipsoid's center and shape are the sample
mean and covariance matrix, respectively, of the elements of $D_T$.

\item Perform a hypothesis test where the null hypothesis is that a
bootstrap moment will be contained in this ellipsoid with probability
at least $q$. This requires computing the number of bootstrap moment resamples, denoted
by $B_1$, and the constant $c$ that defines the rejection region to
attain the desired Type I error and power for the test. 

\item Generate $B_{1}$ additional independent bootstrap resamples from
$\mathbf{z}_{\mathbf{m}}^{(0)}$ and compute the moments
$\widehat{\mathbf{X}}_{\mathbf{m}}^{(b)},
b=B_0+1,B_0+2,\ldots,B_0+B_1$.  If more than $c$ of these $B_1$
resamples are contained in the ellipsoid, then accept the current $E$
as the design space. Otherwise, add these bootstrap resamples to
$D_T$, let $B_0 \leftarrow B_0 + B_1$ and go to Step~2 to update the
ellipsoid.

\item Generate $k$ space-filling design points in the ellipsoid $E$.
To place design points into this space, we employ an algorithm due to
\cite{SunFarooq_2002}, \S3.2.1, for generating points uniformly
distributed in an ellipsoid. The algorithm first generates the polar
coordinates of a point uniformly distributed in a hypersphere, then
transforms it to Cartesian coordinates, and finally transforms it
again to a point uniformly distributed in an ellipsoid.  The advantage
of this approach is that each element of the initial polar coordinates
are independently distributed, allowing them to be generated
coordinate by coordinate via their inverse cumulative distribution
function. Rather than use randomly chosen points, however, we begin
with a Latin hypercube sample on $(0, 1)^d$.

\item  Assign $n=N/k$ replications to each design point,
where $N$ denotes total computational budget. Together the
transformed Latin hypercube design points and the number of
replications $n$ define the experiment design $\mathcal{D}$.

\end{enumerate}
In our experiments we set Type I error of the hypothesis test to
$0.005$ and its power to $0.95$ when the true probability is $q = 0.97$.

\color{black}

\section{Queueing Network Example}
\label{subsec:queueingNetworkExample}

\begin{figure*}[h!]
{
\centering
\includegraphics[scale=0.25]{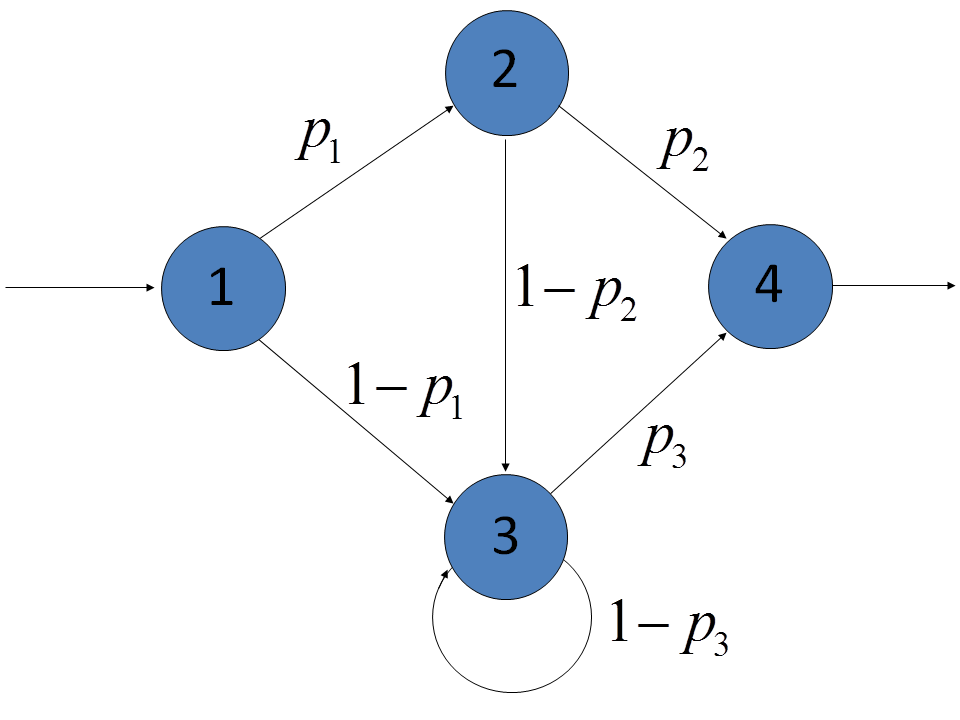}
\vspace{-0.1in}
\caption{Queueing network example.
\label{fig:Figure1_3}}
}
\end{figure*}

In this section we use an queueing network example in Figure~\ref{fig:Figure1_3} to evaluate the performance of our uncertainty analysis framework. 
Consider estimating
the steady state expected number of customers in this network. The interarrival times follow a gamma distribution, $A\sim \mbox{gamma}(\alpha_A,\beta_A)$, and the service times at the $i$th station also follow a gamma distribution, $S_i\sim\mbox{gamma}(\alpha_{S_i},\beta_{S_i})$. Customers finishing service at stations $1,2,3$ must make decisions about their next station. These routing decisions follow Bernoulli distributions $P_i\sim\mbox{Ber}(p_i), i=1,2,3$. The parameters of the input distributions, $\alpha_A,\beta_A$, $\{(\alpha_{S_i},\beta_{S_i}),i=1,2,3,4\}$ and $\{p_i,i=1,2,3\}$ are all unknown and estimated from real-world data. 
Our goal is to 
estimate the steady-state expected number of customers in the system when the input parameters assume their true but unknown values. 

Both interarrival and service times follow gamma distributions and the routing decisions follow Bernoulli distributions. Thus, it is a 13-dimensional problem with $L=8$ input processes that include both continuous and discrete distributions. The true model parameters 
are $\alpha_A=1,\beta_A=0.25$, $\alpha_{S_i}=1$, $\beta_{S_i}=0.2$ for $i=1,2,3,4$ and $p_1=p_2=0.5, p_3=0.75$. These parameter values imply a tractable Jackson network with steady-state number of customers in system $\mu(\bx_c) = 12.67$. 

In the experiments we assume that all input model parameters 
are unknown and are estimated from a finite sample of real-world data. Notice that $\alpha_A,\beta_A$, $\alpha_{S_i}$, $\beta_{S_i}$ for $i=1,2,3,4$ are estimated from continuous measurements, while the routing probabilities $p_1,p_2,p_3$ are estimated from 0 or 1 observations that would correspond to customer routing decisions.  The model with \textit{estimated} input parameters is almost surely not a Jackson network and it could be unstable. Our measure of uncertainty is a $95\%$ CI for $\mu(\bx_c)$ \textcolor{black}{as defined by (\ref{eq.CI_1}) because this is the objective desired in practice.}

To evaluate the robustness of the metamodel-assisted bootstrapping approach, we systematically examine the effect of the quantity of
real-world data and the number of design points and replications per design point used to fit the metamodel; 
We consider a wide range for the quantity of real-world data $m=50,500,5000$, letting $m_\ell=m$ for
$\ell=1,2,\ldots L$. The levels for the number of design points are $k=20,40,80,130$. For a 13-dimensional problem $k=20$ is a very small
design. The studies by \cite{Jones_1998} and \cite{Loeppky_2009} recommend that the number of design points should be $10$ times the
dimension of the problem for kriging; we take this as the maximum number of design points. The same number of replications are assigned
to all design points and we try $n=10,50,100$.  


\cite{barton_nelson_xie_2011} demonstrated that $\mbox{CI}_0$ has good performance when the impact of metamodel or simulation uncertainty is negligible. In this empirical study we focus on situations where metamodel uncertainty may be significant. However, rather than creating a problem that actually takes hours or days to run, we instead construct a problem with high metamodel uncertainty by using short run lengths for each replication: 20 time units after the warm up, which is roughly equivalent to 80 finished customers. To avoid the influence from initial bias, all simulations start loaded with the number of customers at each station being their steady-state expected values (rounded) under $\bx_c$.  Furthermore, a long warmup period of 200 time units is used. The net effect is that the point estimators of the steady-state number in the network have low bias, but may be quite variable.


\subsection{Performance of CIs}
\label{subSec:empirical_CI}

A fundamental assumption of simulation is that the expectation $\mu(\bx_c)$ exists. This assumption does not imply, however, that it exists for \textit{all} possible values of $\bx$, $\bX_\bm$ or $\widehat{\bX}_\bm^{(b)}$ that might be realized. The prototype example is a congestion-related performance measure of a queueing system as time goes to infinity when congestion increases without bound for some values of its interarrival-time and service-time parameters. We refer to systems for which $\mu(\bx)$ is $\pm \infty$ for some values of $\bx$ as potentially \textit{unstable}. 
The conditional
probability that a bootstrap resampled moment
$\widehat{\mathbf{X}}_{\mathbf{m}}^{(b)}$ is located in the unstable
region, denoted by $U$, given the real-world data is
\begin{equation}
 P_U \equiv \mbox{Pr}\left\{\left.
\widehat{\mathbf{X}}_{\mathbf{m}}^{(b)} \in U \right| \mathbf{z}_{\mathbf{m}}^{(0)}
\right\}.
\label{eq.infP}
\end{equation} 
Since $P_U$ only depends on $m$ and $\bx_c$, we ran a side experiment to estimate it using
\begin{equation}
\label{eq.infP_hat}
\widehat{P}_U=\frac{1}{B}\sum_{b=1}^B\mbox{I}
\left(\widehat{\mathbf{X}}_{\mathbf{m}}^{(b)} \in U\right),
\end{equation}
where $\mbox{I}(\cdot)$ is the indicator function. The means and standard deviations (SD) of $\widehat{P}_U$ for $m=50,500,5000$ were
estimated based on 1000 macro-replications and are displayed in Table~\ref{table:infP}. In each macro-replication we independently
generated a sample of size $m$ of ``real-world data.''  Then, conditional on these data, we drew $B=2000$ bootstrap resampled
moments.

\begin{table}[] 
\small
\caption{Percentage of unstable bootstrap resampled moments.}
\label{table:infP}
\begin{center}
\begin{tabular}{|c|c|c|c|} 
\hline
& $m=50$ & $m=500$ & $m=5000$  \\
 \hline
mean of $\widehat{P}_U$  & 44.4\% & 2.3\% & 0 \\ \hline
SD of $\widehat{P}_U$  & 31.7\% & 7.9\% & 0 
\\ \hline
\end{tabular}
\end{center}
\end{table}

As $m$ increases the bootstrap resampled moments become more closely centered around $\bx_c$.  Thus, both the mean and SD of $\widehat{P}_U$ decrease with increasing $m$ as shown in Table~\ref{table:infP}. When $m=50$, ${P}_U$ appears to be much larger
than $\alpha/2$ so the bootstrap moments $\widehat{\mathbf{X}}_{\mathbf{m}}^{(b)}$ that correspond to the upper confidence bound are located in the unstable region $U$ with high probability. When $m=500$, ${P}_U$ appears to be close to $\alpha/2=2.5\%$, while when $m=5000$ there is little chance of getting unstable bootstrap moments.

Tables~\ref{table:HDm50}--\ref{table:HDm500-5000} show the results for $\mbox{CI}_0$ and $\mbox{CI}_+$ when $m=50,500,5000$, including the probability of covering $\mu(\bx_c)$, and the mean and SD of the interval widths. All results are based on $1000$ macro-replications. 
When $m=50$, $P_U$ is much greater than $\alpha/2$ according to Table~\ref{table:infP}. This explains the very large CI widths in Table~\ref{table:HDm50}. Nevertheless, both $\mbox{CI}_0$ and $\mbox{CI}_+$ have reasonable coverage overall.
Notice that $\mbox{CI}_0$ does exhibit undercoverage when we use a very small experiment design of $k=20$ points, while the coverage of $\mbox{CI}_+$ is much closer to the nominal value of $95\%$ in this case. If we fix the number of replications $n$ and increase the number of design points $k$, the coverage of $\mbox{CI}_0$ improves. For a fixed $k$ the effect of increasing $n$ is not as obvious.

Table~\ref{table:HDm500-5000} shows the results for $m=500,5000$. Compared with the results for $m=50$, the mean and SD of the interval
widths drop dramatically.  The effects of $k$ and $n$ are easier to
discern especially when $m=5000$, which has no unstable bootstrap
moments.  Specifically, for a fixed quantity of real-world data $m$,
if either the number of design points $k$ or replications per design
point $n$ is small then $\mbox{CI}_0$ tends to have undercoverage
because it fails to account for substantial simulation uncertainty.
The most troubling observation about $\mbox{CI}_0$ is that, for fixed
$(n,k)$, as the amount of input data $m$ increases its undercoverage becomes more serious.  The diminished coverage occurs because as $m\rightarrow\infty$ the width of $\mbox{CI}_0$ shrinks to zero, which
is not appropriate when there is still simulation uncertainty. 
As the interval, $\mbox{CI}_+$, is able to account for the effect of the remaining simulation estimation error, it can work under
more general situations where the simulated systems are complex and
the simulation budget is tight.
As $n$ and $k$
increase, the coverages of $\mbox{CI}_0$ and $\mbox{CI}_+$ become
closer to each other.


\begin{table}[] 
\small
\caption{Results for $\mbox{CI}_{0}$, $\mbox{CI}_{+}$ and $\widehat{\sigma}_I/\widehat{\sigma}_T$
when $m=50$. } \label{table:HDm50}
\begin{center}
\begin{tabular}{|c|c|c|c|c|c|c|} 
\hline
$m=50$ & \multicolumn{3}{c|}{$k=20$} & \multicolumn{3}{c|}{$k=40$} \\ \cline{2-7}
 &  $n=10$ & $n=50$  & $n=100$ & $n=10$ & $n=50$  & $n=100$  \\
 \hline
Coverage of $\mbox{CI}_0$ & 91.9\% & 92.3\% & 91.5\% & 93.8\%  & 94.4\% & 93.4\%\\ \hline
Coverage of $\mbox{CI}_{+}$ & 93.9\% & 94.9\% & 93.7\% & 94.9\% & 95.6\% & 95.9\% \\ \hline
$\mbox{CI}_0$ Width (mean)& 326.4 & 332.4 & 339.5 & 319.1 & 328.6 & 326.5 
\\ \hline 
$\mbox{CI}_{+}$ Width (mean) & 344.1 & 348.8 & 357.1 & 332.3 & 342.3 & 341.2 
\\ \hline 
$\mbox{CI}_0$ Width (SD) & 183.1 & 173.6 & 180.7 & 176.4 & 167.6 & 175  
\\ \hline 
$\mbox{CI}_{+}$ Width (SD) & 188 & 175.7 & 183.8 & 178.2 & 169.2 & 176.1   
\\ \hline 
$\widehat{\sigma}_I/\widehat{\sigma}_T$ & 0.963 & 0.965 & 0.964 & 0.973 & 0.973 & 0.971
\\ \hline\hline
$m=50$ & \multicolumn{3}{c|}{$k=80$} & \multicolumn{3}{c|}{$k=130$} \\ \cline{2-7}
 &  $n=10$ & $n=50$  & $n=100$ & $n=10$ & $n=50$  & $n=100$  \\
 \hline
Coverage of $\mbox{CI}_0$ & 94.6\% & 96.3\% & 95.4\% & 94.2\%  & 95.1\% & 95.4\%\\ \hline
Coverage of $\mbox{CI}_{+}$ & 95.9\% & 96.7\% & 96.1\% & 94.5\% & 96\% & 96.1\% \\ \hline
$\mbox{CI}_0$ Width (mean) & 312.1 & 314.8 & 322.7 & 322 & 321.86 & 320
\\ \hline  
$\mbox{CI}_{+}$ Width (mean) & 322 & 325.7 & 334 & 330.2 & 331 & 329.4
\\ \hline 
 $\mbox{CI}_0$ Width (SD) & 169.7 & 159.1 & 164.7 & 171.5 & 169.3 & 172.3  
\\ \hline 
$\mbox{CI}_{+}$ Width (SD) & 171.2 & 159.4 & 165 & 172.7 & 169.5 & 172.7 
\\ \hline
$\widehat{\sigma}_I/\widehat{\sigma}_T$ & 0.982 & 0.98 & 0.978 & 0.985 & 0.985 & 0.983
\\ \hline
\end{tabular}
\end{center}
\end{table}

\begin{table}[] 
\caption{Results of the queueing network example for $\mbox{CI}_{0}$, $\mbox{CI}_{+}$ and $\widehat{\sigma}_I/\widehat{\sigma}_T$
when $m=500$ and $m=5000$. }
\small
\label{table:HDm500-5000}
\begin{center}
\begin{tabular}{|c|c|c|c|c|c|c|} 
\hline
$m=500$ & \multicolumn{3}{c|}{$k=20$} & \multicolumn{3}{c|}{$k=40$} \\ \cline{2-7}
 &  $n=10$ & $n=50$  & $n=100$ & $n=10$ & $n=50$  & $n=100$  \\
 \hline
Coverage of $\mbox{CI}_0$ & 90.5\% & 94.6\% & 95.1\% & 94.9\%  & 96.7\% & 96.4\%\\ \hline
Coverage of $\mbox{CI}_{+}$ & 95.7\% & 97.7\% & 97.8\% & 96.6\% & 98.3\% & 97.8\% \\ \hline
$\mbox{CI}_0$ Width (mean)& 24.8 & 28.1 & 29.4 & 27.1 & 28.5 & 28.7
\\ \hline 
$\mbox{CI}_{+}$ Width (mean) & 28.9 & 30.8 & 32.2 & 29.6 & 30.3 & 30.5
\\ \hline
$\mbox{CI}_0$ Width (SD) & 19.9 & 19.4 & 20.6 & 19.1 & 19.2 & 19.9 \\\hline
$\mbox{CI}_+$ Width (SD) & 20.6 & 20.4 & 21.7 & 19.7 & 19.9 & 20.6 \\ \hline 
$\widehat{\sigma}_I/\widehat{\sigma}_T$ & 0.88 & 0.932 & 0.933 & 0.932 & 0.957 & 0.958
\\ \hline\hline
$m=500$ & \multicolumn{3}{c|}{$k=80$} & \multicolumn{3}{c|}{$k=130$} \\ \cline{2-7}
 &  $n=10$ & $n=50$  & $n=100$ & $n=10$ & $n=50$  & $n=100$  \\
 \hline
Coverage of $\mbox{CI}_0$ & 96.5\% & 97.5\% & 95.8\% & 95.4\%  & 96.5\% & 95.9\%\\ \hline
Coverage of $\mbox{CI}_{+}$ & 98\% & 98.3\% & 97.3\% & 97.5\% & 97.1\% & 96.9\% \\ \hline
$\mbox{CI}_0$ Width (mean) & 26.3 & 28 & 28.7 & 26.4 & 27.9 & 27.6
\\ \hline 
$\mbox{CI}_{+}$ Width (mean) & 28 & 29 & 29.7 & 27.9 & 28.6 & 28.2
\\ \hline
$\mbox{CI}_0$ Width (SD) & 17.4 & 18 & 19.3 & 18.8 & 19.6 & 19.3 \\ \hline
$\mbox{CI}_+$ Width (SD) &  17.7 & 18.4 & 19.6 & 18.9 & 19.9 & 19.5 \\ \hline
$\widehat{\sigma}_I/\widehat{\sigma}_T$ & 0.952 & 0.977 & 0.978 & 0.957 &0.984 & 0.987
\\ \hline\hline\hline
 $m=5000$ & \multicolumn{3}{c|}{$k=20$} & \multicolumn{3}{c|}{$k=40$} \\ \cline{2-7}
 &  $n=10$ & $n=50$  & $n=100$ & $n=10$ & $n=50$  & $n=100$  \\
 \hline
Coverage of $\mbox{CI}_0$ & 70.7\% & 89.2\% & 93.1\% & 81.5\%  & 94.3\% & 94.8\%\\ \hline
Coverage of $\mbox{CI}_{+}$ & 91.3\% & 96.3\% & 95.6\% & 96.5\% & 96.1\% & 96.3\% \\ \hline
$\mbox{CI}_0$ Width (mean)& 3.29 & 3.97 & 4.14 & 3.93 & 4.23 & 4.3
\\ \hline 
$\mbox{CI}_{+}$ Width (mean) & 5.85 & 4.8 & 4.56 & 6.08 & 4.64 & 4.52
\\ \hline 
$\mbox{CI}_0$ Width (SD) & 1.89 & 1.2 & 1 & 1.64 & 0.87 & 0.83 \\ \hline
$\mbox{CI}_+$ Width (SD) & 2.12 & 1.13 & 1 & 1.52 & 0.89 & 0.85 \\ \hline
$\widehat{\sigma}_I/\widehat{\sigma}_T$ & 0.588 & 0.85 & 0.924 & 0.664 & 0.924 & 0.959
\\ \hline\hline
$m=5000$ & \multicolumn{3}{c|}{$k=80$} & \multicolumn{3}{c|}{$k=130$} \\ \cline{2-7}
 &  $n=10$ & $n=50$  & $n=100$ & $n=10$ & $n=50$  & $n=100$  \\
 \hline
Coverage of $\mbox{CI}_0$ & 88.9\% & 93.6\% & 94.9\% & 89.5\%  & 93.7\% & 94.8\%\\ \hline
Coverage of $\mbox{CI}_{+}$ & 98.1\% & 95\% & 96\% & 98\% & 95.6\% & 95.5\% \\ \hline
$\mbox{CI}_0$ Width (mean) & 4.54 & 4.29 & 4.29 & 4.52 & 4.35 & 4.32
\\ \hline 
$\mbox{CI}_{+}$ Width (mean) & 6.1 & 4.56 & 4.42 & 5.98 & 4.64 & 4.45
\\ \hline
$\mbox{CI}_0$ Width (SD) & 1.37 & 0.85 & 0.77 & 1.28 & 0.9 & 0.79 \\ \hline
$\mbox{CI}_+$ Width (SD) & 1.27 & 0.85 & 0.78 & 1.13 & 0.87 & 0.77 \\ \hline
$\widehat{\sigma}_I/\widehat{\sigma}_T$ & 0.757 & 0.946 & 0.974 & 0.766 & 0.945 & 0.974
\\ \hline
\end{tabular}
\end{center}
\end{table}

\subsection{Performance of $\widehat{\sigma}_I/\widehat{\sigma}_T$}
\label{subSec:empirical_Ratio}

\begin{figure*}[tb]
\vspace{0.2in}
{
\centering
\includegraphics[scale=0.5]{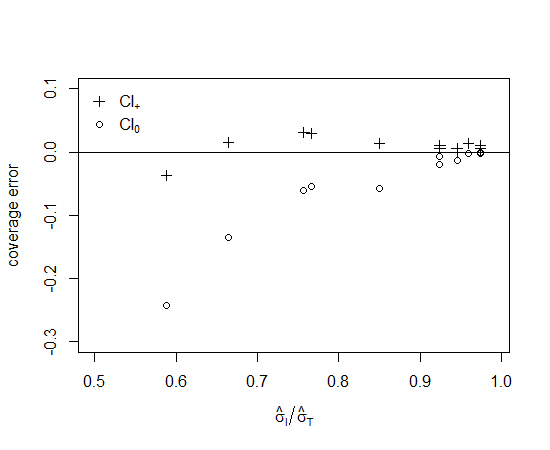}
\vspace{-0.2in}
\caption{The coverage errors for $\mbox{CI}_0$ and $\mbox{CI}_+$ 
vs.\ $\widehat{\sigma}_I/\widehat{\sigma}_T$ when $m=5000$ across
all values of $n$ and $k$.
\label{fig:coverageErr_ratio_m5000}}
}
\end{figure*}

Tables~\ref{table:HDm50}--\ref{table:HDm500-5000} also
demonstrate that $\widehat{\sigma}_I/\widehat{\sigma}_T$ provides a good measure of the relative contribution of model uncertainty to overall uncertainty.
For a fixed amount of real-world data $m$, increasing the number of design points and replications $(n,k)$ drives $\widehat{\sigma}_I/\widehat{\sigma}_T$ toward 1, indicating a decrease in simulation uncertainty.
For fixed simulation effort $(n,k)$, increasing the amount of real-world data $m$ decreases $\widehat{\sigma}_I/\widehat{\sigma}_T$, indicating that there is relatively less model uncertainty. Notice, however, that the relationship is not simple because as $m$ increases the design space over which we fit the metamodel becomes smaller, so that even with the same simulation effort the absolute level of simulation uncertainty will decrease somewhat. 
When $\widehat{\sigma}_I/\widehat{\sigma}_T$ is near $1$, the behaviors (coverage and width) of $\mbox{CI}_0$ and $\mbox{CI}_{+}$ are similar and both have coverage close to the nominal level; this is illustrated in Figure~\ref{fig:coverageErr_ratio_m5000}. Recall that $\mbox{CI}_0$ does not account for simulation uncertainty, and that $\widehat{\sigma}_I/\widehat{\sigma}_T \approx 1$ indicates that model uncertainty is large relative to simulation uncertainty, which is when $\mbox{CI}_0$ will do best. Figure~\ref{fig:coverageErr_ratio_m5000} also illustrates the general robustness of $\mbox{CI}_+$.

\end{document}